\documentclass[sigconf]{acmart}
\usepackage{subfigure}
\usepackage{color}
\usepackage{multirow}
\usepackage{balance}
\graphicspath{{./jpg/}}
\AtBeginDocument{%
  \providecommand\BibTeX{{%
    \normalfont B\kern-0.5em{\scshape i\kern-0.25em b}\kern-0.8em\TeX}}}

\copyrightyear{2021}
\acmYear{2021}
\setcopyright{acmcopyright}\acmConference[MM '21]{Proceedings of the 29th ACM
International Conference on Multimedia}{October 20--24, 2021}{Virtual Event, China}
\acmBooktitle{Proceedings of the 29th ACM International Conference on Multimedia
(MM '21), October 20--24, 2021, Virtual Event, China}
\acmPrice{15.00}
\acmDOI{10.1145/3474085.3475408}
\acmISBN{978-1-4503-8651-7/21/10}

\begin{document}
\fancyhead{}

\title{Perception-Oriented Stereo Image Super-Resolution}


\author{Chenxi Ma}
\affiliation{
  \institution{Shanghai Key Laboratory of Intelligent Information Processing, School of Computer Science, Fudan University} %
  \city{}
  \country{}
}
\email{17210240039@fudan.edu.cn}

\author{Bo Yan}
\authornote{Corresponding Author. This work is supported by NSFC (Grant No.: U2001209, 61902076, 61772137) and Natural Science Foundation of Shanghai (21ZR1406600).}
\affiliation{
  \institution{Shanghai Key Laboratory of Intelligent Information Processing, School of Computer Science, Fudan University} %
  \city{}
  \country{}
}
\email{byan@fudan.edu.cn}

\author{Weimin Tan}
\affiliation{
  \institution{Shanghai Key Laboratory of Intelligent Information Processing, School of Computer Science, Fudan University} 
  \city{}
  \country{}
}
\email{wmtan14@fudan.edu.cn}

\author{Xuhao Jiang}
\affiliation{
  \institution{Shanghai Key Laboratory of Intelligent Information Processing, School of Computer Science, Fudan University} 
  \city{}
  \country{}
}
\email{20110240011@fudan.edu.cn}


\renewcommand{\shortauthors}{Chenxi, et al.}

\begin{abstract}
  Recent studies of deep learning based stereo image super-resolution (StereoSR) have promoted the development of StereoSR. However, existing StereoSR models mainly concentrate on improving quantitative evaluation metrics and neglect the visual quality of super-resolved stereo images. To improve the perceptual performance, this paper proposes the first perception-oriented stereo image super-resolution approach by exploiting the feedback, provided by the evaluation on the perceptual quality of StereoSR results. To provide accurate guidance for the StereoSR model, we develop the first special stereo image super-resolution quality assessment (StereoSRQA) model, and further construct a StereoSRQA database. Extensive experiments demonstrate that our StereoSR approach significantly improves the perceptual quality and enhances the reliability of stereo images for disparity estimation.
\end{abstract}




\maketitle

\section{Introduction}
Stereo image super-resolution (StereoSR), aiming to increase the spatial resolution of stereo image pairs, becomes a growing research direction and benefits a lot from the deep-learning. Though remarkable advances have been achieved, convolutional neural network (CNN) based StereoSR methods~\cite{stereosr, passr, pscassr, our} are mainly designed to improve the common objective evaluation metrics (peak signal to noise ratio PSNR, structural similarity index SSIM~\cite{ssim}). These StereoSR models are trained with the pixel-level mean square error (MSE) or mean absolute error (MAE) loss. It is well known that MSE/MAE loss forces the restored pixels to be the average of all possible solutions, which leads to high PSNR/SSIM accompanied by poor perceptual performance. 
\begin{figure}[tb]
\centering
\subfigure{
    \begin{minipage}[t]{4.20cm}  \centering  \includegraphics[width=4.20cm]{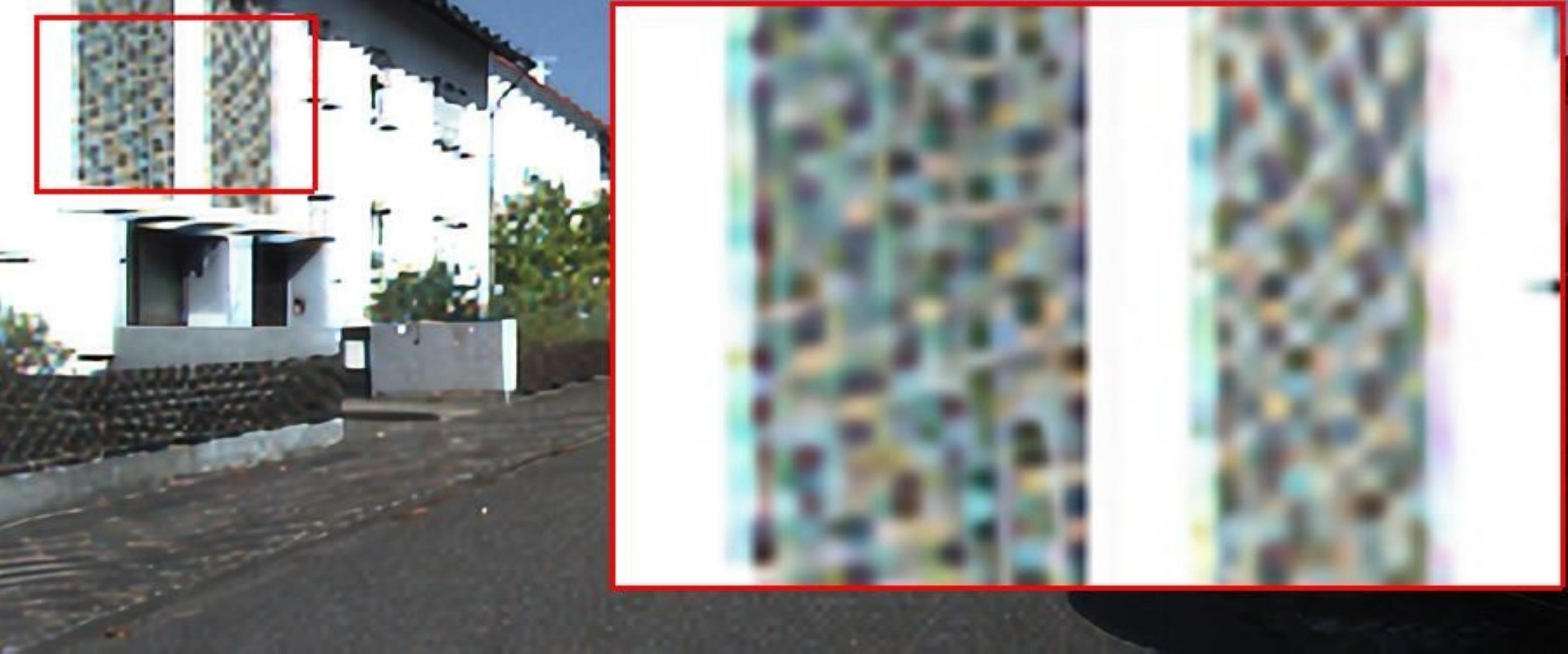} \\ \centering{\small{MSE-based SISR (DRN)}} \end{minipage}
\begin{minipage}[t]{4.20cm}  \centering \includegraphics[width=4.20cm]{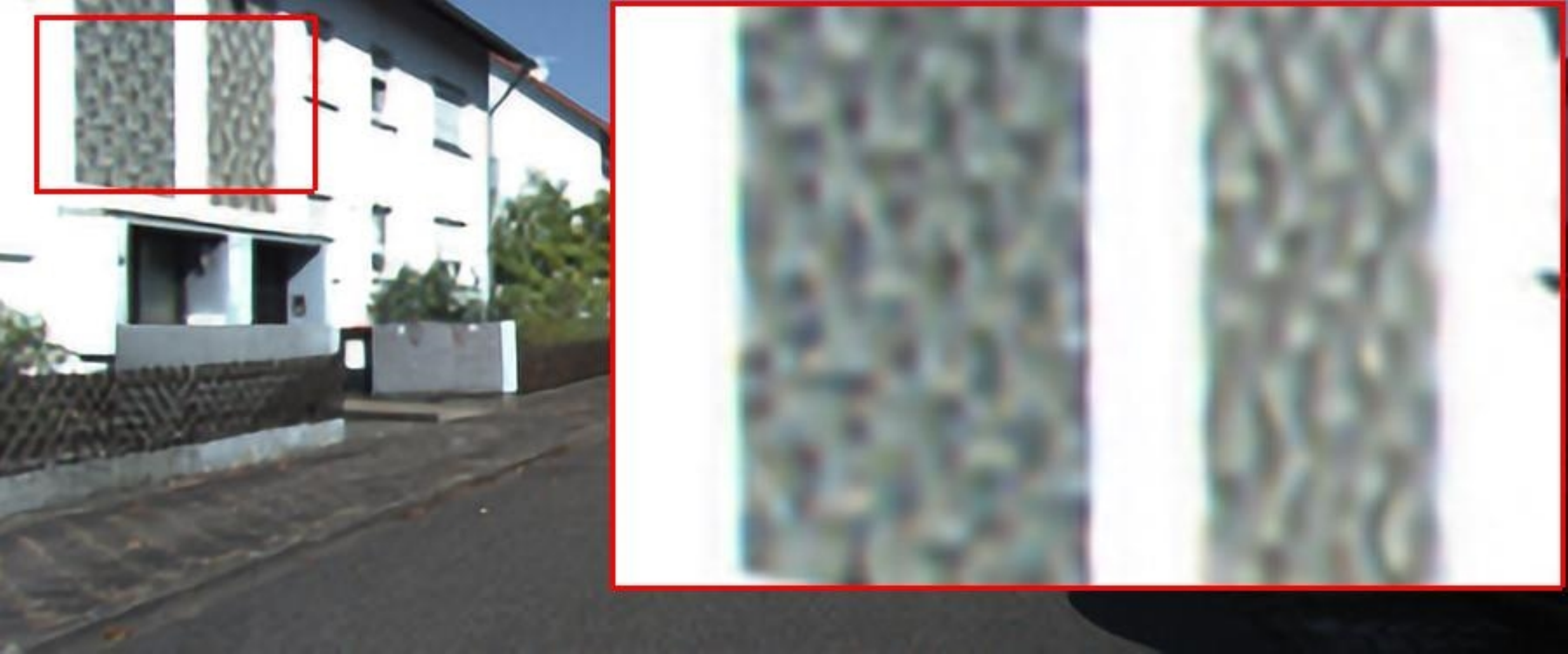} \\ \centering{\small{StereoSR (PASSR)}} \end{minipage}
 }\
\centering
\subfigure{
     \begin{minipage}[t]{4.20cm}  \centering  \includegraphics[width=4.20cm]{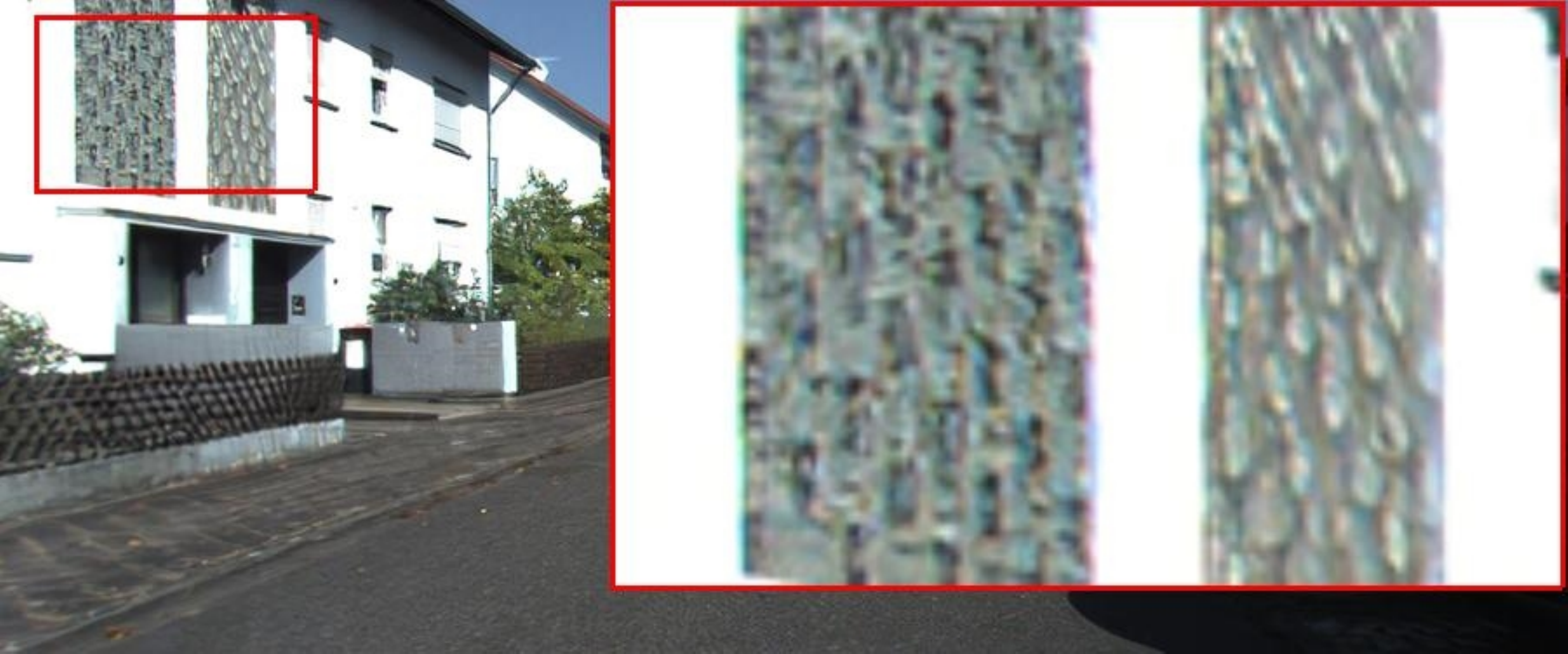} \\ \centering{\small{GAN-based SISR (SPSR)}}  \end{minipage}
     \begin{minipage}[t]{4.20cm}  \centering \includegraphics[width=4.20cm]{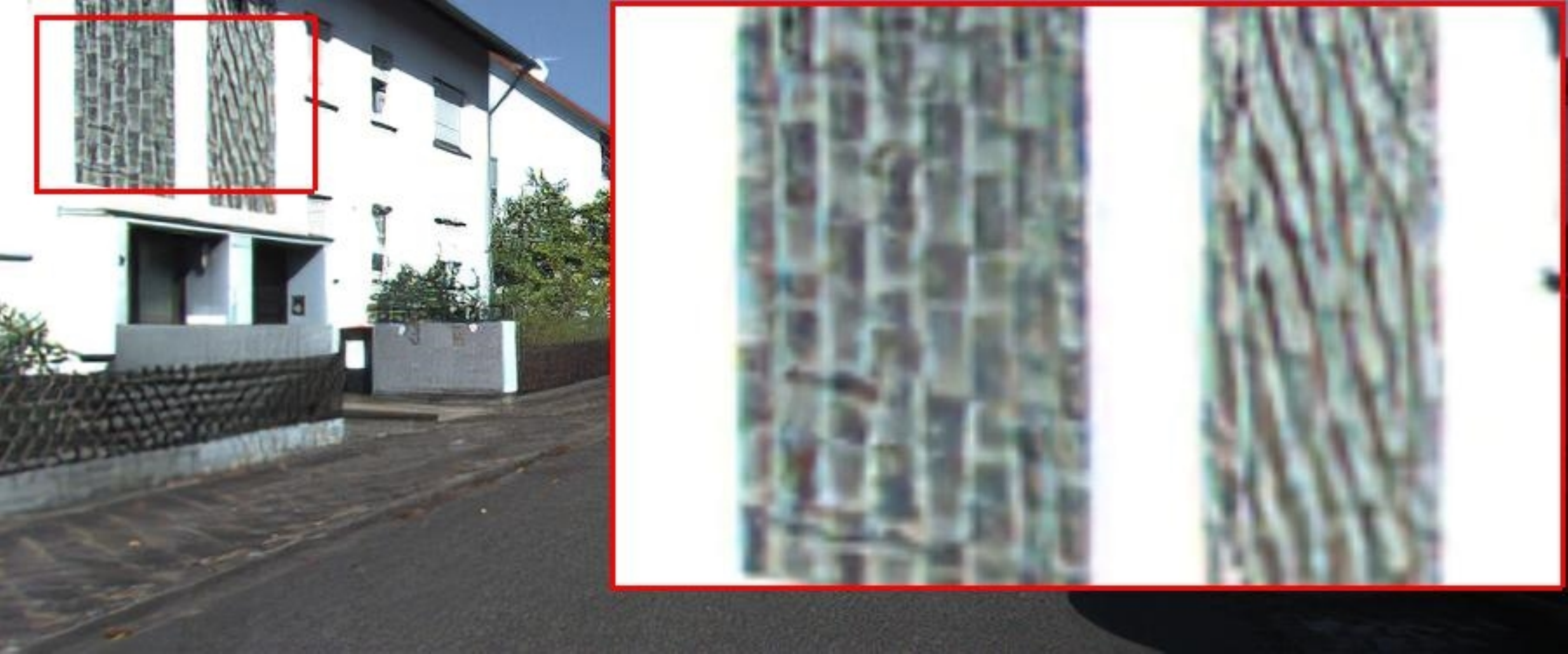} \\ \centering{\small{DASSR-GAN}} \end{minipage}
    }\
\centering
\subfigure{
   \begin{minipage}[t]{4.20cm}  \centering \includegraphics[width=4.20cm]{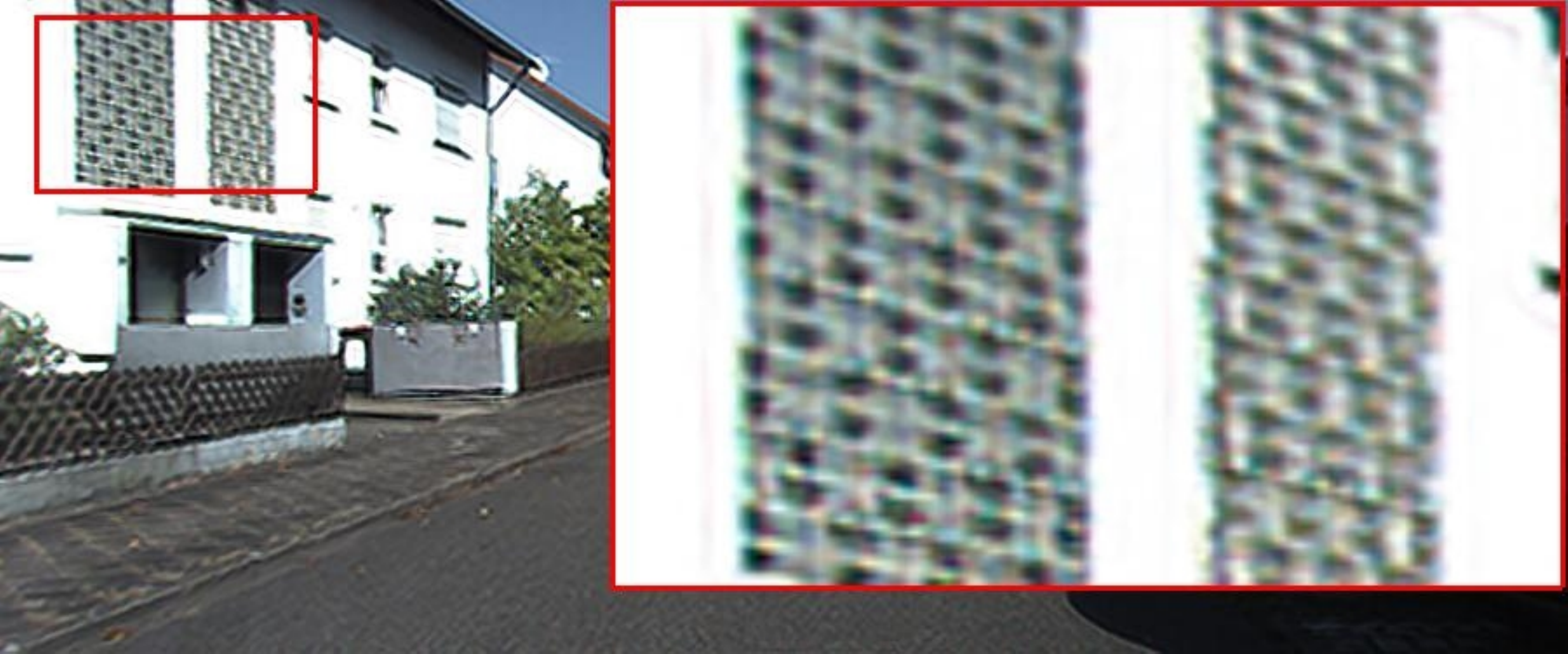} \\ \centering{\small{DASSR-IQP}} \end{minipage}
   \begin{minipage}[t]{4.20cm}  \centering \includegraphics[width=4.20cm]{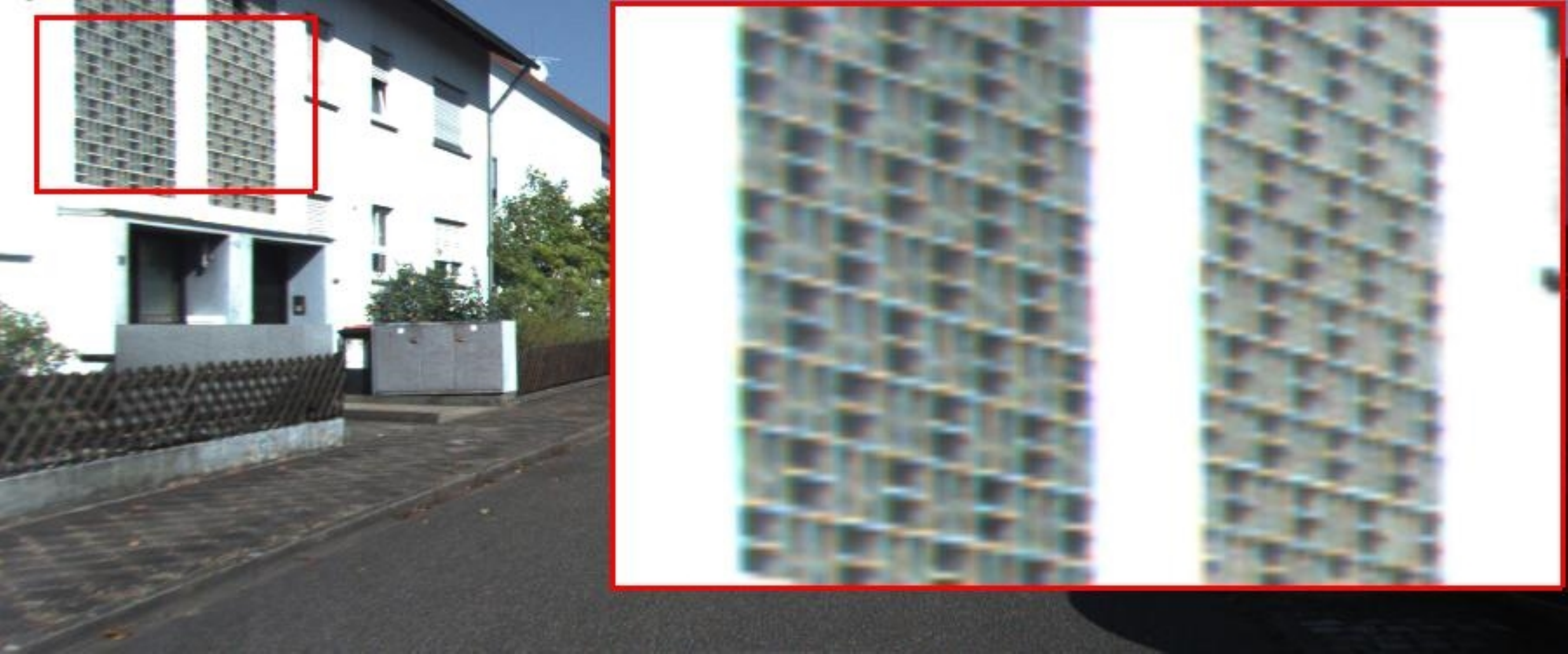} \\ \centering{\small{GroundTruth}} \end{minipage}
}\  \vspace{-0.8em}
\caption{Comparison of MSE-based SISR (DRN~\cite{drn}), GAN-based SISR (SPSR~\cite{spsr}), StereoSR (PASSR~\cite{passr}), and StereoSR model (DASSR~\cite{our}) trained with GAN loss~\cite{srgan} and the proposed image quality perceptual (IQP) approach respectively, on image `000000\_10' in KITTI2012.}
\label{f0} \end{figure}

In single image super-resolution (SISR), perceptual loss~\cite{perceptual} and generative adversarial network (GAN) based loss~\cite{srgan, esrgan} are proposed to restore better visual effect. Due to the characteristic of GAN, some visual artifacts are generated. As shown in Figure~\ref{f0}, the StereoSR (PASSR~\cite{passr}), MSE-based SISR (DRN~\cite{drn}), and GAN-based SISR (SPSR~\cite{spsr}) methods output non-ideal visual effect with over-smooth or unrealistic texture of the wall. 

To restore stereo images with better visual experience, we propose to achieve better perceptual StereoSR results by utilizing the feedback from the evaluation of the perceptual quality. Therefore, how to effectively assess the subjective quality of the super-resolved stereo images becomes the key issue for StereoSR, and is still an unsolved problem.

\begin{figure}[tb]
\centering
\subfigure{
\begin{minipage}[t]{2.8cm}  \centering  \includegraphics[width=2.9cm]{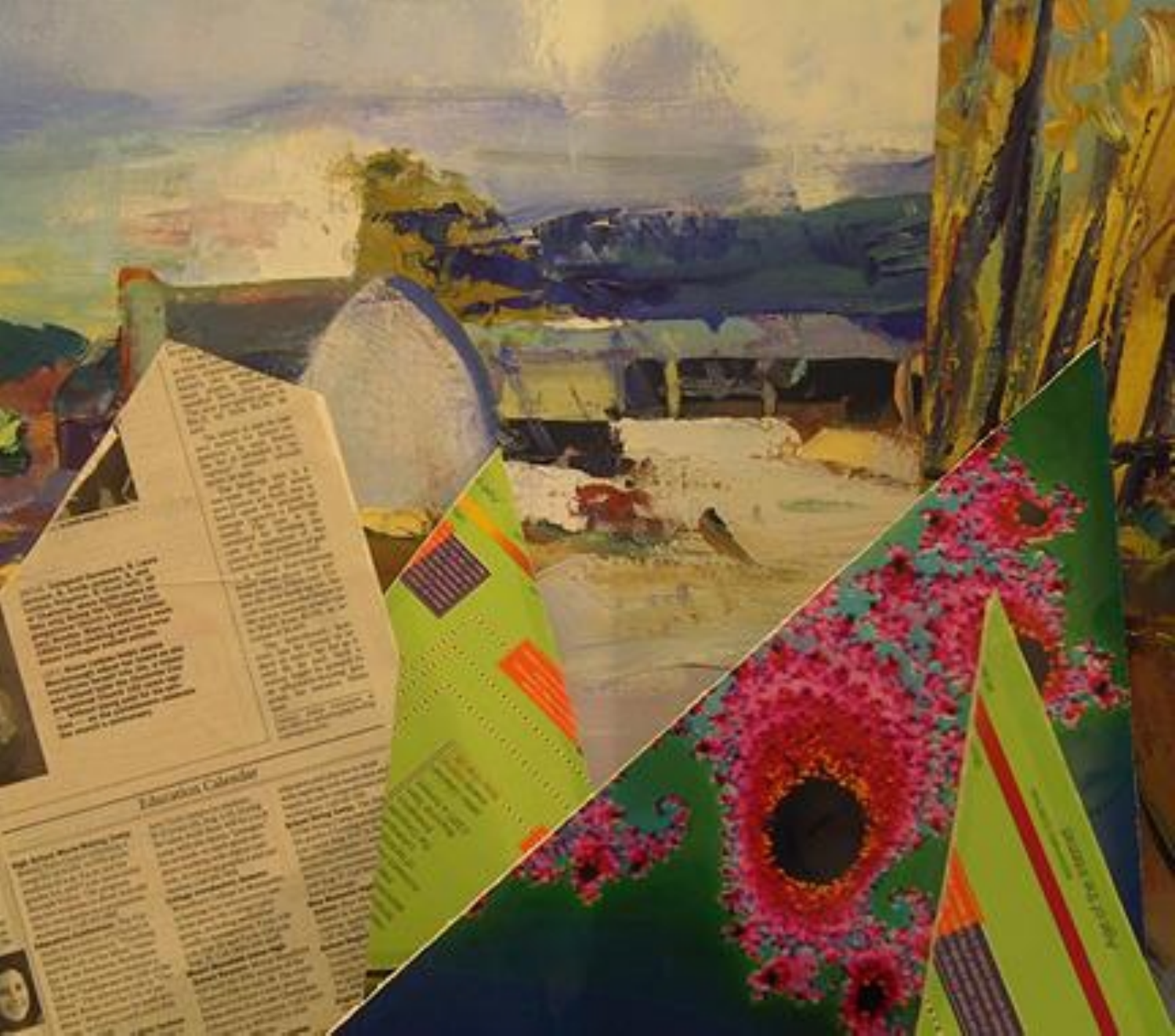} \\ \centering{\small{PSNR/SSIM\\rankMOS}} \end{minipage}
\begin{minipage}[t]{2.8cm}  \centering  \includegraphics[width=2.9cm]{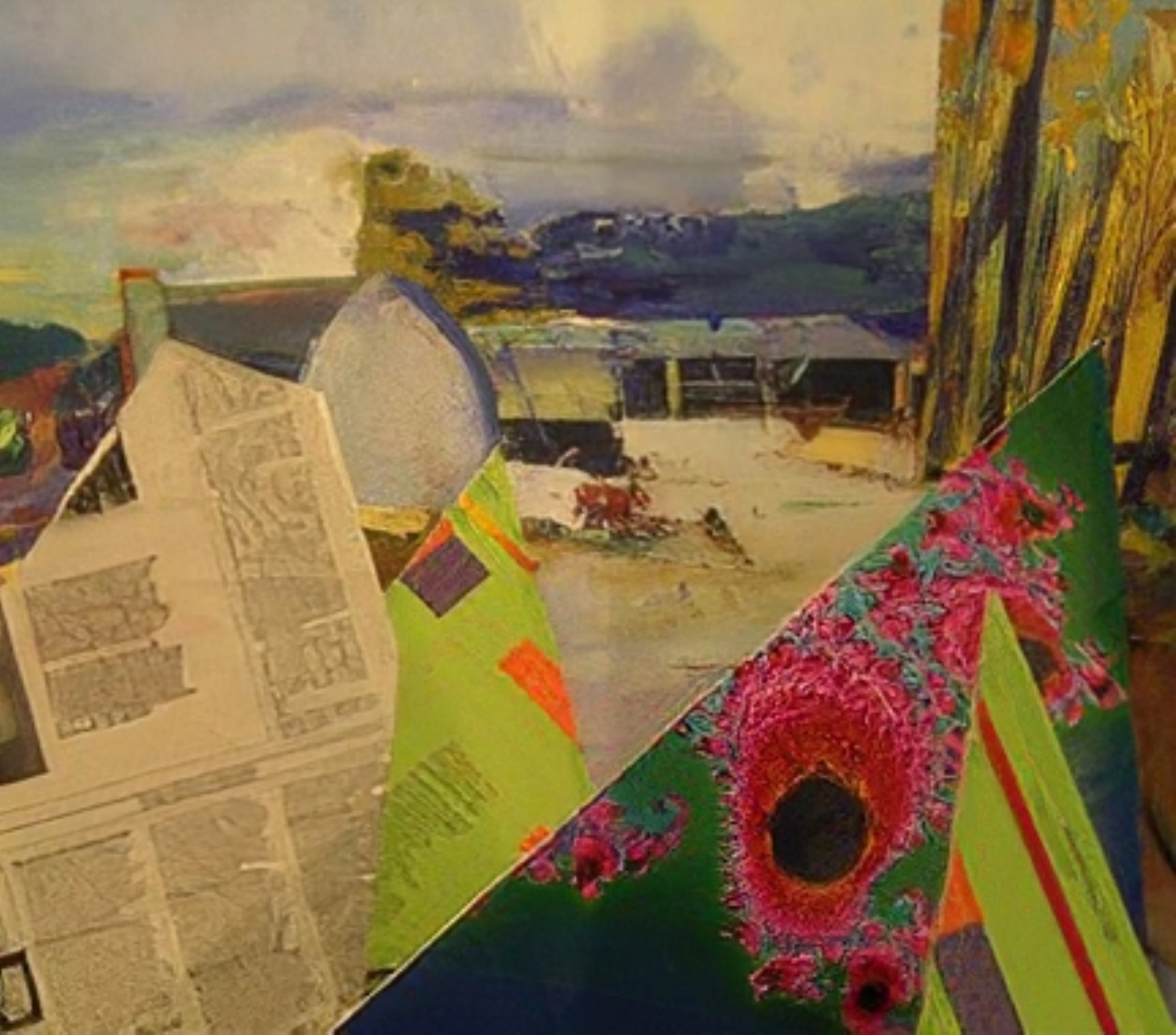} \\ \centering{\small{23.336/0.7063 \\8.5584}} \end{minipage}
\begin{minipage}[t]{2.8cm}  \centering \includegraphics[width=2.9cm]{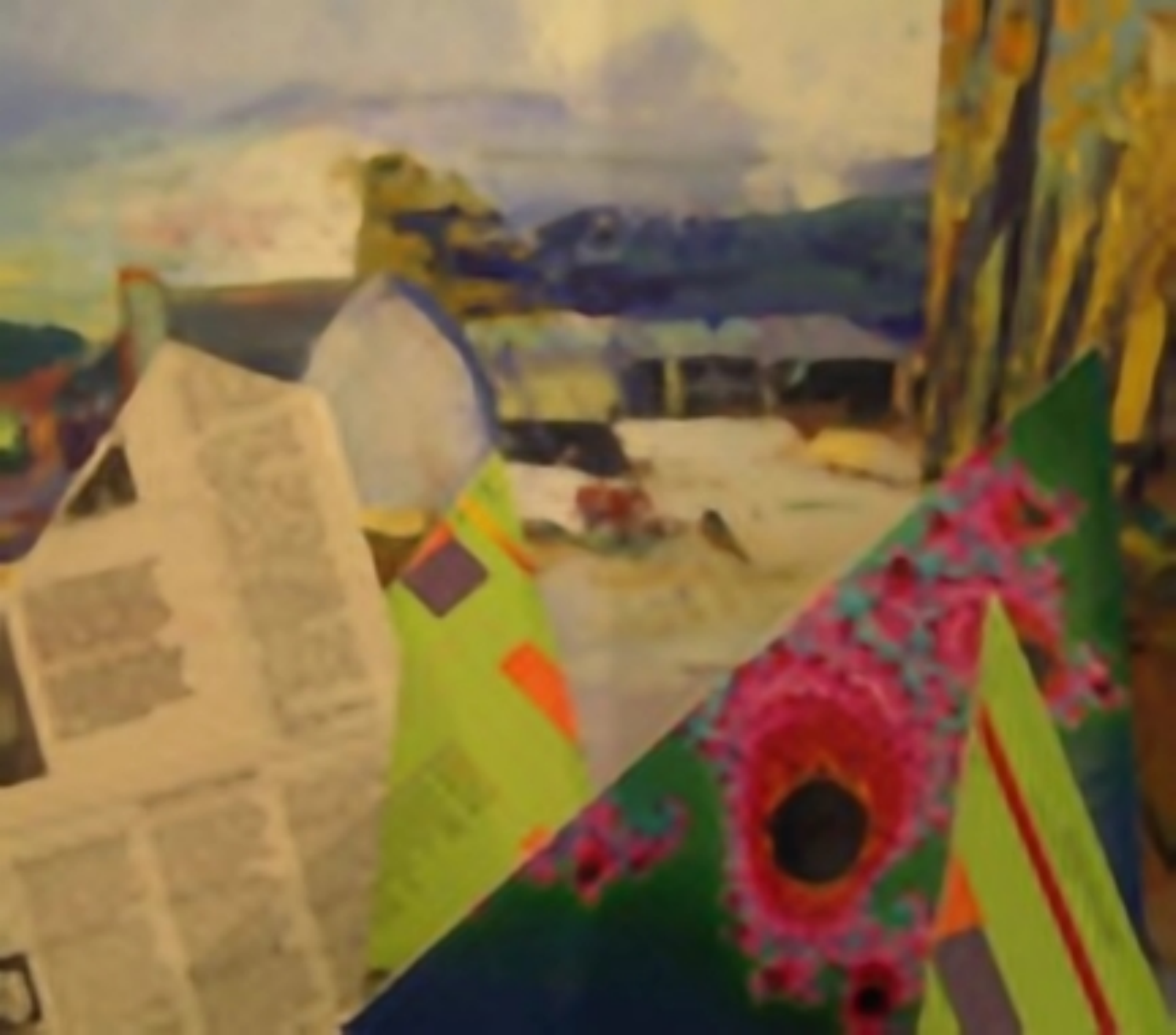} \\ \centering{\small{24.690/0.7436 \\5.1142}} \end{minipage} }\
\centering \subfigure{
\begin{minipage}[t]{2.8cm}  \centering \includegraphics[width=2.9cm]{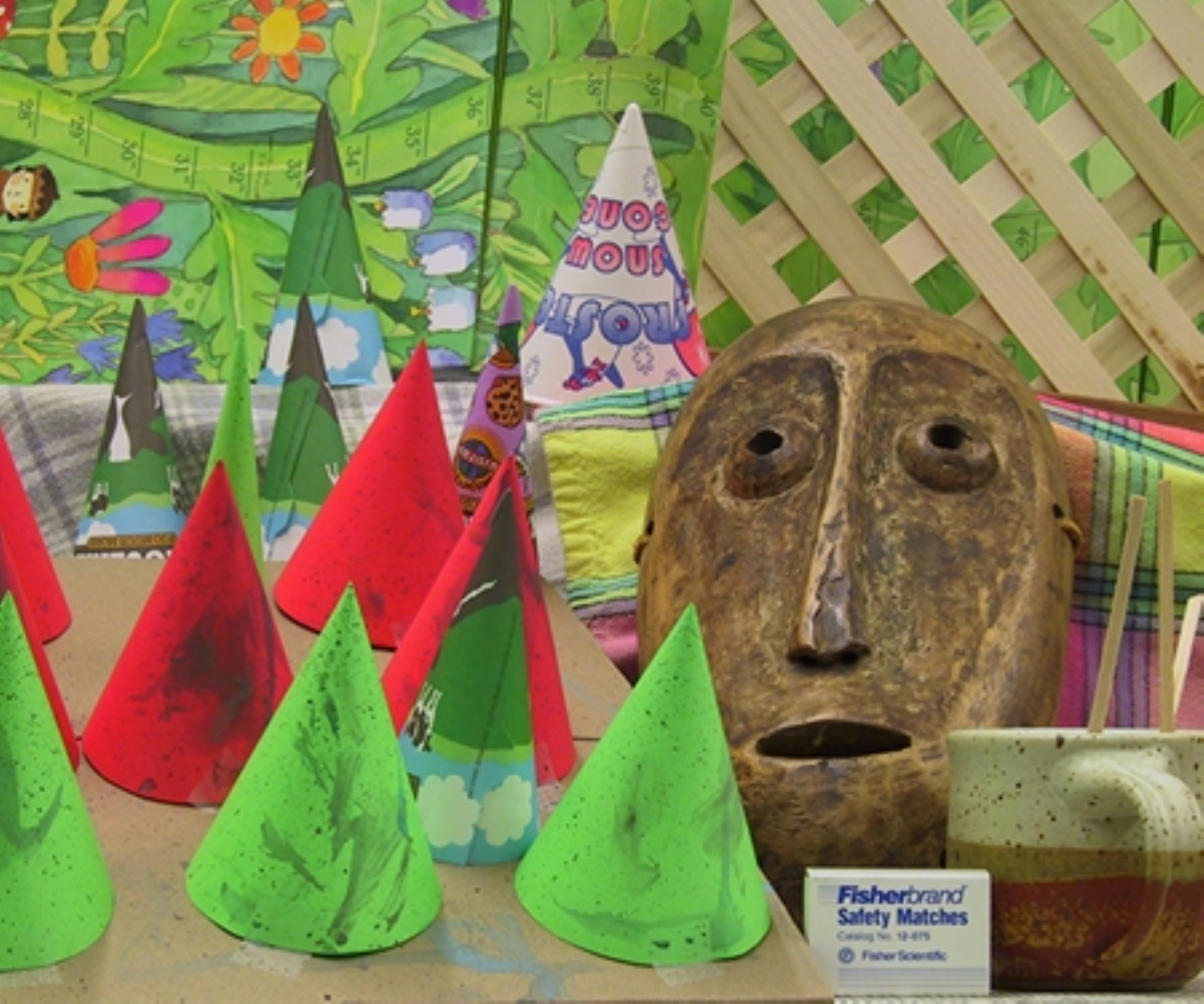} \\ \centering{\small{PSNR/SSIM\\rankMOS}} \end{minipage}
\begin{minipage}[t]{2.8cm}  \centering \includegraphics[width=2.9cm]{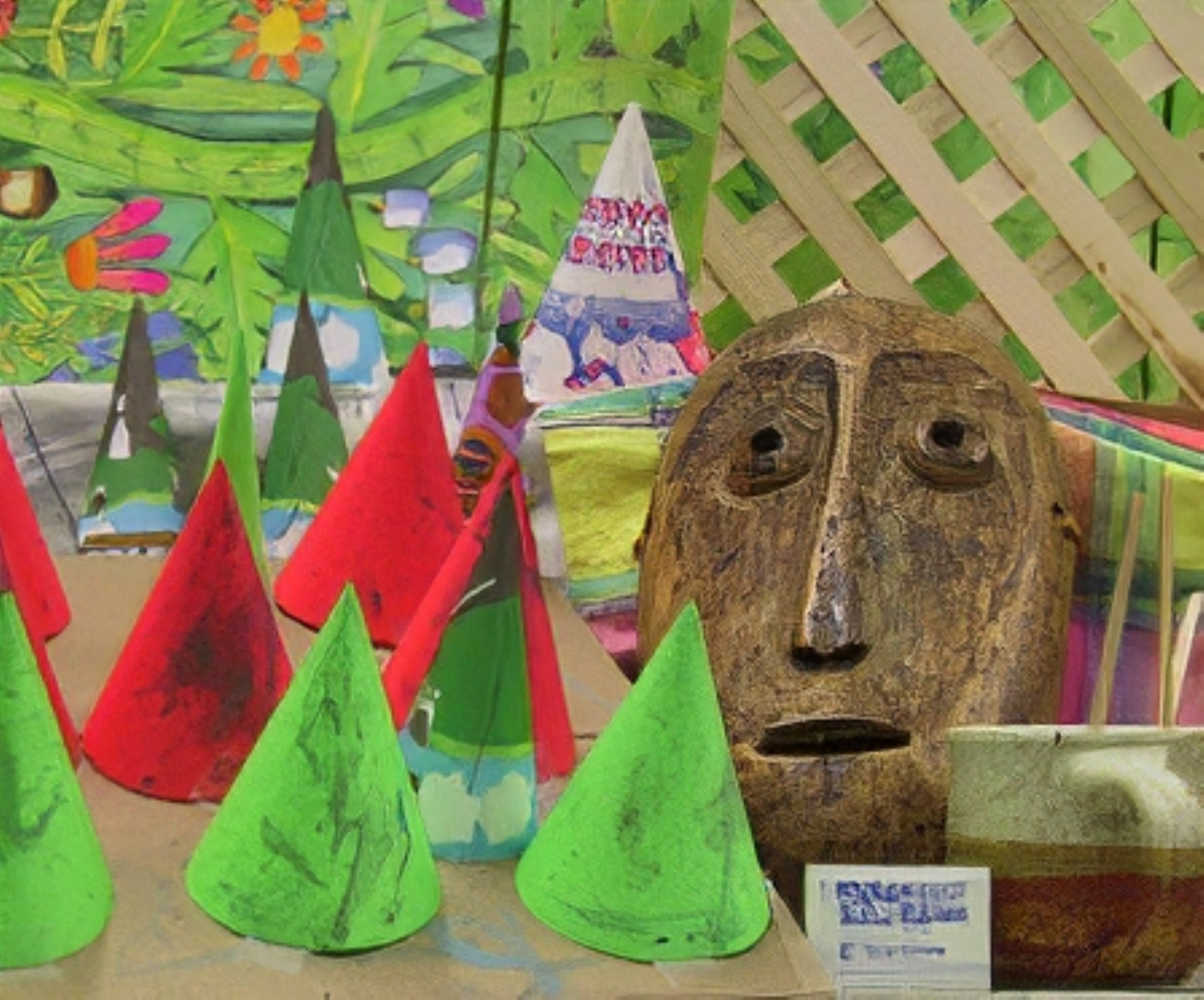} \\ \centering{\small{20.279/0.6063 \\9.4014}} \end{minipage}
\begin{minipage}[t]{2.8cm}  \centering \includegraphics[width=2.9cm]{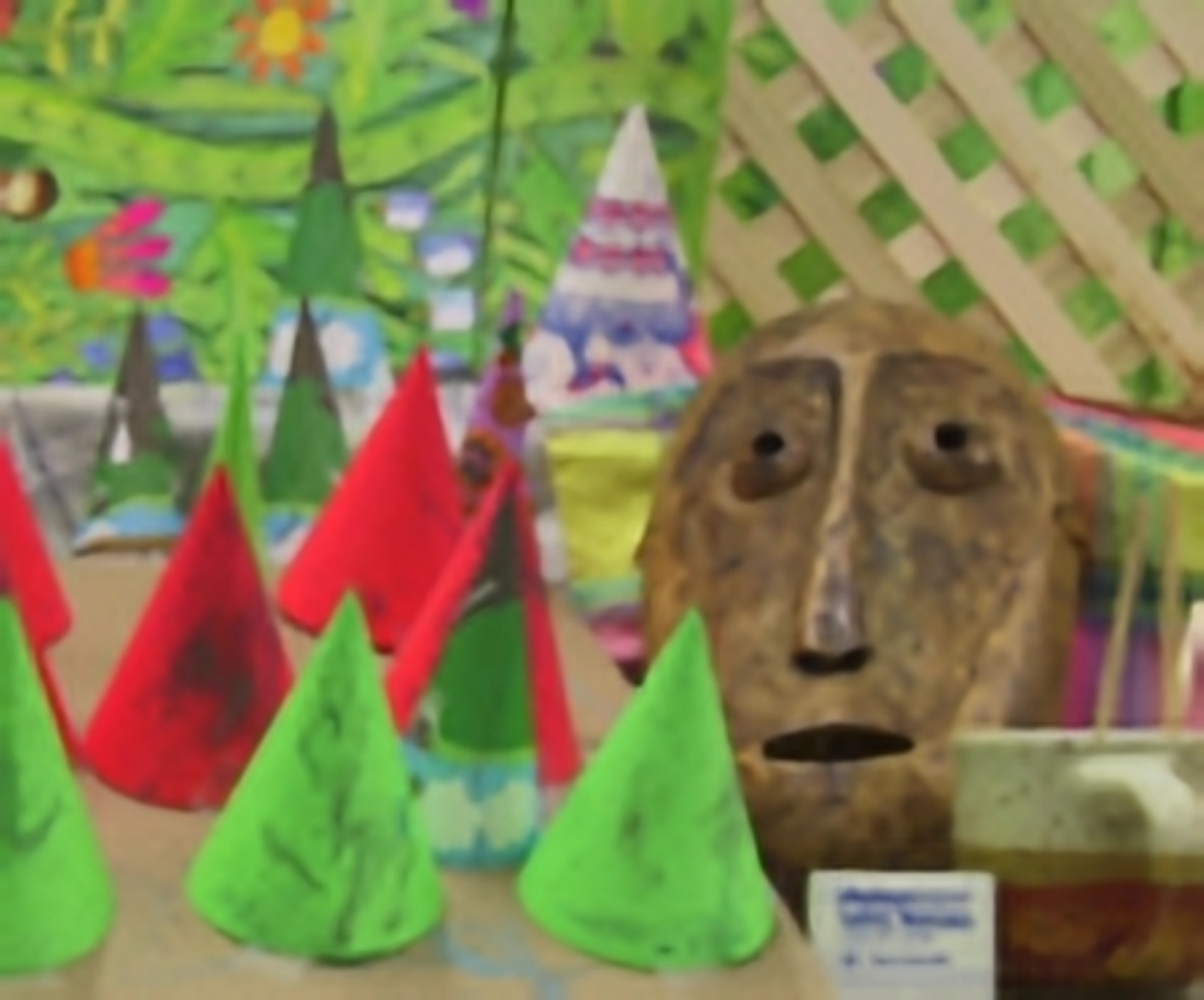} \\ \centering{\small{24.479/0.7678 \\5.4229}} \end{minipage}
}\ \vspace{-0.2em} \caption{The PSNR/SSIM and rankMOS scores (higher is better) of stereo images super-resolved by ESRGAN~\cite{esrgan} (middle) and EDSR~\cite{edsr} (right). The left images denote the groundtruth. Though PSNR/SSIM of right images are higher than middle images, middle images are sharper and more consistent with human visual system (HVS), which can be figured out by our quality assessment score (rankMOS).}
\label{fpsnr1} \end{figure}

As the most common evaluation metrics, PSNR and SSIM measure the distortion degree of images rather than human visual perception. Figure~\ref{fpsnr1} shows the inconsistency between PSNR/SSIM and subjective quality of two stereo images, super-resolved by a generative adversarial network (GAN) based SR method (middle) and a MSE-based SR method (right). Traditional image quality assessment (IQA)~\cite{cnniqa, sfa, sisrqa} and stereo image quality assessment (StereoQA)~\cite{stereoqa} methods focus on distortion types, including JPEG compression, noise, Gaussian blur. These distortions are different from artifacts produced by different super-resolution methods. Therefore, we first propose a special stereo image super-resolution quality assessment (StereoSRQA) model, which focuses on degradation of both image structure, texture, and stereo effect. 

This paper develops a universal perception-oriented StereoSR framework (PSSR), which learns the LR-to-HR mapping by utilizing the feedback information from the proposed quality assessment model. In specific, the proposed PSSR introduces a novel perception-oriented training constraint based on the prior knowledge from the StereoSRQA to enhance the visual quality and the stereo effect of the super-resolved stereo images. Note that, the proposed PSSR can be applied to different StereoSR networks. The main contributions of this paper are summarized as follows:

\begin{itemize}
\item We propose a perception-oriented StereoSR framework to restore stereo image with better subjective quality and stereo effect by employing the feedback from the quality assessment of the StereoSR results.
\item We develop the first quality assessment model for StereoSR, which not only provides an unified method for assessing subjective StereoSR results, but also accurately constrains the optimization of StereoSR model. 
\item We construct the first StereoSRQA database (Mid3D\_QA), covering a wide range of super-resolved stereo images. The quality score (rankMOS) of each stereo image pair can fairly reflect the comprehensive visual quality and stereo performance.
\end{itemize}

Extensive experiments demonstrate the proposed PSSR can significantly improve the subjective quality and practicability of the StereoSR results. 

\section{Related Work}
\label{secrel}
\subsection{Super-Resolution}

\textbf{CNN-based Single Image Super-Resolution.}
Deep learning based methods have dominated the development of SISR. SRCNN~\cite{srcnn} first introduces convolution neural network (CNN) in SISR to learn LR-to-HR mapping. Based on this work, following deep learning based SISR methods achieve continuous breakthroughs by proposing deeper, wider networks and other novel structures (VDSR \cite{vdsr}, EDSR~\cite{edsr}, SRFBN \cite{srfbn}, DRN~\cite{drn}).

To generate visually pleasing results, Johnson \emph{et al.}~\cite{perceptual} proposed a perceptual loss for SISR, which calculates the feature similarity between SR and HR images based on the VGG network~\cite{vgg19}. SRGAN~\cite{srgan} exploits the generative adversarial network (GAN) to reconstruct more high-frequency details. SPSR \cite{spsr} introduces gradient loss into GAN-based SR method to guide the restoration of image by employing structure prior of gradient map. 

\textbf{Stereo Image Super-Resolution.}
StereoSR~\cite{stereosr}, the first CNN-based StereoSR methods, achieves impressive results by taking the parallax prior of stereo image into consideration. In order to deal with various parallax, PASSRnet~\cite{passr} and PSCASSRnet~\cite{pscassr} present parallax-based spatial and spatial-channel attention mechanisms, which can extract the correspondence between two views of a stereo pair. To increase the correspondence between two views in a stereo image pair, DASSR~\cite{our} first proposes to align these two views based on the disparity map before restoring the degraded images.

\subsection{CNN-based Image Quality Assessment}
Mean Opinion Score (MOS), the common image quality assessment (IQA) score, is obtained by manually scoring to evaluate the image quality from the perspective of the human sense. Many IQA metrics (PSNR, SSIM~\cite{ssim}, BRISQUE~\cite{brisque}, NIQE~\cite{niqe}) use a mathematical model to evaluate the image quality. Deep learning has also promoted the development of IQA. Kang \emph{et al.}~\cite{kang} first constructed a no reference (NR) IQA CNN, which does not require the clean image as reference. Li \emph{et al.}~\cite{cyclopean} exploited ensemble learning and saliency map for stereo image quality assessment. StereoQA-Net~\cite{stereoqa} presents an end-to-end dual-stream interactive network for no-reference stereo image quality assessment. 

To evaluate the image quality of SISR methods instead of traditional distortion types (blur, compression, noise, spatial warping), Yan \emph{et al.}~\cite{sisrqa} proposed a FR SISR quality assessment (SISRQA) CNN for assessing the visual perception of the super-resolved images.

Since generating subjective scores (MOS) requires many human judgement, the size of IQA database is limited. Wu \emph{et al.} \cite{tip20} built a large-scale IQA database, and synthesized pseudo-MOS to represent the subjective quality of the distorted image by combining several classical IQA methods. RankIQA~\cite{rankiqa} also constructs a large IQA database without human annotation by generating distorted images with image processing transformation of different levels and types.

\section{Perception-Oriented StereoSR}
\label{secstereosr}
This section mainly introduces the proposed perception-oriented StereoSR (PSSR) approach. Note that our goal is not proposing a novel StereoSR network structure but an effective framework, which can constrain the StereoSR network to generate promising visual quality. Therefore, our StereoSR network can adopt any structure, and we choose the DASSR~\cite{our} in this paper. 

Traditional StereoSR methods treat pixel-wise mean square error (MSE) or mean absolute error (MAE) as their optimization objective, which forces the model output images with higher PSNR but over-smooth textures. To have a better understanding of the artifacts in StereoSR problem, we first propose a special quality assessment network for StereoSR, which will be introduced in detail latter. Given a well-trained StereoSRQA model, which is able to predict the perceptual quality of StereoSR results, we consider that an ideal StereoSR network can generate visually promising results with high StereoSRQA scores.

Different from the MSE or MAS loss, which constrains the pixel similarity between the SR and HR images, we aim to constrain the perceptual similarity between the SR and HR images by utilizing the guidance of the StereoSRQA model. In this way, the StereoSR results enjoy promising visual effect, that is close to groundtruth and satisfies the human perception.

\begin{figure}[tb]
\centering \includegraphics[width=0.49\textwidth]{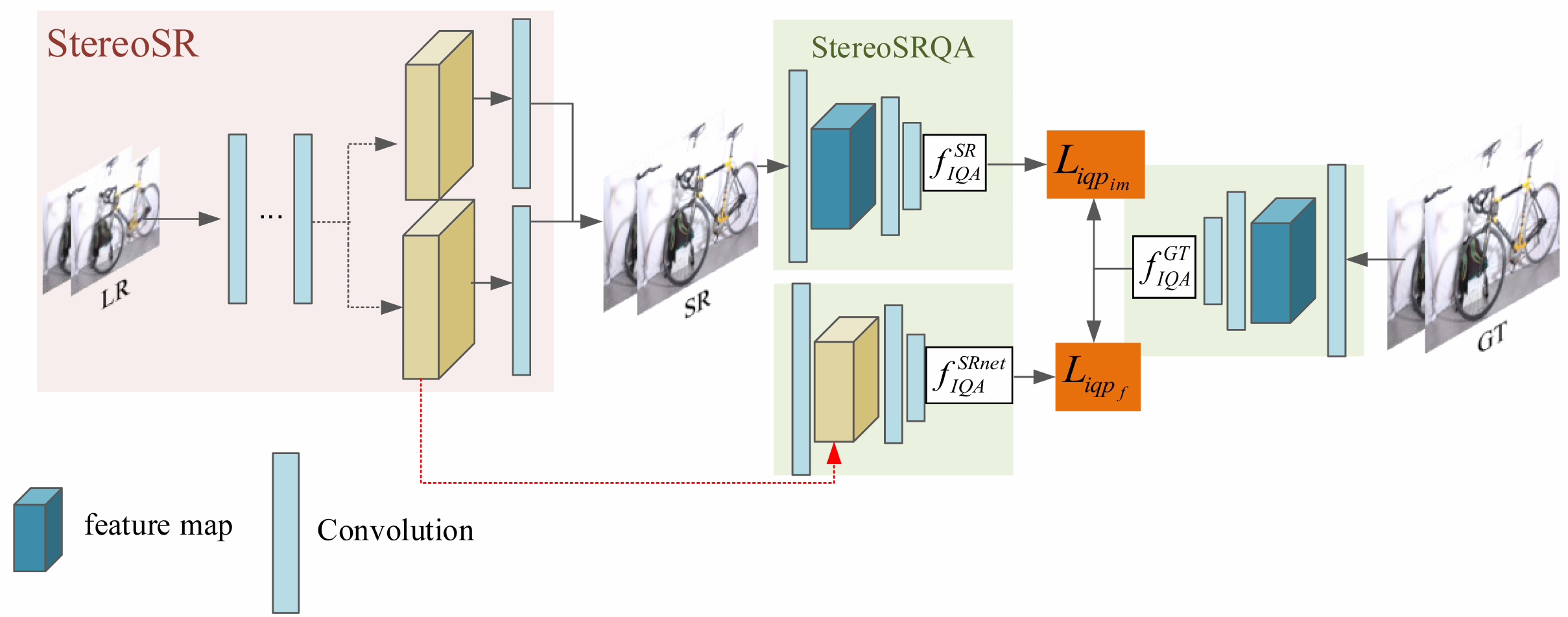} \vspace{-0.2em}
\caption{The proposed perception-oriented StereoSR algorithm. The super-resolved stereo image (SR), the last feature of StereoSR model, and the groundtruth stereo image (GT) are passed to the StereoSRQA model to obtain features $f_{IQA}^{SR}, f_{IQA}^{SRnet}, f_{IQA}^{GT}$, which are used to calculate the IQP losses.}
\label{fstereosrtrain} \end{figure}

As shown in Figure~\ref{fstereosrtrain}, we propose the image quality perceptual (IQP) constraint to measure the feature similarity between groundtruth and super-resolved stereo images based on the StereoSRQA network. In specific, we design two IQP losses, an image-level IQP loss ($\mathcal{L}_{IQP_{im}}$), and a feature-level IQP loss ($\mathcal{L}_{IQP_{f}}$). The two IQP losses calculate the similarity at image and feature levels respectively by feeding the output image and the middle feature from the StereoSR network to the StereoSRQA network.

To obtain the \textbf{image-level IQP loss ($\mathcal{L}_{IQP_{im}}$)}, the super-resolved stereo image $I^{SR}$ and the ground-truth stereo image $I^{GT}$ are passed to the StereoSRQA network. Therefore, two features from the first convolution layers in StereoSRQA network ($f_{IQA}^{SR}$, $f_{IQA}^{GT}$) are obtained. Then, we adopt L2 norm function to calculate the feature similarity between $f_{IQA}^{SR}$ and $f_{IQA}^{GT}$. In this way, the $\mathcal{L}_{IQP_{im}}$ is able to constrain the StereoSRQA feature of super-resolved stereo image to be similar to that of the groundtruth. 
\begin{equation}
\mathcal{L}_{IQP_{im}} = ||f_{IQA}^{SR}, f_{IQA}^{GT}||_{2}.
\label{eq3}
\end{equation}

The \textbf{feature-level IQP loss ($\mathcal{L}_{IQP_{f}}$)} constrains features in StereoSR network to be able to reconstruct visually promising image. Specifically, as shown in Figure~\ref{fstereosrtrain}, we replace the first layer feature of StereoSRQA network with the last layer feature of StereoSR network to output middle feature $f_{IQA}^{SRnet}$. Then, we also calculate the L2 norm between $f_{IQA}^{SRnet}$ and $f_{IQA}^{GT}$.
\begin{equation}
\mathcal{L}_{IQP_{f}} = ||f_{IQA}^{SRnet}, f_{IQA}^{GT}||_{2}.
\label{eq4}
\end{equation}

The feature-level IQP loss is proposed based on following considerations: First, feature extraction of the first layer in StereoSRQA shares same information with the image reconstruction feature of the last convolution layer in StereoSR network, and the feature-level IQP loss enhances the constraint strength of the feature similarity; Second, compared with the three-channel image, the feature map in convolutional neural network generally has more channels and consequently covers more information for image reconstruction. Therefore, optimizing the accuracy based on feature instead of image may enable stronger constraint; Third, since the IQP loss is calculated on the features of CNN, we believe that directly putting the output feature rather than the output image of the StereoSR network to the StereoSRQA network can effectively bridges the gap between image and feature.

In this way, the proposed IQP constraints make sure the feature similarity related to the quality assessment, and further enforce the super-resolved stereo image share similar spatial details with the groundtruth, especially the details that are sensitive for StereoSRQA network to evaluate the visual quality of stereo images. In other words, the super-resolved stereo image has similar visual nature with the groundtruth image. The StereoSR network is trained on the combination of L2 and IQP losses.
\begin{equation}
\begin{array}{lr}
\mathcal{L}_{IQP} = \lambda_{1} \mathcal{L}_{IQP_{im}} + \lambda_{2} \mathcal{L}_{IQP_{f}}, \\
\mathcal{L} = \lambda_{0} ||I^{SR} - I^{GT}||_{2} + \mathcal{L}_{IQP},
\end{array} \label{eq5}
\end{equation}
where $\lambda_{0}, \lambda_{1}, \lambda_{2}$ are the weights to balance different losses. We set $\lambda_{0}$ = 1 and $\lambda_{1}$ = $\lambda_{2}$ = 0.1.

By optimizing the StereoSR network with the IQP constraint in training stage, the knowledge from the StereoSRQA network can be back-propagated to the StereoSR network.

\section{StereoSR Quality Assessment} 
Our perceptual-oriented StereoSR method requires a well-performing StereoSRQA model. To our best knowledge, such a model has never been proposed before. This section constructs a model and a database for StereoSRQA.

\subsection{StereoSRQA Network}
\label{secstereosrqa}
First, we train a StereoSRQA network, which simultaneously considers spatial distortion inside each view and disparity-related quality degradation cross different views, to predict the perceptual quality of different SR results. Different from IQA methods only considering spatial content of image, assessing a stereo image should additionally refer to the correlation between two views. Zhou \emph{et al.} \cite{stereoqa} indicated that integrating left and right views by summation and subtraction is consistent with the fusion and disparity information in human visual system (HVS). Inspired by their work, our StereoSRQA network takes the left view, the right view, and their difference map as input. Note that the main purpose of this paper is not proposing a novel QA network, but a suitable QA model that can provide helpful guidance for the StereoSR model. Therefore, our StereoSRQA model can flexibly adopt the structure of different StereoQA models.

\begin{figure}[tb]
\centering \includegraphics[width=0.49\textwidth]{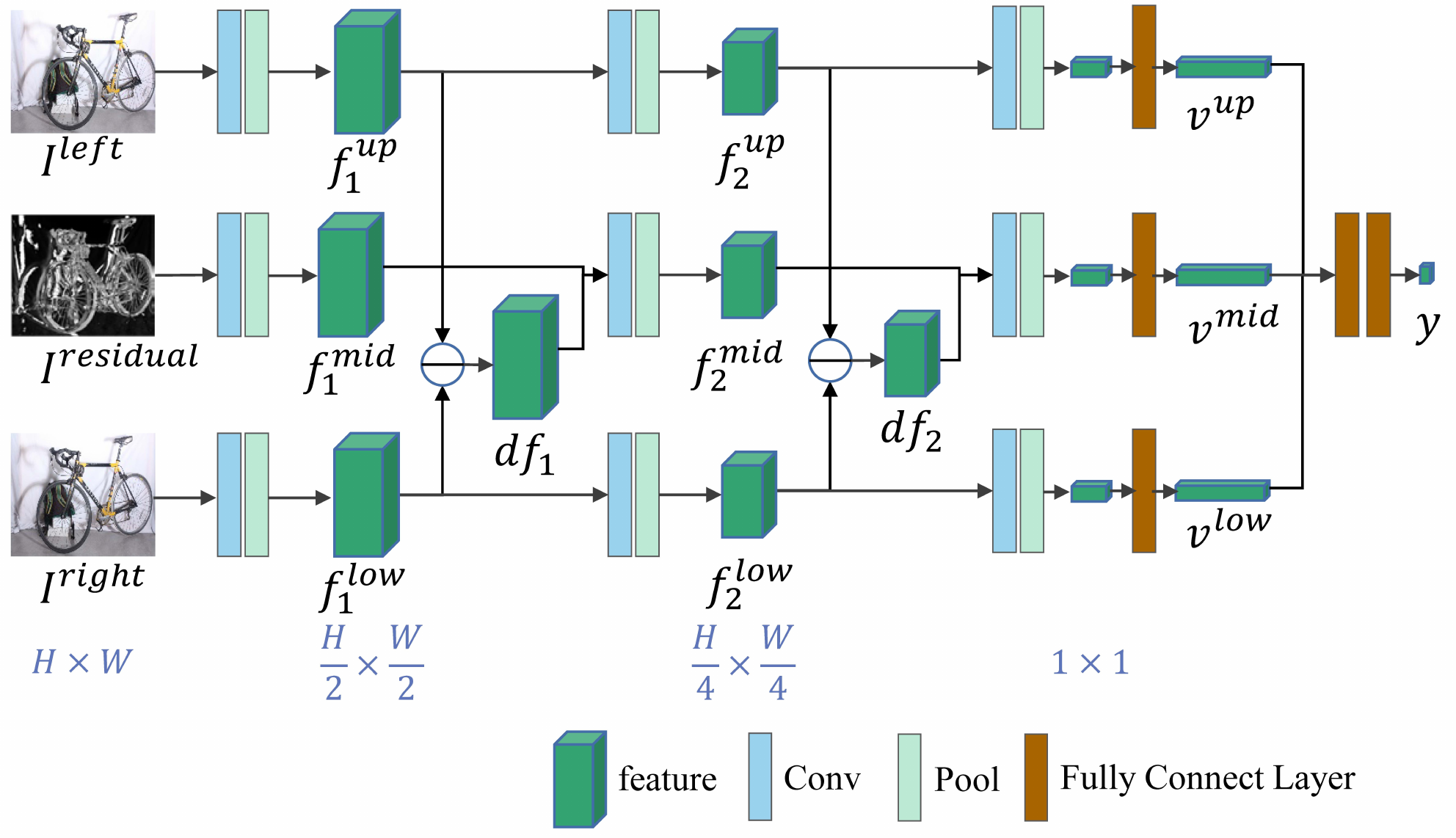} \vspace{-0.2em}
\caption{Architecture of the StereoSRQA network.}
\label{fstereosrqa}
\end{figure}
As shown in Figure~\ref{fstereosrqa}, the proposed StereoSRQA model is composed of three branches with similar structure. The upper and lower branches learn spatial features of two views by taking the left and the right views as input respectively.

The only difference between the middle and the upper/lower branches is the input at the beginning and middle convolution layers. The middle branch extracts the cross-view correlation by working on the difference map of image and the difference map of feature. In specific, we get the residual image of the left and right views, and feed it to the middle branch. Similarly, the difference map $df_{i}$ can be obtained from the middle features ($f^{up}_{i}$, $f^{low}_{i}$) of all convolution layers in upper and lower branches. The concatenation of $df_{i}$ and feature $f^{mid}_{i-1}$ of previous convolution layer is regarded as the input of following convolution layer in the middle branch.

After obtaining the vectors ($v^{up}$, $v^{low}$, $v^{mid}$) of three branches, two fully connected layers are processed on the concatenation of $v^{up}$, $v^{low}$, $v^{mid}$ to predict the final score $y$. Following most IQA methods, our StereoSRQA network is trained with MSE loss between the predicted score $y$ and the groundtruth score $z$.
\begin{equation}
\begin{array}{lr}
\mathcal{L}_{StereoSRQA} = ||y - z||_{2}.
\end{array} \label{eq2-1}
\end{equation}

Note that, to calculate the IQP loss, the feature similarity is calculated based on the concatenation of features from the first convolution layers of the three branches.

\subsection{StereoSRQA Database with RankMOS}
\label{sec:data}
Existing stereo image quality assessment databases (\emph{e.g.}, LIVE Phase I~\cite{live1}, LIVE Phase II~\cite{live2}, NBU~\cite{nbu}, MCL\_3D~\cite{mcl} etc.) focus on distortion types, including white noise, Gaussian blur, JPEG compressing, which differ from artifacts in SR. Toward this end, we build the first database, named Mid3D\_QA, specifically for assessing the quality of StereoSR. The Mid3D\_QA database contains 66 HR stereo image pairs from Middlebury stereo database~\cite{middlebury} as reference, and each HR stereo pairs are degraded by 51 distortion types (including 13 SR methods under different scales, blur and noise levels) to generate 3366 (66$\times$51) super-resolved versions served as the distorted stereo image pairs. Then, we split 3009 (59$\times$51) stereo pairs for training and 357 (7$\times$51) stereo pairs for testing.

As stated by Wu \emph{et al.} \cite{tip20}, collecting MOS for images by subjective experiment is laborious and needs highly controlled conditions. Therefore, inspired by them, we generate quality scores of distorted images by measuring the quality of the SR results in three aspects to simulate the MOS. Consequently, the labeled data can be leveraged to train a stable StereoSRQA network. 

\textbf{Distorted Images.}
First, 66 high quality stereo image pairs from Middlebury stereo database~\cite{middlebury} are used as reference. To generate diverse LR stereo images, we downsample the HR stereo images with scale factors (2, 3, 4, 5, 6, 8) by bicubic interpolation, convolve them with Gaussian blur ($\sigma \in [0.2, 1.2]$), add noise with noise level sampled from [0, 30]. Then, a wide range of super-resolved stereo images with diverse visual effects are generated by applying different SR algorithms, including interpolation-based SISR (bicubic, lanczos, nearest neighbor, bilinear), CNN-based SISR (SRCNN~\cite{srcnn}, EDSR~\cite{edsr}, IDN~\cite{idn}, SRMD~\cite{srmd}), GAN-based SISR (SRGAN~\cite{srgan}, ESRGAN~\cite{esrgan}) and StereoSR (StereoSR~\cite{stereosr}, PASSRnet~\cite{passr}, DASSR~\cite{our}) methods. There are 51 public SR models for all scales in total. That is, each stereo image pair have 51 SR versions corresponding to 51 distortion levels.

\textbf{Generation of Subjective Score.}
Inspired by the generation of the quality assessment data in \cite{tip20} and RankIQA~\cite{rankiqa}, we propose to synthesize the quality scores for distorted stereo image pairs by combining existing quality assessment metrics. In order to obtain a comprehensive evaluation score, we measure the quality of the super-resolved stereo image pairs from three aspects: the spatial quality inside single view, the 3D effect, and the practicality.

Supposing $I^{SR}_{i,j}$ denotes the results of $j$-th SR method for $i$-th stereo image pair, where $i=1:66, j=1:51$, its quality score $rankMOS_{i,j}$ is generated by voting, ordering, and merging operations, as specified in Algorithm~1:

\textbf{1. Voting} First, we treat the SISRQA model~\cite{sisrqa} and the StereoQA model~\cite{stereoqa} as two voters to evaluate the spatial and stereo effect respectively. Since the disparity estimation is a significant application of stereo image, we also measure the disparity estimation accuracy to represent the practicality of the super-resolved stereo image pair $I^{SR}_{i,j}$. The disparity, estimated by the StereoNet~\cite{stereonet}, is measured with endpoint-error (EPE),  which serves as the third voter. Thus, each distorted stereo image pair has three values from three voters, which measure the image quality from different perspectives. The $k$-th voter gives a score value $v_{i,j,k}$ for $I^{SR}_{i,j}$.

\begin{figure}[tb]
\centering  \includegraphics[width=0.46\textwidth]{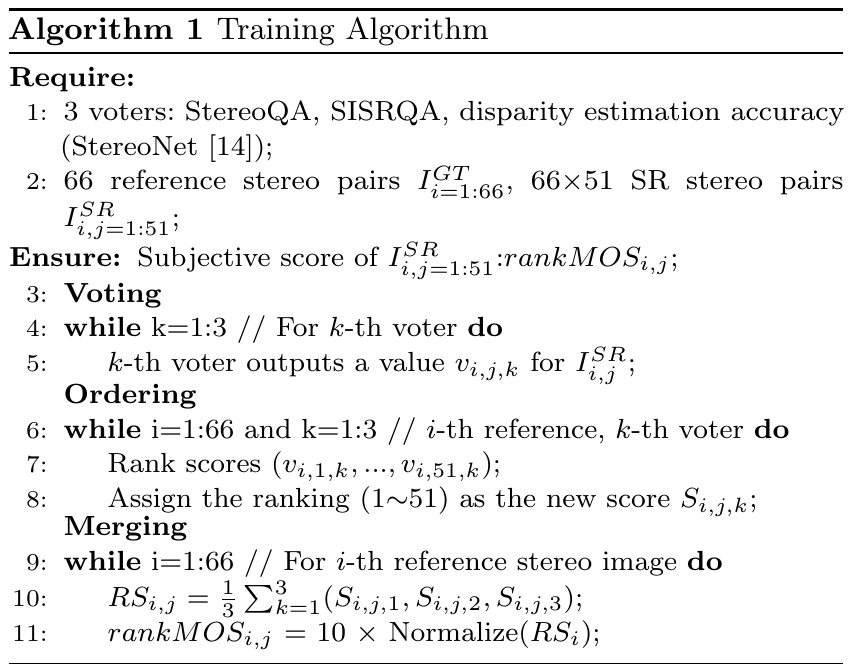} \vspace{-0.2em}
\label{fiqamos}\end{figure}

\begin{figure*}[tb]
\centering
\subfigure{
\begin{minipage}[t]{4.5cm}  \centering \includegraphics[width=4.65cm]{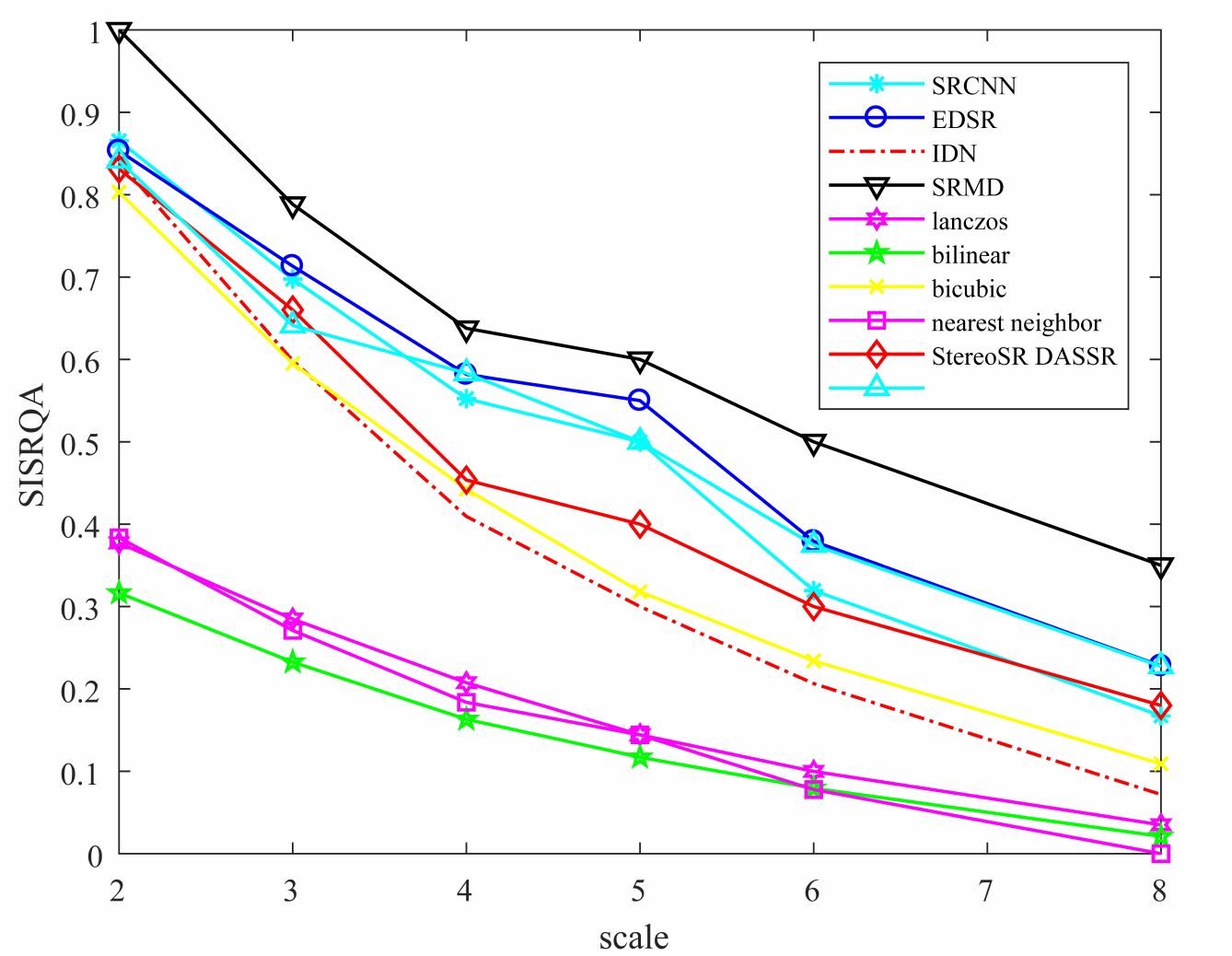}   \end{minipage}
\begin{minipage}[t]{4.5cm}  \centering \includegraphics[width=4.65cm]{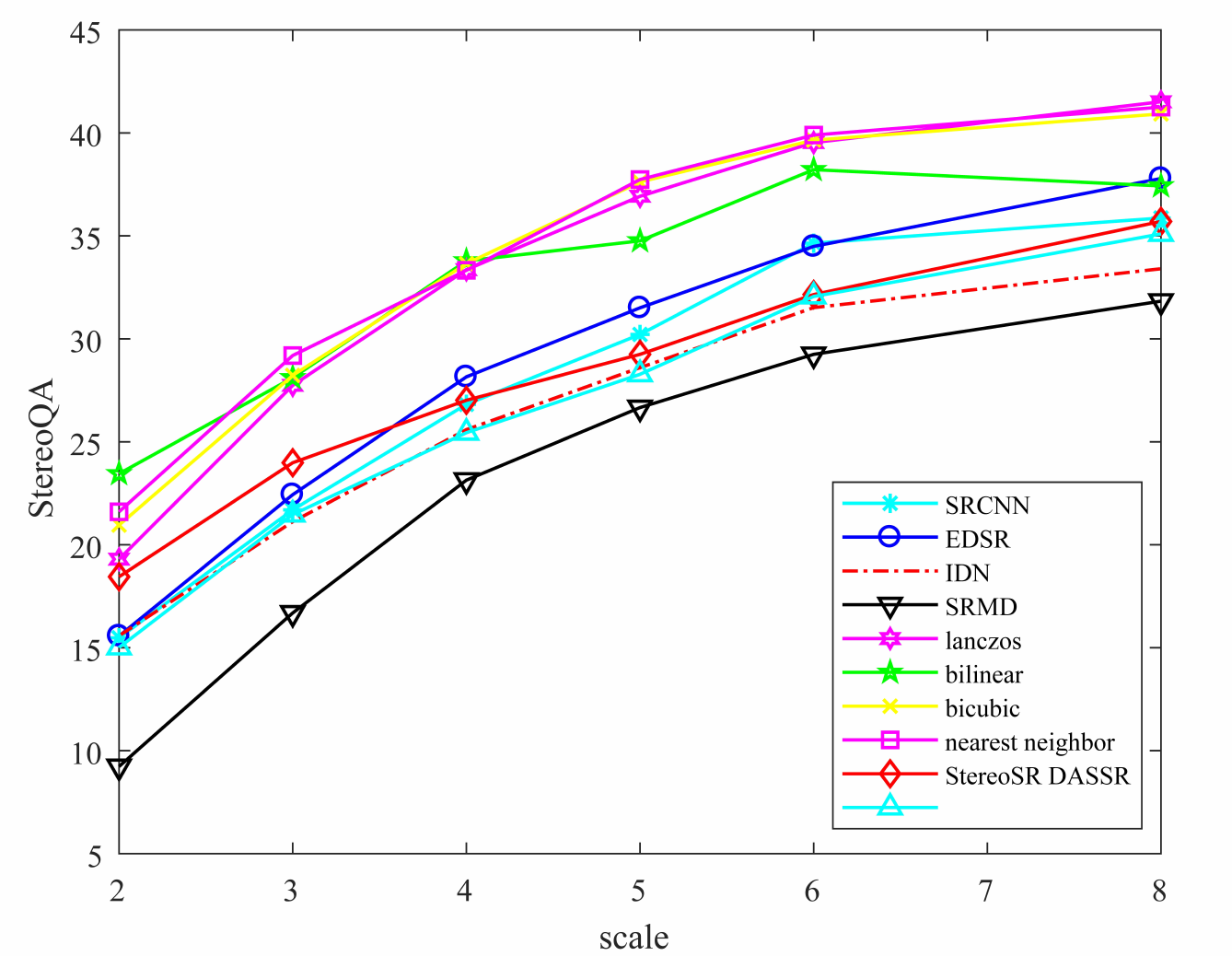}   \end{minipage}
\begin{minipage}[t]{4.5cm}  \centering \includegraphics[width=4.65cm]{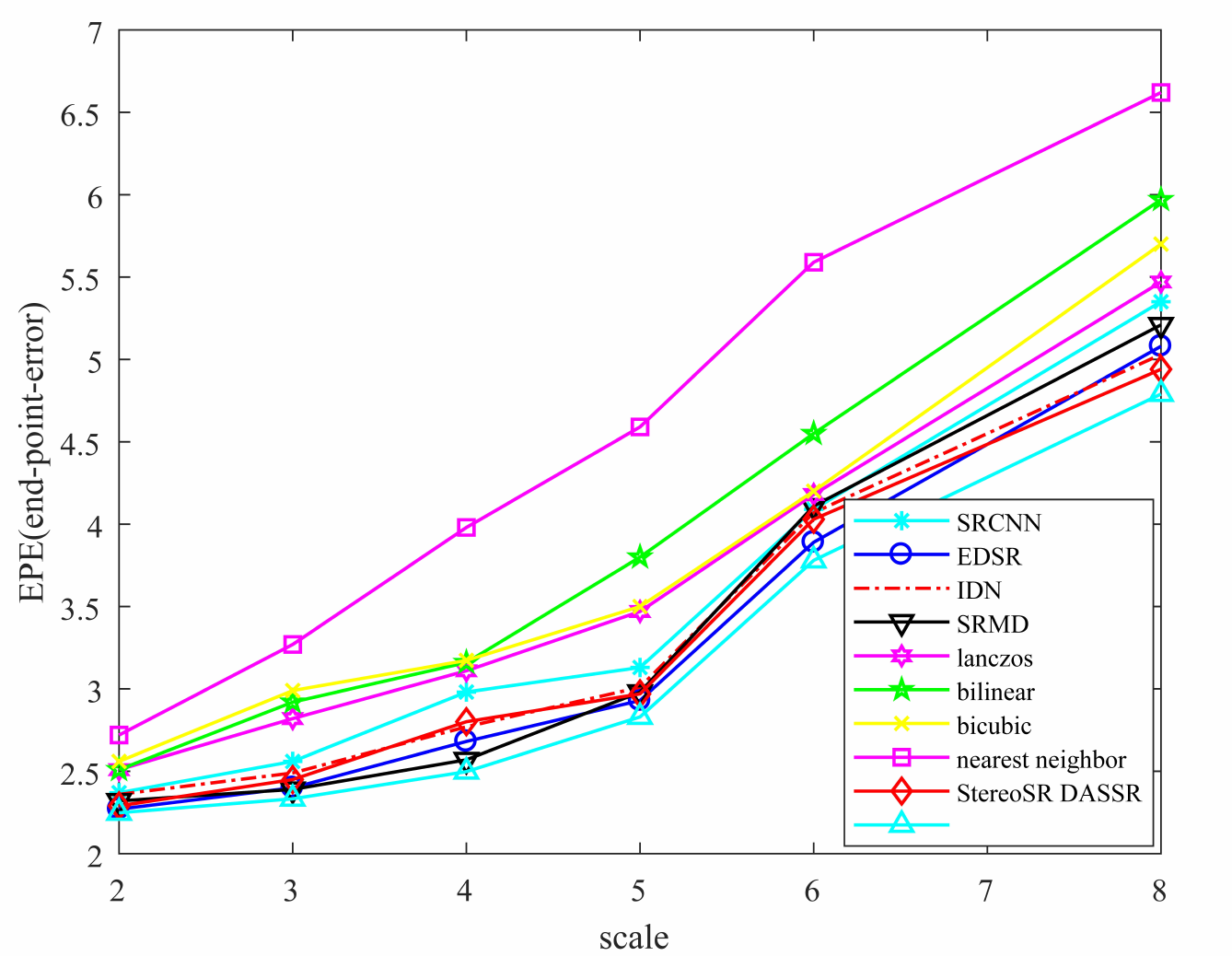} \end{minipage}
\begin{minipage}[t]{4.5cm}  \centering \includegraphics[width=4.65cm]{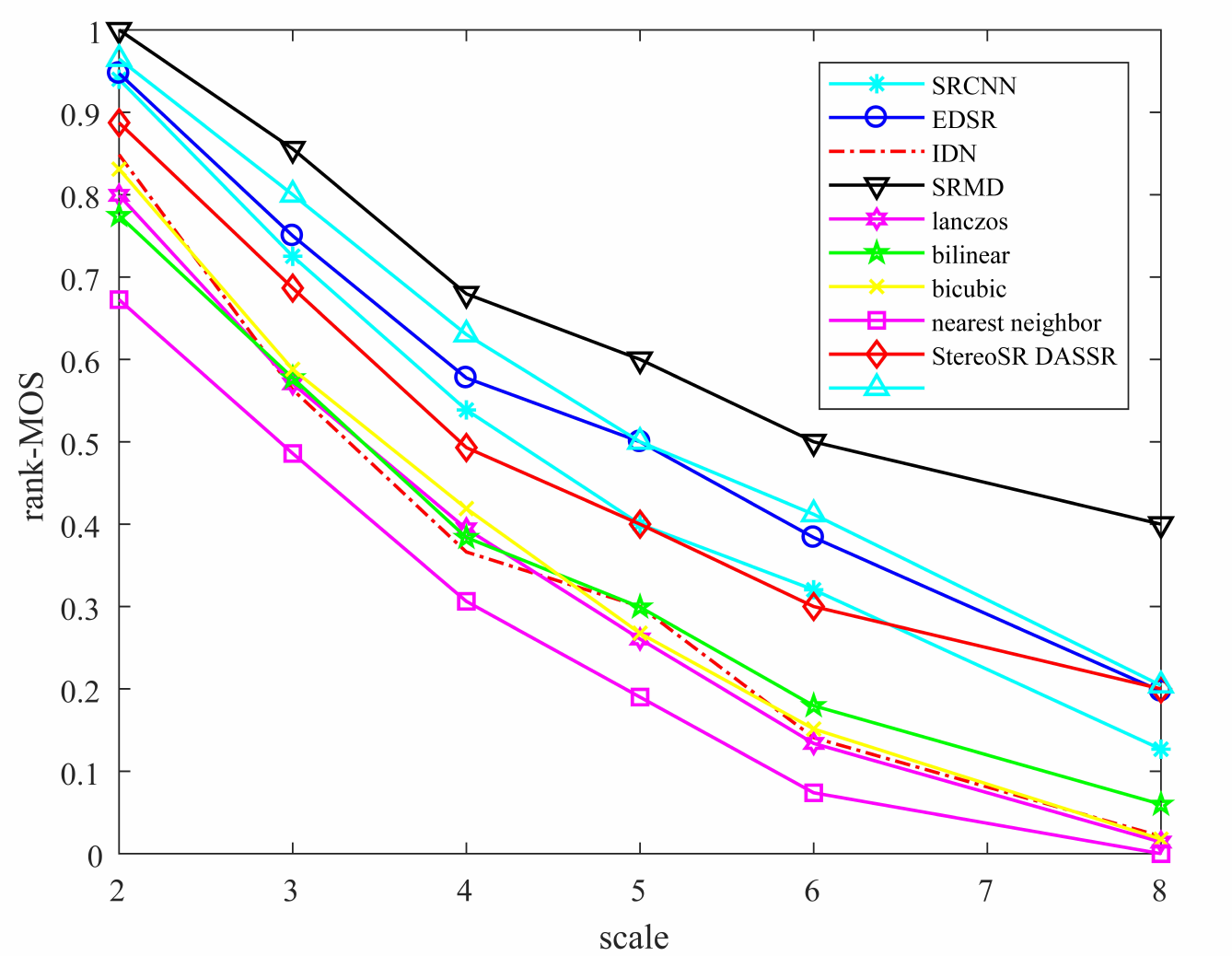}    \end{minipage}
}\  \vspace{-0.2em}
\caption{Distributions of different quality scores (StereoQA~\cite{stereoqa}, SISRQA~\cite{sisrqa}, disparity accuracy (EPE)), and our rankMOS across different scales. The rankMOS keeps a better balance among different SR results.}
\label{fpsnrmos} \end{figure*}

\begin{figure}[tb]
\centering  \includegraphics[width=0.4\textwidth]{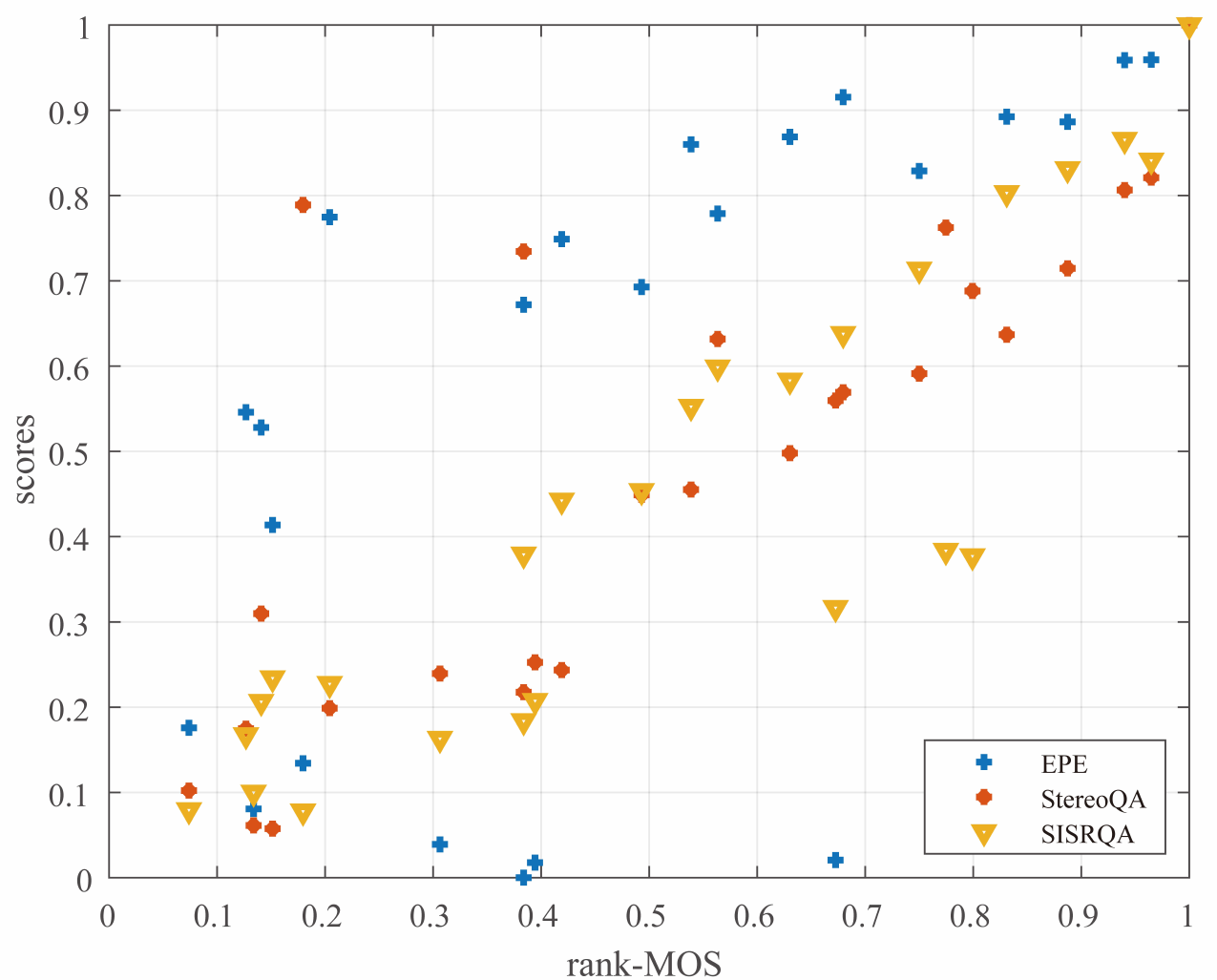} 
\vspace{-0.2em} \caption{Distributions of different IQA methods relative to rankMOS. Different colors represent distributions of different IQAs and points with same color are normalized IQA results of different SR stereo pairs.}
\label{fiqamos} \end{figure}

\textbf{2. Ordering} 
Since quality values from different voters have disparate scales and variation. The quality values $v_{i,j,k=1:3}$ need to be normalized into a unified range and re-scaled to the same distribution. Therefore, we adopt relative ranking value instead of the absolute output value as the new score, $S_{i,j,k}$. In specific, for $i$-th stereo image pair and $k$-th voter, we order scores of all SR versions (51 in total) from worst to best and get rank 1 to rank 51. Rank 1 denotes the worst and rank 51 denotes the best. Then, the rank value of $j$-th distorted stereo image ($I^{SR}_{i,j}$) is assigned to the new score $S_{i,j,k}$ from $k$-th voter.

\textbf{3. Merging} The single ranking score ($RS_{i,j}$) for current SR stereo image $I^{SR}_{i,j}$ is obtained by averaging the three scores ($S_{i,j,k=1:3}$). We further normalize the $RS_{i,j}$ and multiply it by 10 to generate the final quality score, $rankMOS_{i,j} \in [1,10]$. 
\begin{equation}
\begin{array}{lr}
RS_{i,j} = \frac{1}{3}\sum_{k=1}^{3}(S_{i,j,k}),\\
rankMOS_{i,j} = 10 \times Norm(RS_{i,j}).
\end{array}
\label{eq1}
\end{equation}

Note that we do not adopt more CNN-based IQA models. Since existing CNN-based IQA methods are trained on same databases, their predicted scores are similar for same image, especially the rank of the images. Therefore, adopting more CNN-based IQA models makes our rankMOS be closer to results of existing IQA databases and lack concentration to SR features. We consider that a good score should merge more IQAs, that pay attention to different image distortion types.

Figure~\ref{fpsnrmos} demonstrates distributions of three voters and our generated score (rankMOS) cross different scales. Different lines denote results of different SR methods. As shown in Figure~\ref{fpsnrmos}, the rankMOS can effectively incorporate three voters with different value ranges and clearly distinguish artifacts among different SR results under different scales. Furthermore, the rankMOS is able to reduce interference of metrics that do not match the subjective quality and better represent the subjective quality.

To more intuitively show the distribution of different voters (StereoQA~\cite{stereoqa}, SISRQA~\cite{sisrqa}, disparity accuracy (EPE)) relative to the synthetic rankMOS, we normalize these scores and demonstrate relations in Figure~\ref{fiqamos}, where different colors represent different voters. Points with same color correspond to results of different super-resolved stereo image pairs. We see that scores of different quality assessment methods are linearly related to our rankMOS, indicating the synthetic rankMOS is consistent with existing QA methods in terms of reflecting the super-resolution performance of stereo image. In particular, some outliers can be effectively weakened by ranking and averaging operations.

By adopting relative ranking score rather than absolute score, we can extract the preference of different QA methods from different aspects, and unify the scoring standards. The variance of same voter to different images and the standard difference among all voters can also be suppressed.

\section{Experiments and Analysis}
\label{secexp}
This section first introduces the databases, evaluation metrics, and other experimental details. Next, ablation experiments are conducted to investigate the contributions of each design in our Perception-oriented StereoSR (PSSR) approaches. Then, our PSSR is compared with state-of-the-arts in terms of quantitative, qualitative, practicability, and running efficiency. Finally, we also report the performance of our proposed StereoSRQA.

\subsection{Database and Protocols}
By following the StereoSR works~\cite{stereosr, our}, our PSSR model is trained on 59 stereo image pairs from the Middlebury. Training images are augmented by randomly down-scaling, flipping and rotating. Each mini-batch contains 32 HR stereo image patches of size 120. The StereoSR models are evaluated on the same test sets in \cite{our} (Middlebury~\cite{middlebury}, Tsukuba~\cite{tsukuba}, KITTI2012~\cite{kt12}, KITTI2015~\cite{kt15}, and SceneFlow~\cite{sceneflow}). All SR models are evaluated with various quality evaluation methods and the StereoSRQA model.

The StereoSRQA models are trained and tested on the Mid3D\_QA database, and work on stereo image patches of size 120. In test, suppose a stereo image pair is cropped into $N$ patch pairs, the final score is obtained by averaging output scores of all patches. All StereoSRQA models are measured with common correlation criterions, including Pearson linear correlation (PLCC), Spearman rank order correlation coefficient (SROCC), Kendall rank-order correlation coefficient (KROCC), and Root mean square error (RMSE). The higher SROCC, PLCC, KROCC, and lower RMSE denote the better performance.

All models are based on the TensorFlow~\cite{tensorflow} implementation and optimized by Adam~\cite{adam} with $\beta$1 = 0.9, $\beta$2 = 0.999. The learning rate is 1e-4. We conduct experiments on the machine with Nvidia GTX1080Ti GPU (128G RAM).

\begin{figure}[tb]
\centering \subfigure{
   \begin{minipage}[t]{2.6cm} \centering  \includegraphics[width=2.65cm]{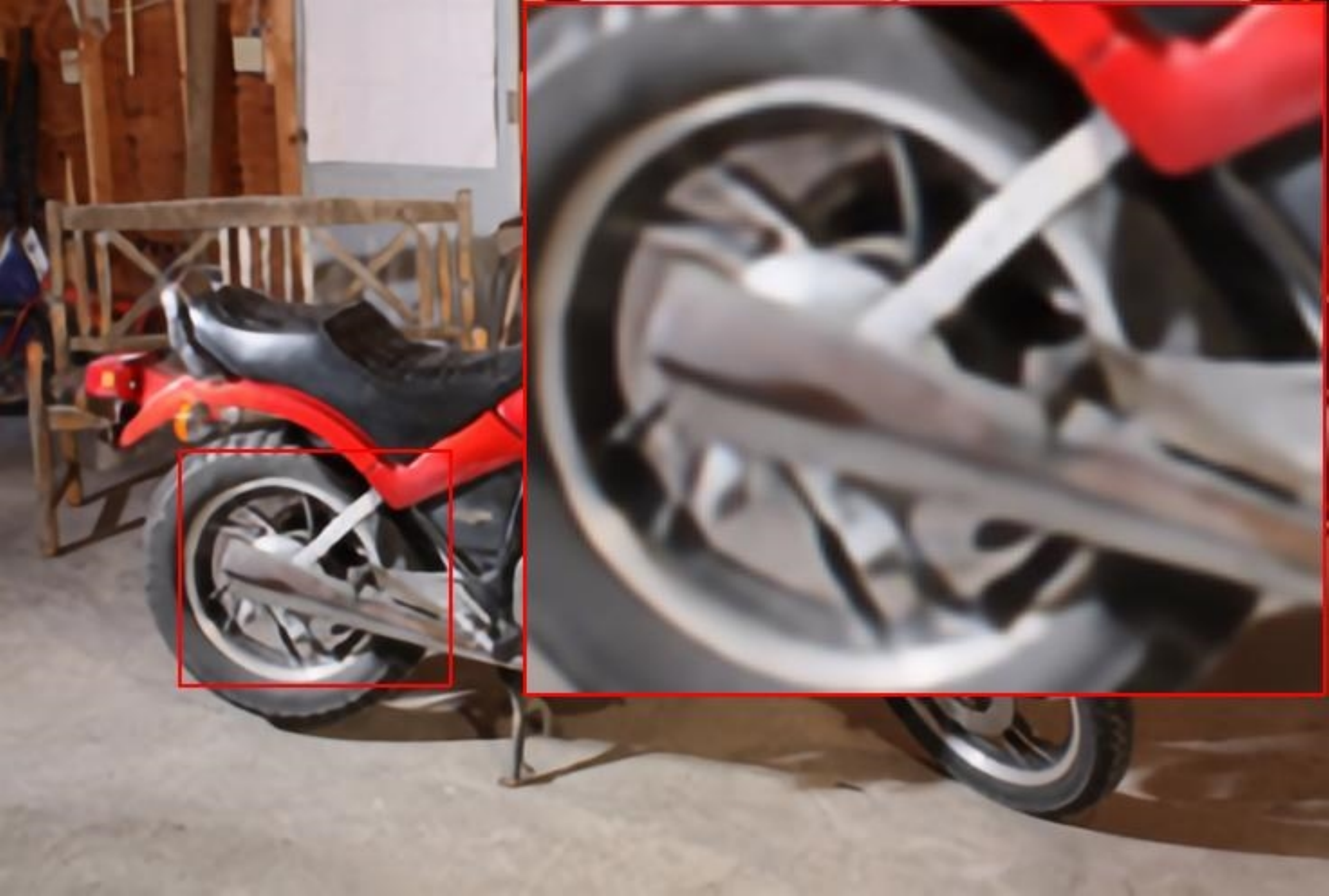} \\ \centering{\small{$\mathcal{L}_{MSE}$}} \end{minipage}
    \begin{minipage}[t]{2.6cm}  \centering  \includegraphics[width=2.65cm]{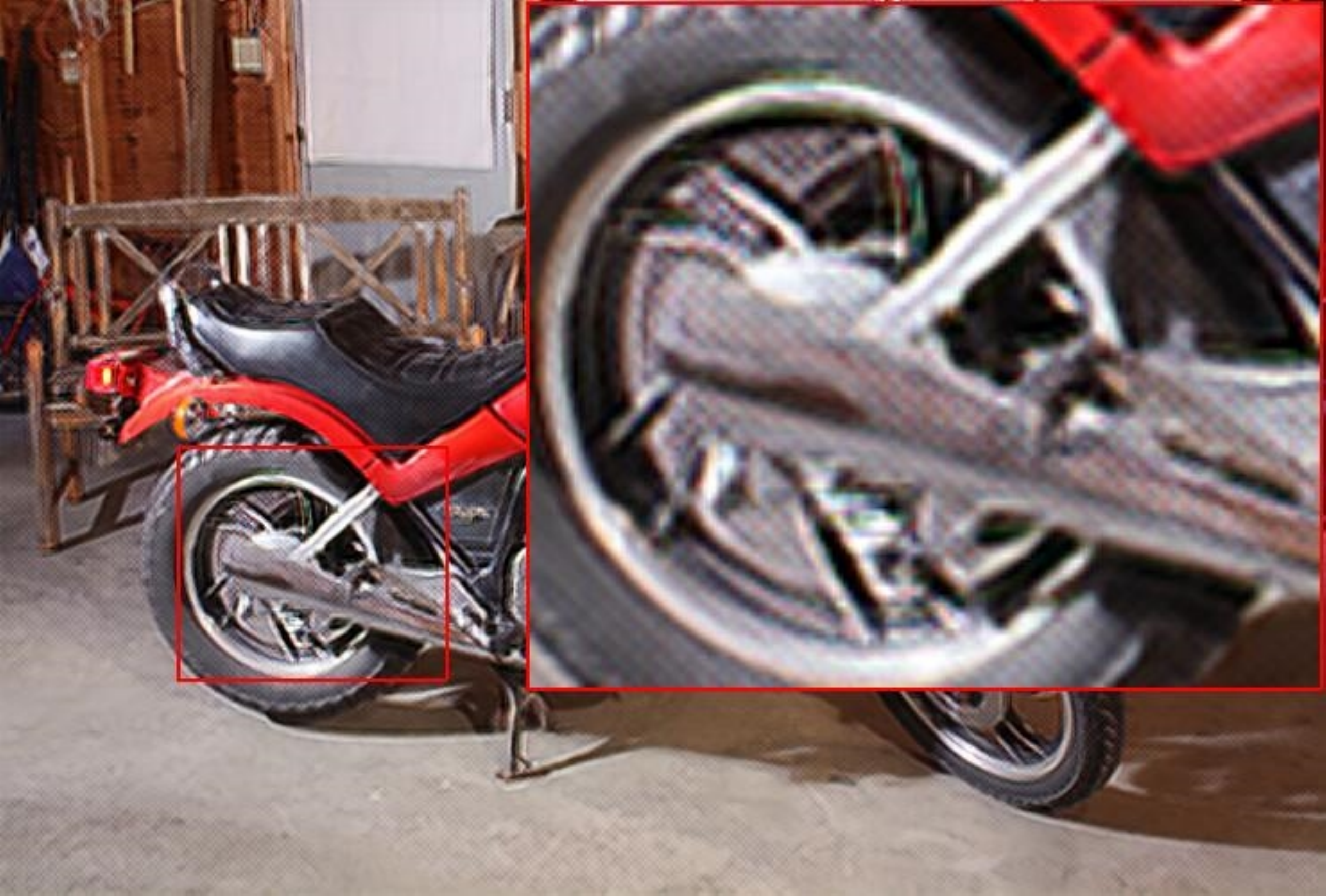} \\ \centering{\small{+ $\mathcal{L}_{VGG}$}}  \end{minipage}
  \begin{minipage}[t]{2.6cm}  \centering \includegraphics[width=2.65cm]{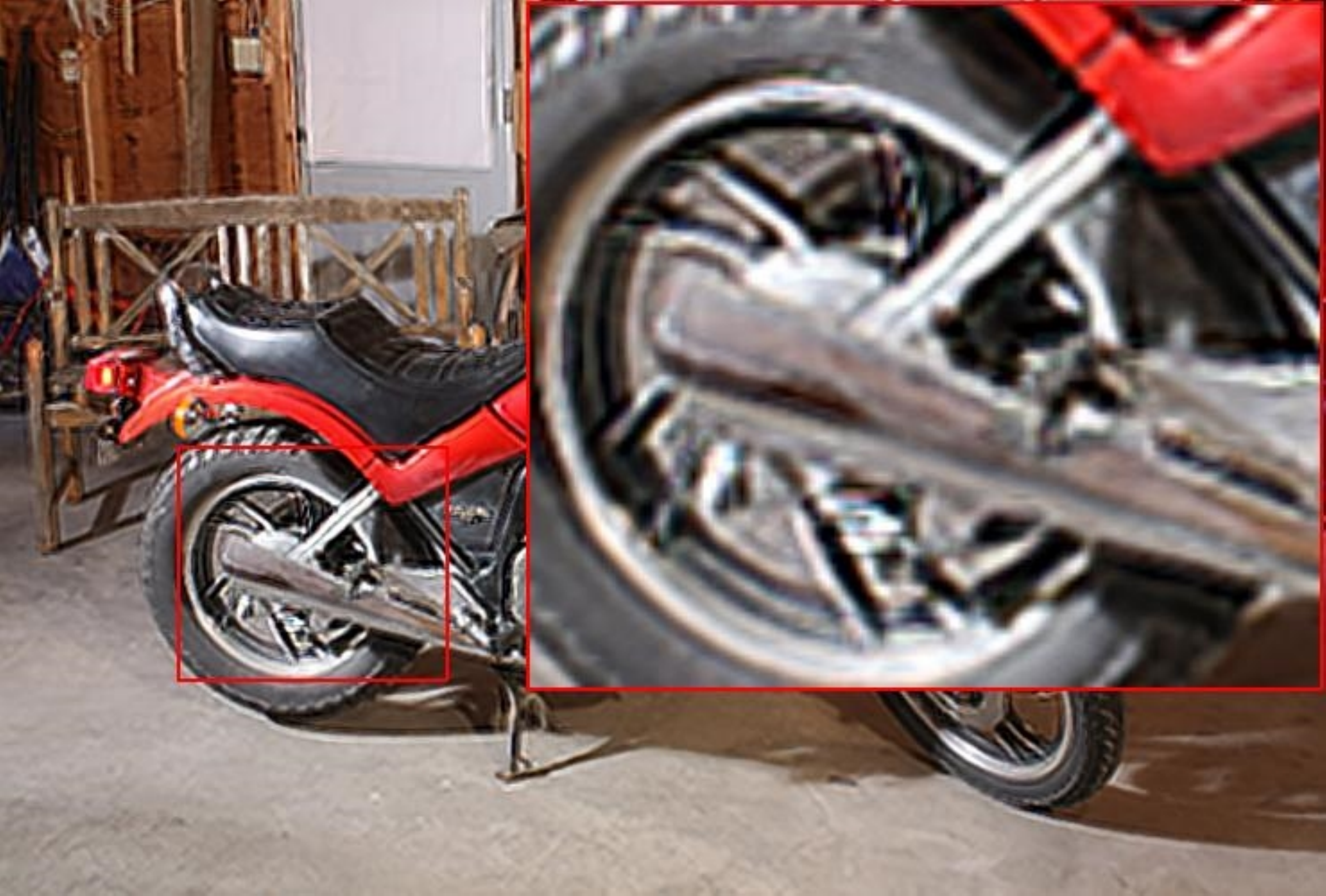} \\ \centering{\small{+ $\mathcal{L}_{IQP_{SISR}}$}} \end{minipage}
   }\
\centering \subfigure{
     \begin{minipage}[t]{2.6cm}  \centering \includegraphics[width=2.65cm]{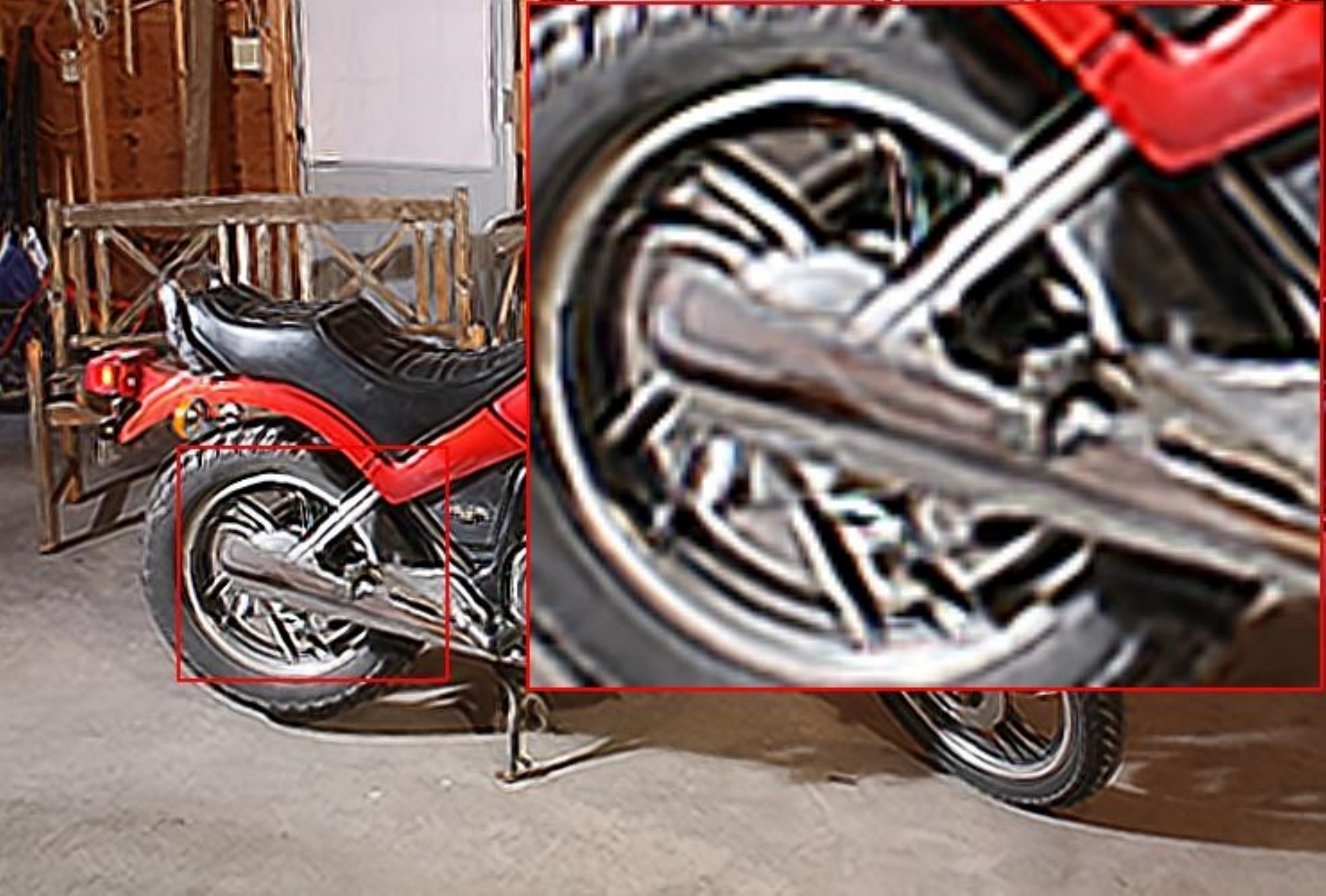} \\ \centering{\small{+ $\mathcal{L}_{IQP_{im}}$}} \end{minipage}
   \begin{minipage}[t]{2.6cm}  \centering \includegraphics[width=2.65cm]{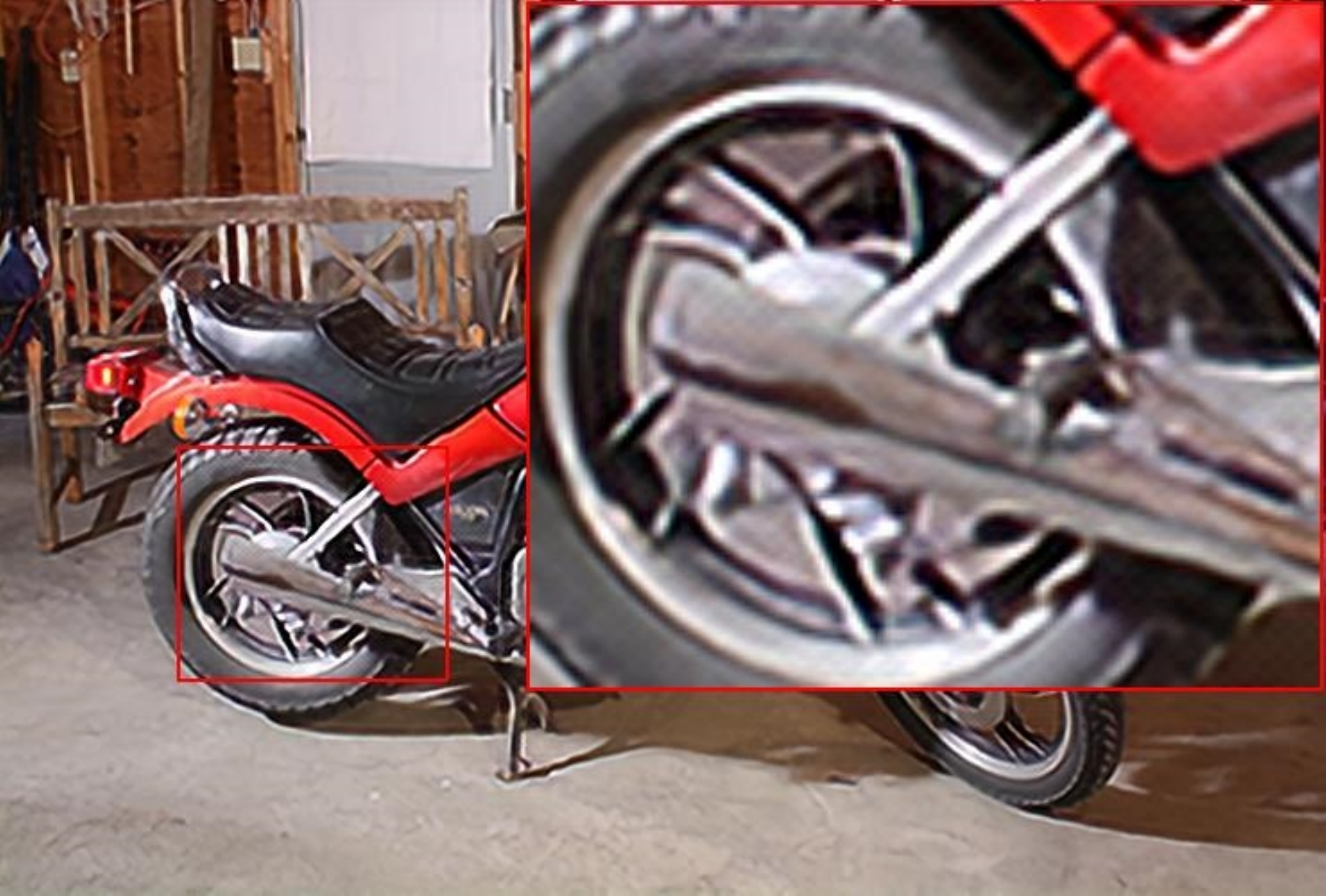}\\ \centering{\small{+ $\mathcal{L}_{IQP}$}} \end{minipage}
  \begin{minipage}[t]{2.6cm} \centering  \includegraphics[width=2.65cm]{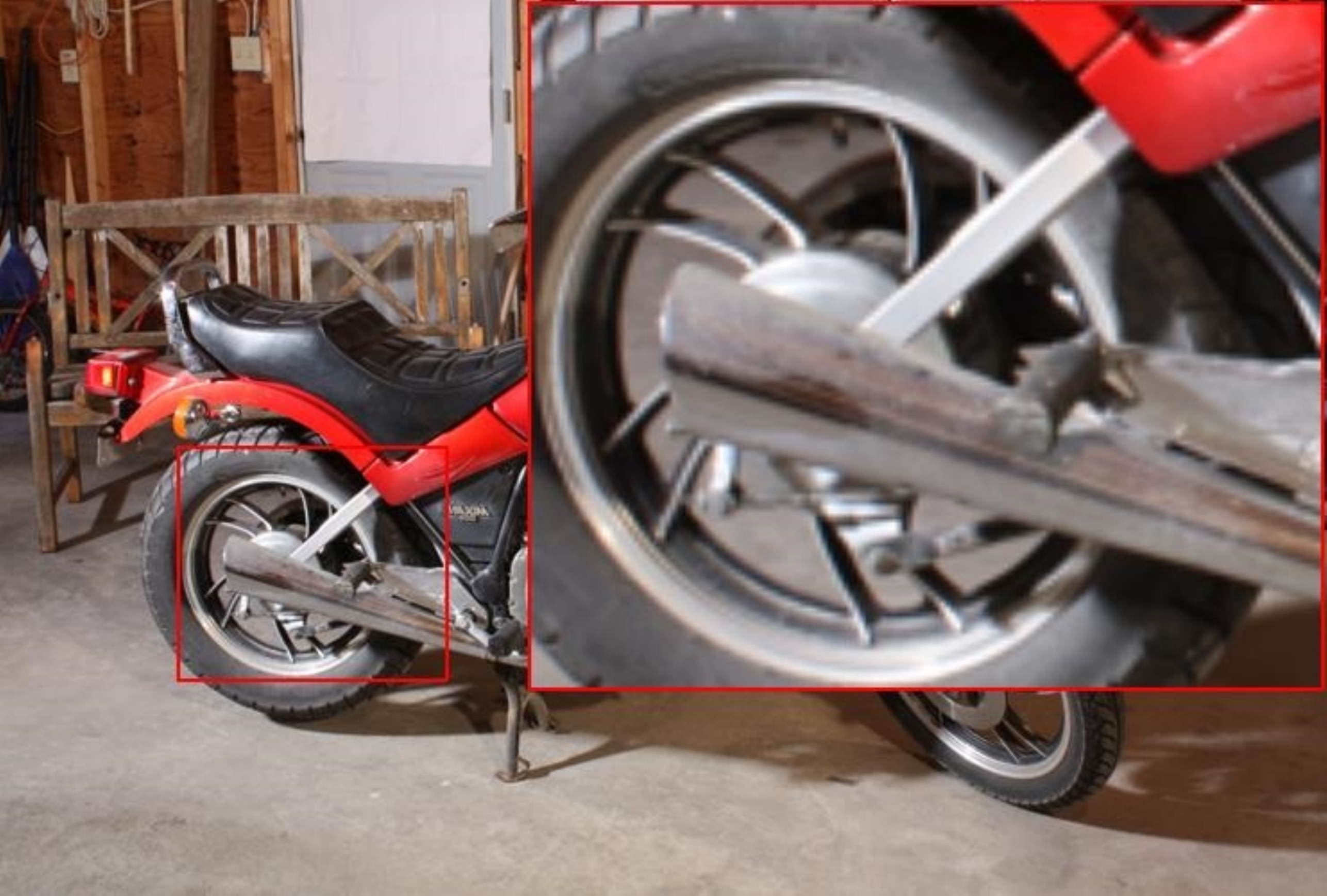} \\ \centering{\small{GT}} \end{minipage}}\  \vspace{-0.2em}
\caption{Visual comparison of StereoSR models trained with different losses. Better zoom in.}
\label{fabsr} \end{figure}

\begin{table}[htb]
\caption{Comparison of StereoSR models trained with different losses. All models are evaluated with both full reference (PSNR, SSIM) and no reference (SFA, CNNIQA, StereoQA, our StereoSRQA) metrics. $\uparrow$ denotes the higher is the better.} \label{tabsr}
\vspace{-0.2em}  \centering \scalebox{0.7}[0.7]{
    \begin{tabular}{c|cccccc}
    \toprule \toprule
 &PSNR$\uparrow$&SSIM$\uparrow$&CNNIQA$\downarrow$&SFA$\downarrow$&StereoQA$\downarrow$&StereoSRQA$\uparrow$\\
\hline
$\mathcal{L}_{MSE}$&\textbf{29.830}&\textbf{0.9090}&41.7167&36.4495& 21.0190&6.7488\\
+ $\mathcal{L}_{VGG}$&28.588&0.889&30.2386&29.1817&18.3689&6.7600\\
+ $\mathcal{L}_{GAN}$&28.993&0.8994&31.2568 &30.8602&19.6349&8.2891\\
+ $\mathcal{L}_{IQP_{SISR}}$&28.855&0.8936&32.3324&\textbf{27.3597}&18.7177&6.6930\\
+ $\mathcal{L}_{IQP_{im}}$&29.099&0.9010&32.2135&28.4532&19.6183&8.3420\\ 
+ $\mathcal{L}_{IQP}$&29.270&0.9015&\textbf{28.3073}&27.8893&\textbf{18.2394}&\textbf{8.7343}\\
 \bottomrule  \end{tabular}  } \end{table}
\begin{table*}[tb]
\caption{Quantitative comparison with state-of-the-art SR approaches in terms of perceptual quality evaluation metrics (NIQE/PI/StereoSRQA). $\uparrow$ denotes the higher is the better. The GAN-based SISR method performs better the groundtruth on NIQE and PI scores, indicating the two IQA metrics can not well reflect the real quality of stereo image.} \vspace{-0.2em} \label{tiqp1}
\centering  \scalebox{0.7}[0.7]{
\begin{tabular}{cc|c|ccc|c|ccc|c|c}
\toprule \toprule
\multicolumn{3}{c}{NIQE$\downarrow$/PI$\downarrow$/StereoSRQA$\uparrow$}&\multicolumn{3}{c}{MSE-based SISR}&\multicolumn{1}{c}{GAN-based SISR}& \multicolumn{3}{c}{Stereo SR}&  \multicolumn{1}{c}{Our}&HR \\
\hline
Database&Scale&Bicubic&EDSR*~\cite{edsr}&SRFBN~\cite{srfbn}&DRN~\cite{drn}&SPSR~\cite{spsr}&StereoSR~\cite{stereosr}& PASSRnet~\cite{passr}&DASSR~\cite{our}&PSSR&\\
\hline \multirow{3}[0]{*}{Middlebury}
    &$\times$4&7.62/7.04/4.89&5.32/5.38/7.26&5.96/5.76/7.43&5.51/3.96/8.16 &\textbf{3.37/2.83}/8.55&5.38/5.24/6.69&5.97/5.72/7.22&6.35/5.99/6.75&4.31/3.88/\textbf{8.73}&\multirow{3}[0]{*}{\textcolor[rgb]{1,0,0}{3.38/2.84/9.20}}\\
    &$\times$3&6.74/6.41/6.49&5.32/5.38/8.44&5.22/4.70/8.65&-&-&4.57/3.91/8.09&-&5.28/4.78/7.77&\textbf{5.52/3.82/9.31}&\\
    &$\times$2&5.13/5.21/8.55&4.17/3.57/9.17&4.59/3.55/9.18&-&-&4.38/3.49/9.08&-&4.43/3.51/9.16&\textbf{5.97/3.86/9.60}&\\
\hline \multirow{3}[0]{*}{KITTI 2012}
    &$\times$4&7.49/6.87/5.10&5.84/5.72/7.17&5.97/5.73/6.42&5.55/4.22/8.11&\textbf{2.49/2.32}/8.18&4.67/4.09/7.14&5.49/5.16/7.55&5.54/5.11/7.64&3.70/3.44/\textbf{8.53}&\multirow{3}[0]{*}{\textcolor[rgb]{1,0,0}{3.42/2.78/9.43}}\\
    &$\times$3&6.67/6.32/6.90&5.86/5.72/8.75&5.16/4.46/7.97&-&-&4.21/3.62/8.64&-&5.33/4.59/8.75&\textbf{4.72/3.52/9.38}&-\\
    &$\times$2&5.04/4.88/8.98&4.12/3.17/9.52&4.18/3.21/8.71&-&-&4.199/3.54/9.22&-&4.20/3.22/9.50&\textbf{4.95/3.61/9.78}&-\\
\hline \multirow{3}[0]{*}{KITTI 2015}
    &$\times$4&7.42/6.86/5.34&5.70/5.88/7.38&6.00/5.96/6.55&5.55/4.22/8.11&\textbf{2.42/2.47}/8.44
    &4.59/4.18/7.31&5.39/5.32/7.82&5.51/5.31/7.64&4.03/3.79/\textbf{8.69}&\multirow{3}[0]{*}{\textcolor[rgb]{1,0,0}{3.239/3.04/9.52}}\\
&$\times$3&7.35/6.75/7.08&5.70/5.88/8.98&5.24/4.72/8.16&-&-&4.10/3.65/8.81&-&5.35/4.82/8.96&\textbf{4.84/3.80/9.52}&-\\
&$\times$2&5.09/5.10/9.15&4.21/3.57/9.66&4.31/3.71/8.79&-&-&4.11/3.59/9.35&-&4.18/3.65/9.64&\textbf{4.56/3.56/9.90}&-\\
\hline \multirow{3}[0]{*}{Tsukuba}
    &$\times$4&7.58/7.24/4.38&6.04/5.68/7.17&5.96/5.62/6.40&5.65/4.44/8.39
    &\textbf{4.17/3.89}/7.26&5.67/5.32/5.64&5.73/5.47/7.34&5.62/5.28/7.64&4.19/3.65/\textbf{7.76}&\multirow{3}[0]{*}{\textcolor[rgb]{1,0,0}{4.87/4.59/8.04}}\\
&$\times$3&6.94/6.54/5.84&6.04/5.68/8.00&5.54/5.10/7.21&-&-&4.61/3.67/6.83&-&5.61/5.20/8.10&\textbf{5.20/3.69/8.58}&-\\
&$\times$2&5.83/5.58/7.62&5.05/4.80/8.34&5.06/4.81/7.51&-&-&4.35/3.00/7.59&-&5.18/4.85/8.27&\textbf{6.01/3.82/8.73}&-\\
\hline \multirow{3}[0]{*}{SceneFlow}
    &$\times$4&6.99/6.70/4.53&5.39/5.05/6.77&5.59/5.04/6.94&5.26/4.24/7.23&\textbf{2.96/2.14}/7.24
    &4.71/3.99/6.38&5.36/4.60/6.95&5.42/4.80/7.02&4.30/3.24/\textbf{7.76}&\multirow{3}[0]{*}{\textcolor[rgb]{1,0,0}{3.10/2.30/8.41}}\\
&$\times$3&7.05/6.23/6.05&5.39/5.05/7.86&4.99/3.95/7.23&-&-&3.96/2.77/6.87&-&5.04/4.13/8.16&\textbf{4.93/3.04/8.57}&-\\
&$\times$2&5.08/4.90/7.81&4.28/3.05/8.50&4.15/2.93/7.71&-&-&4.10/2.85/7.56&-&4.27/2.99/8.44&\textbf{4.49/2.77/8.88}&-\\
\bottomrule \end{tabular}  } \end{table*}

\subsection{Ablation Study}
The ablation study is conducted for $\times$4 SR on the Middlebury database. Table~\ref{tabsr} and Figure~\ref{fabsr} compare different constraints, used to optimize the StereoSR model. We first train a baseline model (`$\mathcal{L}_{MSE}$') with the MSE loss only, and then train other models with different combinations of losses. The `+ $\mathcal{L}_{*}$' denotes that the model is trained with the MSE and $\mathcal{L}_{*}$ loss.

From Figure~\ref{fabsr}, the perceptual loss~\cite{perceptual} makes the model `+ $\mathcal{L}_{VGG}$' generate some visual artifacts, such as checkerboard textures. We replace our StereoSRQA model with SISRQA model~\cite{sisrqa} to compute the feature similarity for the model `+ $\mathcal{L}_{IQP_{SISR}}$', which has better result on SFA. This is mainly because that $\mathcal{L}_{IQP_{SISR}}$ enforces the SR results to have better subjective quality on single view, and the NR IQA method (SFA) pays more attention on quality of single view and prefers results of  `+ $\mathcal{L}_{IQP_{SISR}}$'. In comparison, our `+ $\mathcal{L}_{IQP}$' gets better visual results and StereoSRQA score. 

\begin{figure*}[tb]
\centering \subfigure{
    \begin{minipage}[t]{2.50cm}  \centering  \includegraphics[width=2.50cm]{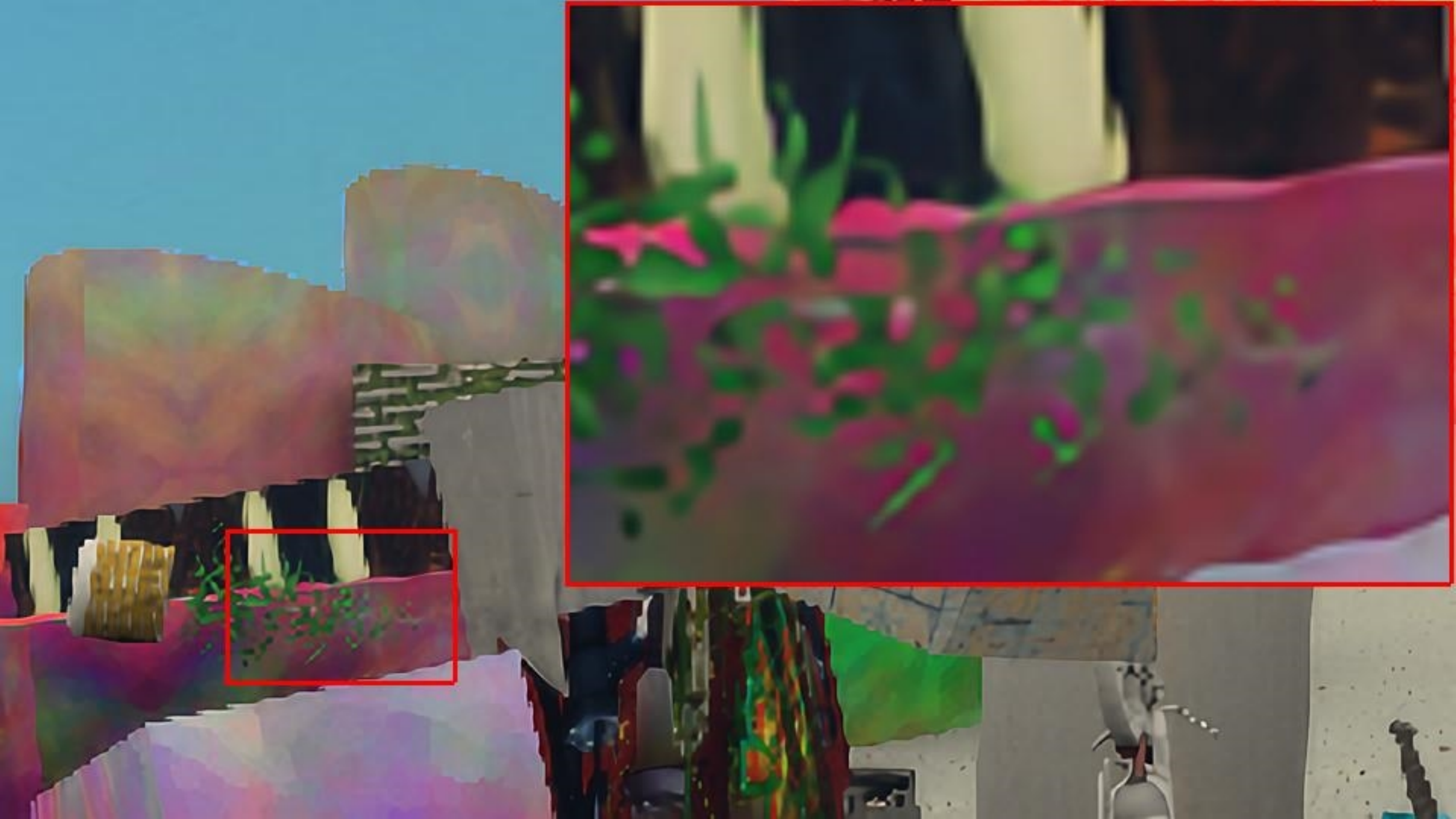}  \end{minipage}
    \begin{minipage}[t]{2.50cm}  \centering  \includegraphics[width=2.50cm]{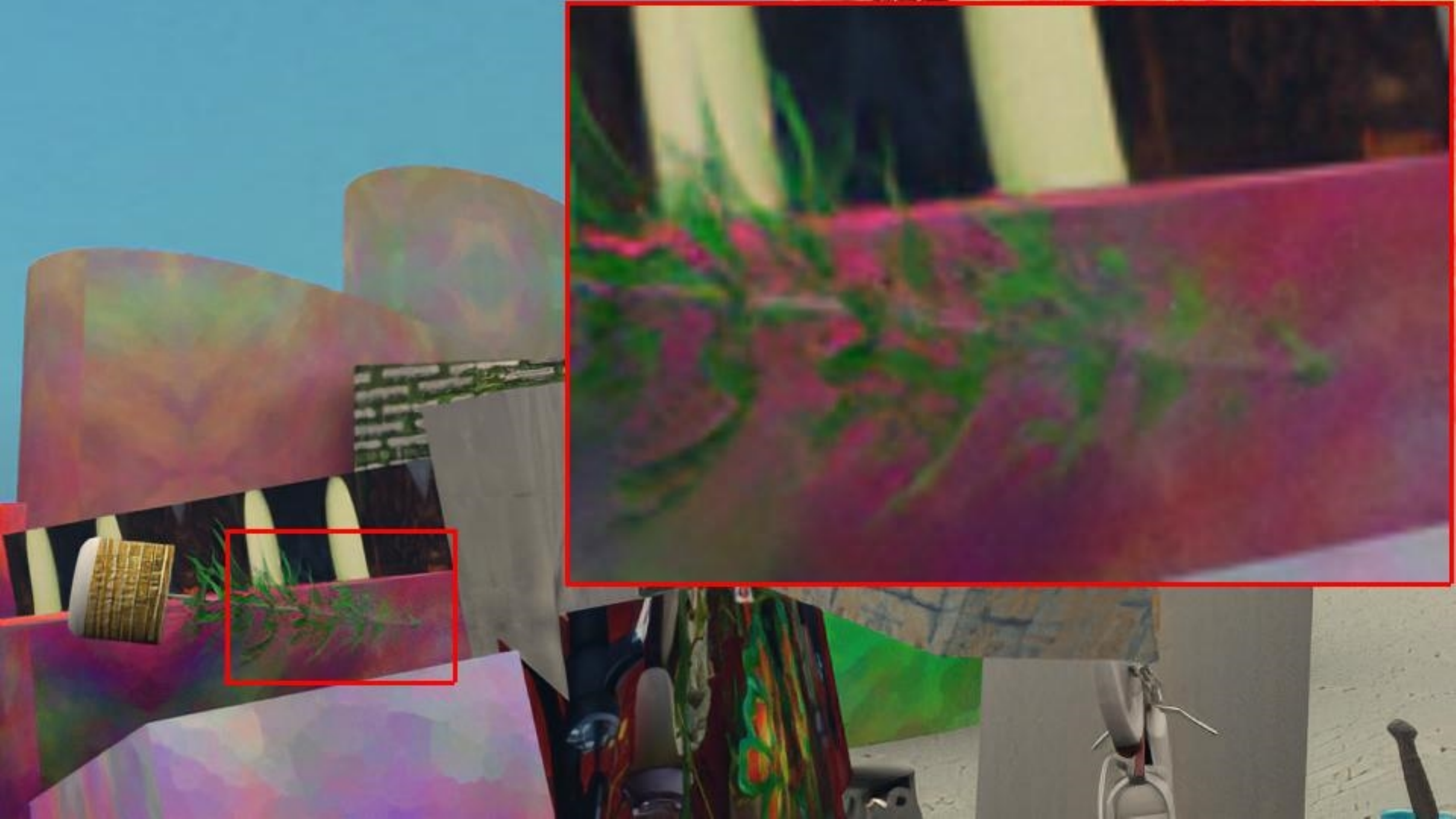}   \end{minipage}
    \begin{minipage}[t]{2.50cm}  \centering \includegraphics[width=2.50cm]{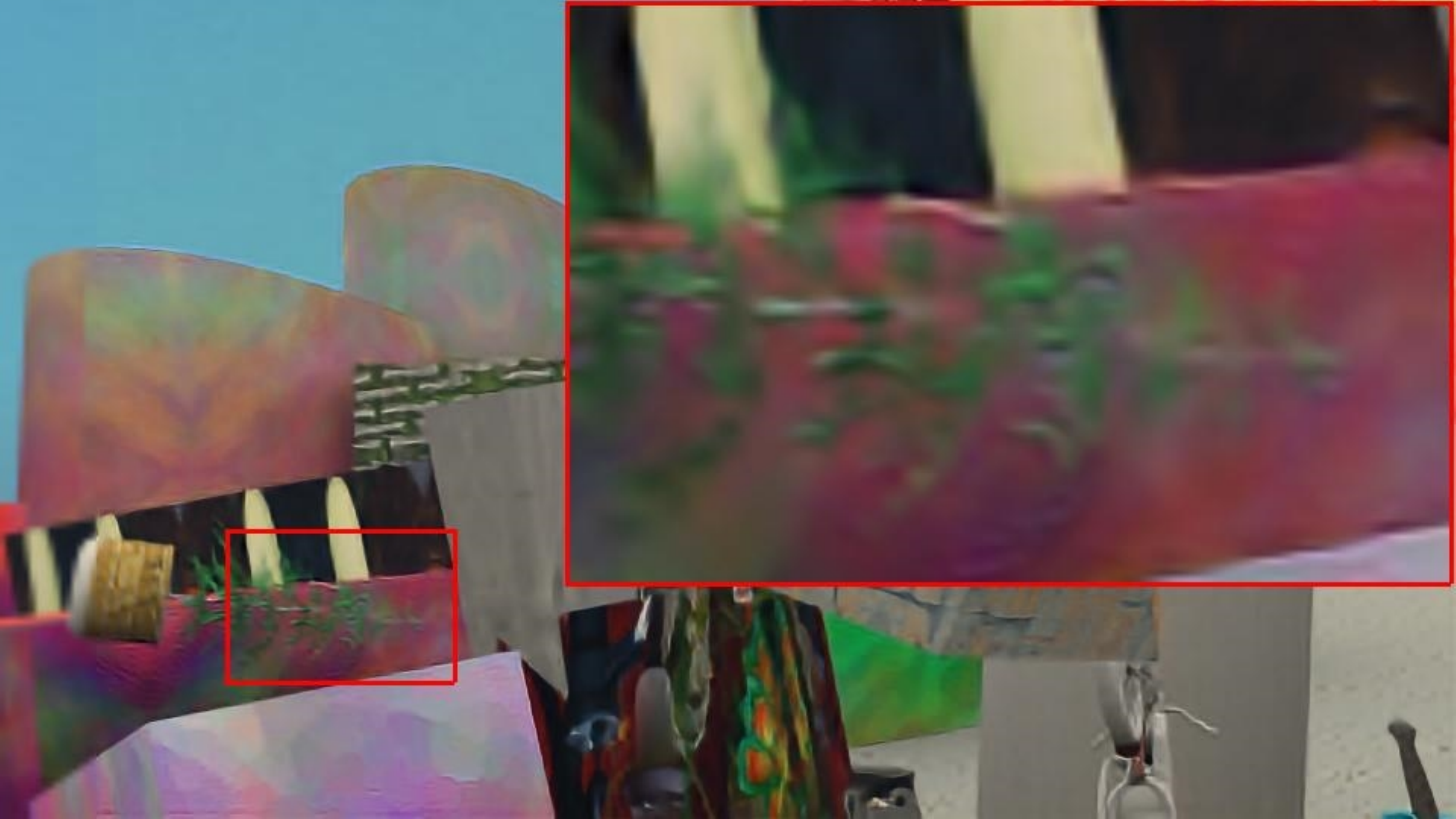}  \end{minipage}
   \begin{minipage}[t]{2.50cm}  \centering \includegraphics[width=2.50cm]{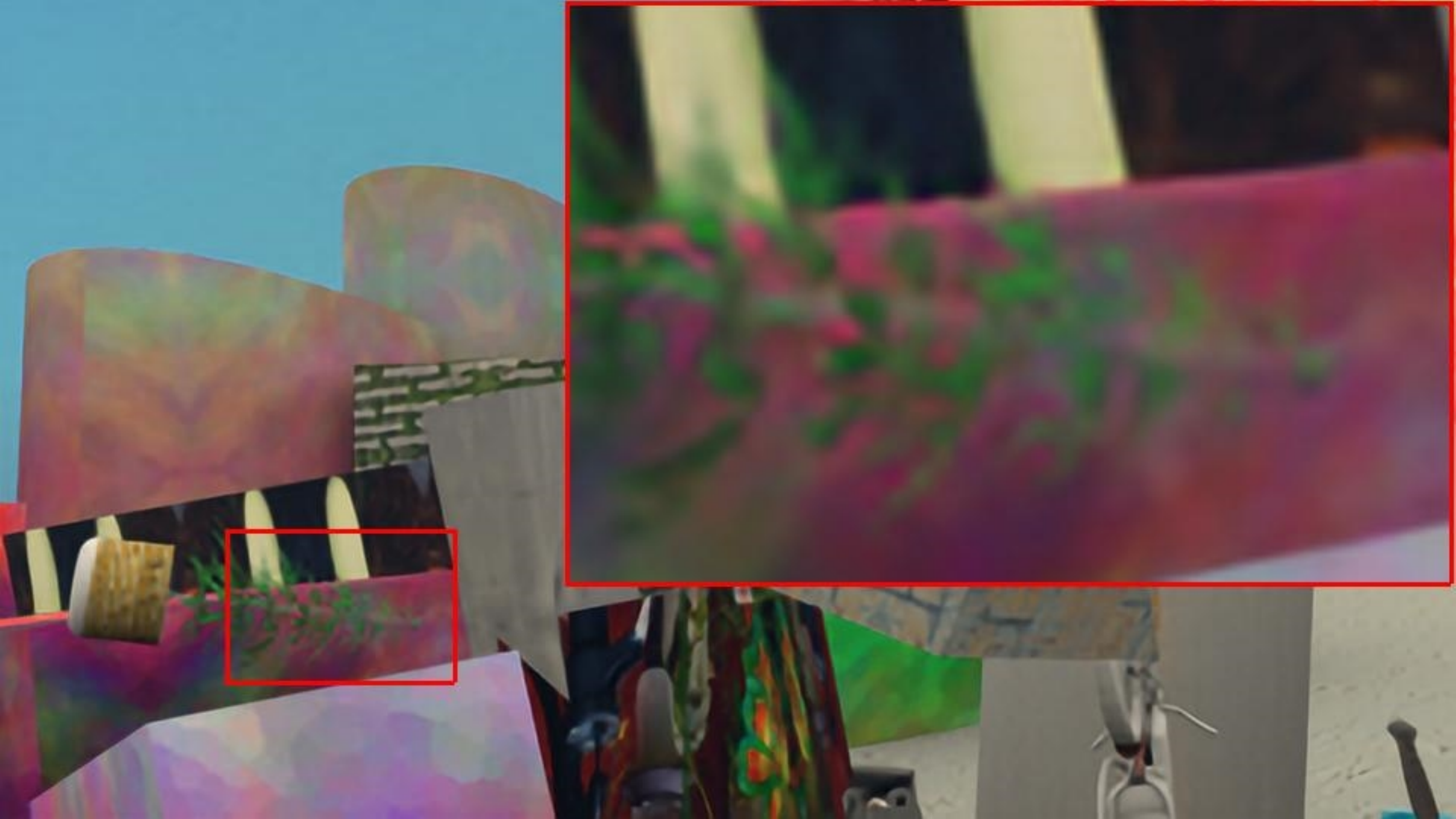}  \end{minipage}
  \begin{minipage}[t]{2.50cm}  \centering \includegraphics[width=2.50cm]{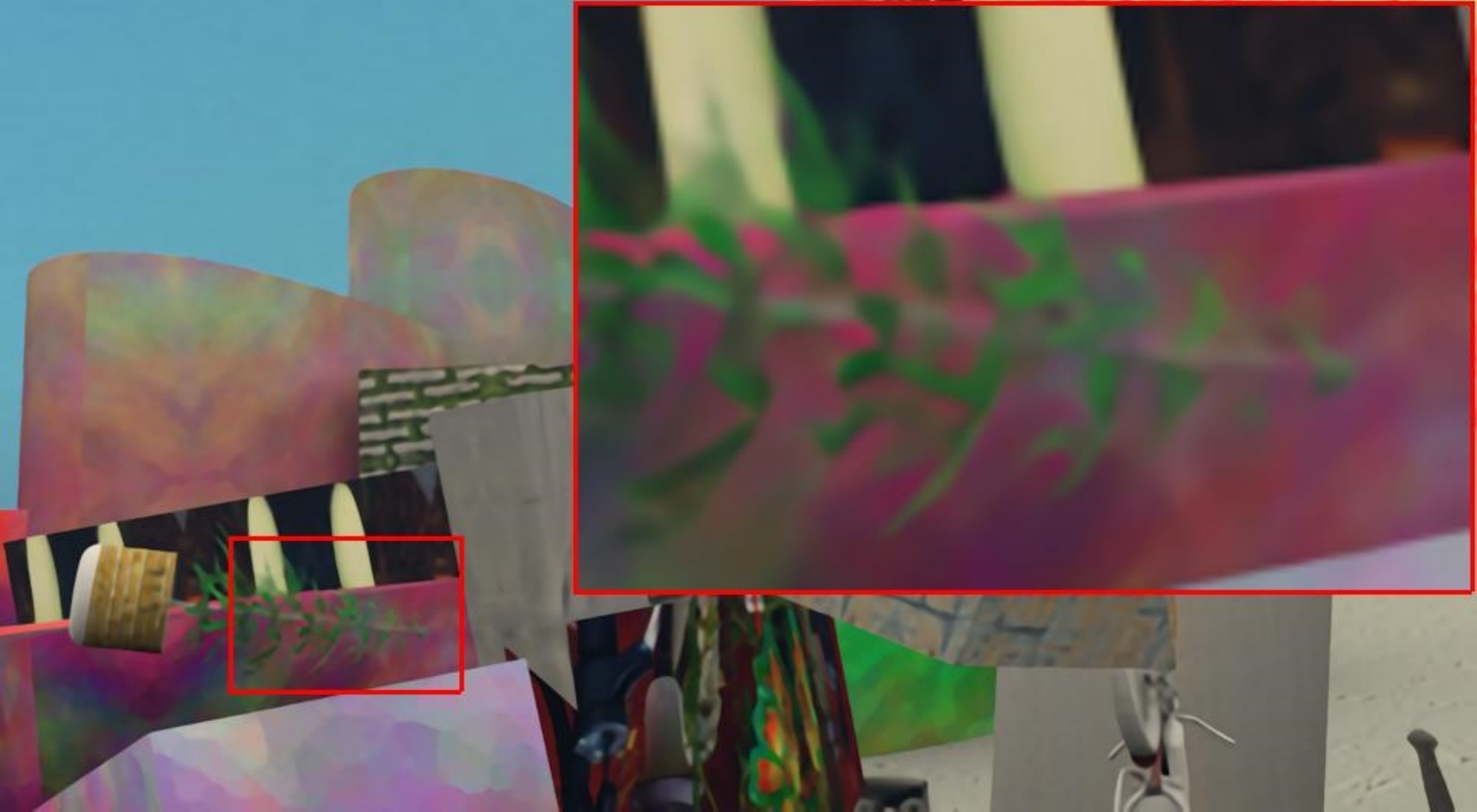}  \end{minipage}
   \begin{minipage}[t]{2.50cm}  \centering \includegraphics[width=2.50cm]{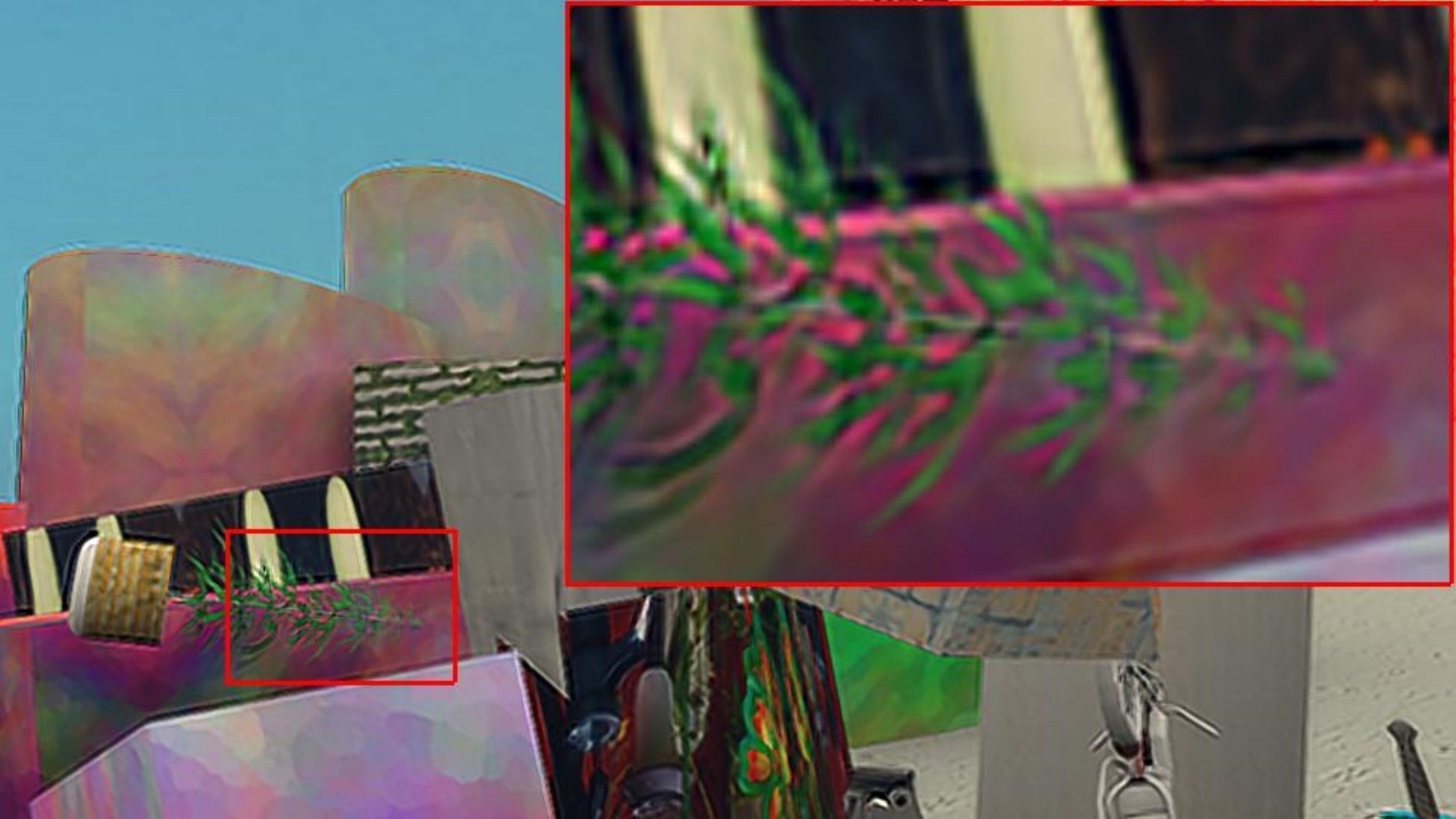} \end{minipage}
   \begin{minipage}[t]{2.50cm}  \centering \includegraphics[width=2.50cm]{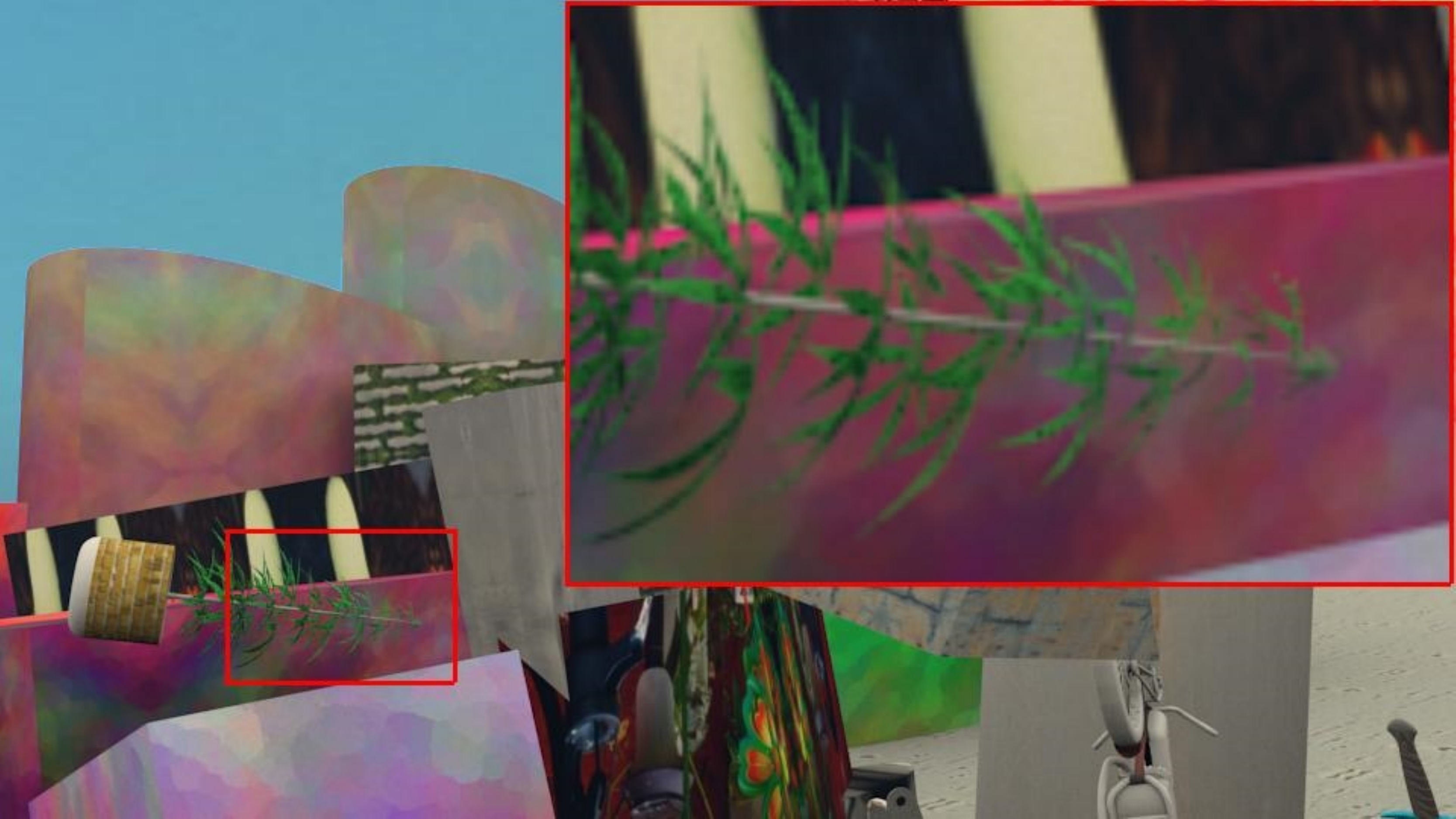} \end{minipage}
}\  \centering
\subfigure{
   \begin{minipage}[t]{2.50cm}  \centering  \includegraphics[width=2.50cm]{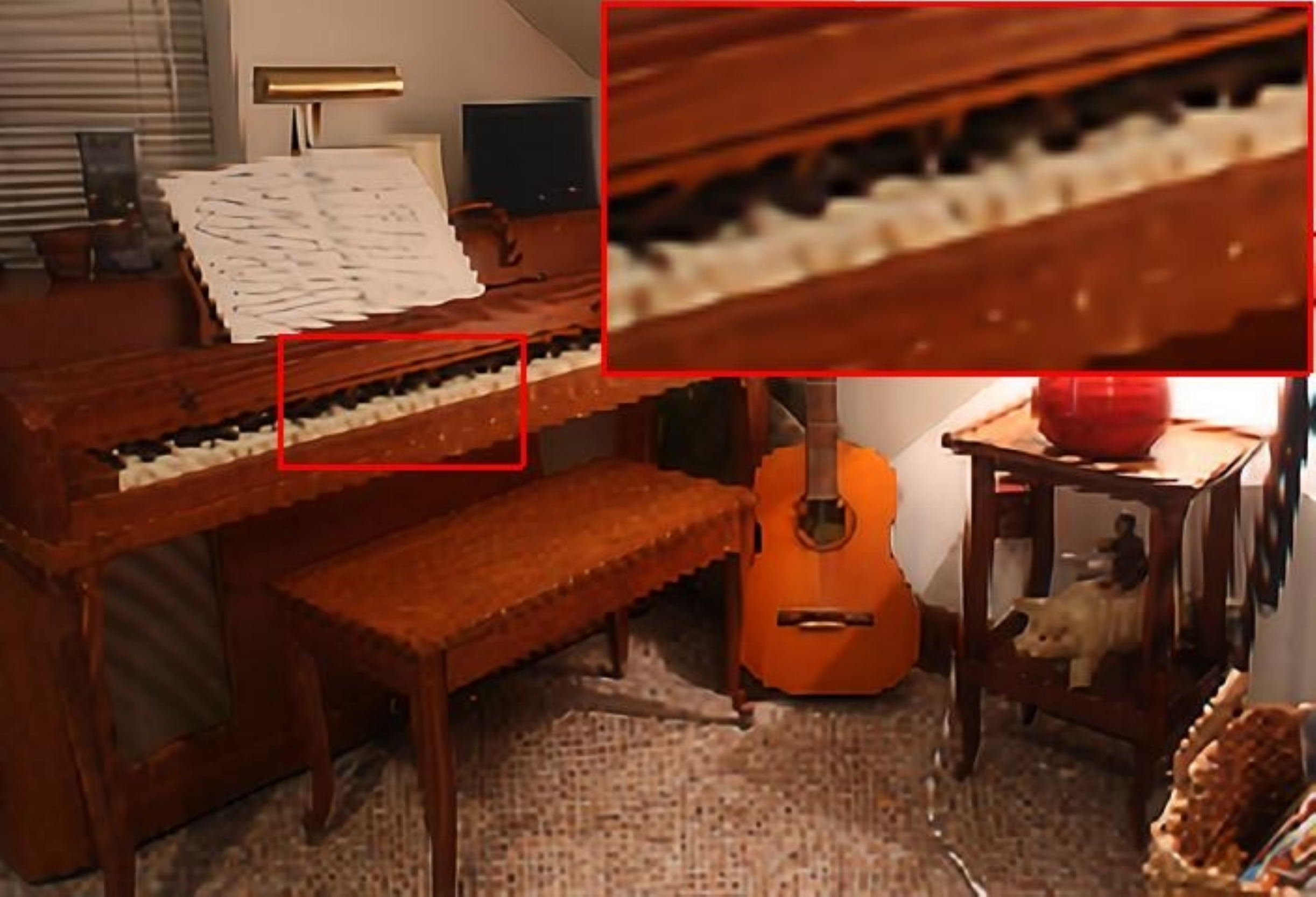} \\ \centering{\small{DRN}~\cite{drn}} \end{minipage}
    \begin{minipage}[t]{2.50cm}  \centering  \includegraphics[width=2.50cm]{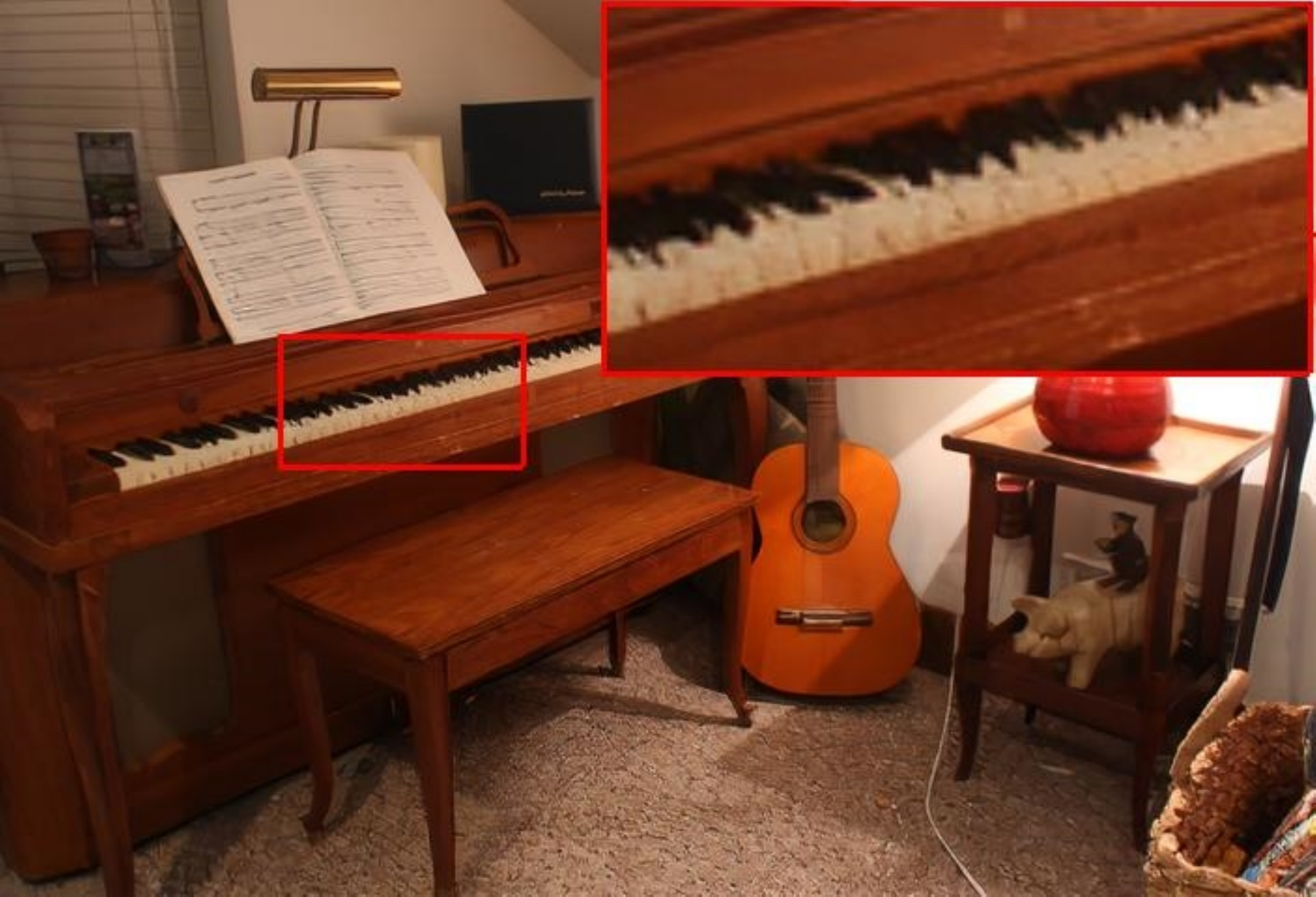} \\ \centering{\small{SPSR~\cite{spsr}}}  \end{minipage}
    \begin{minipage}[t]{2.50cm}  \centering \includegraphics[width=2.50cm]{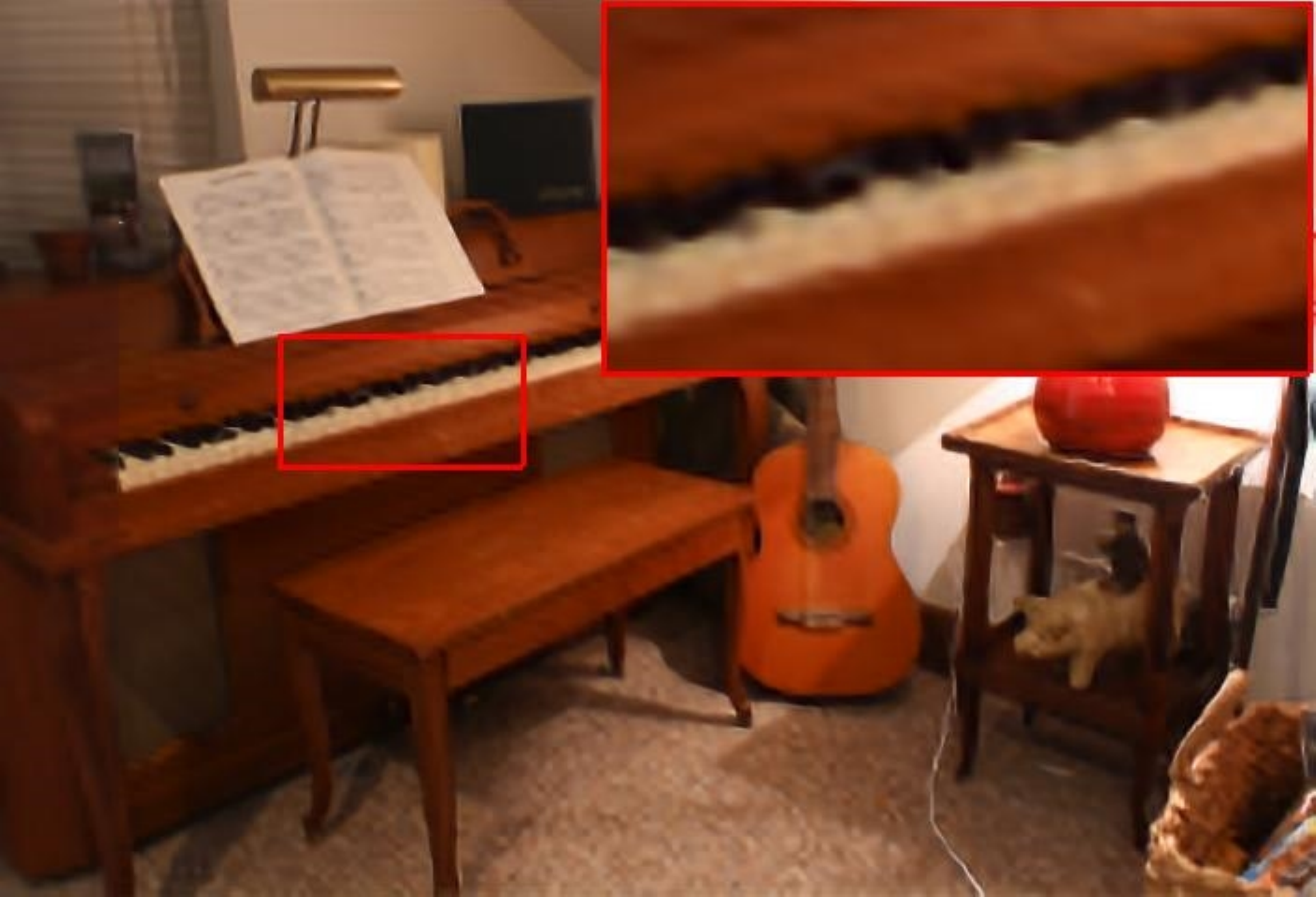} \\ \centering{\small{StereoSR~\cite{stereosr}}} \end{minipage}
   \begin{minipage}[t]{2.50cm}  \centering \includegraphics[width=2.50cm]{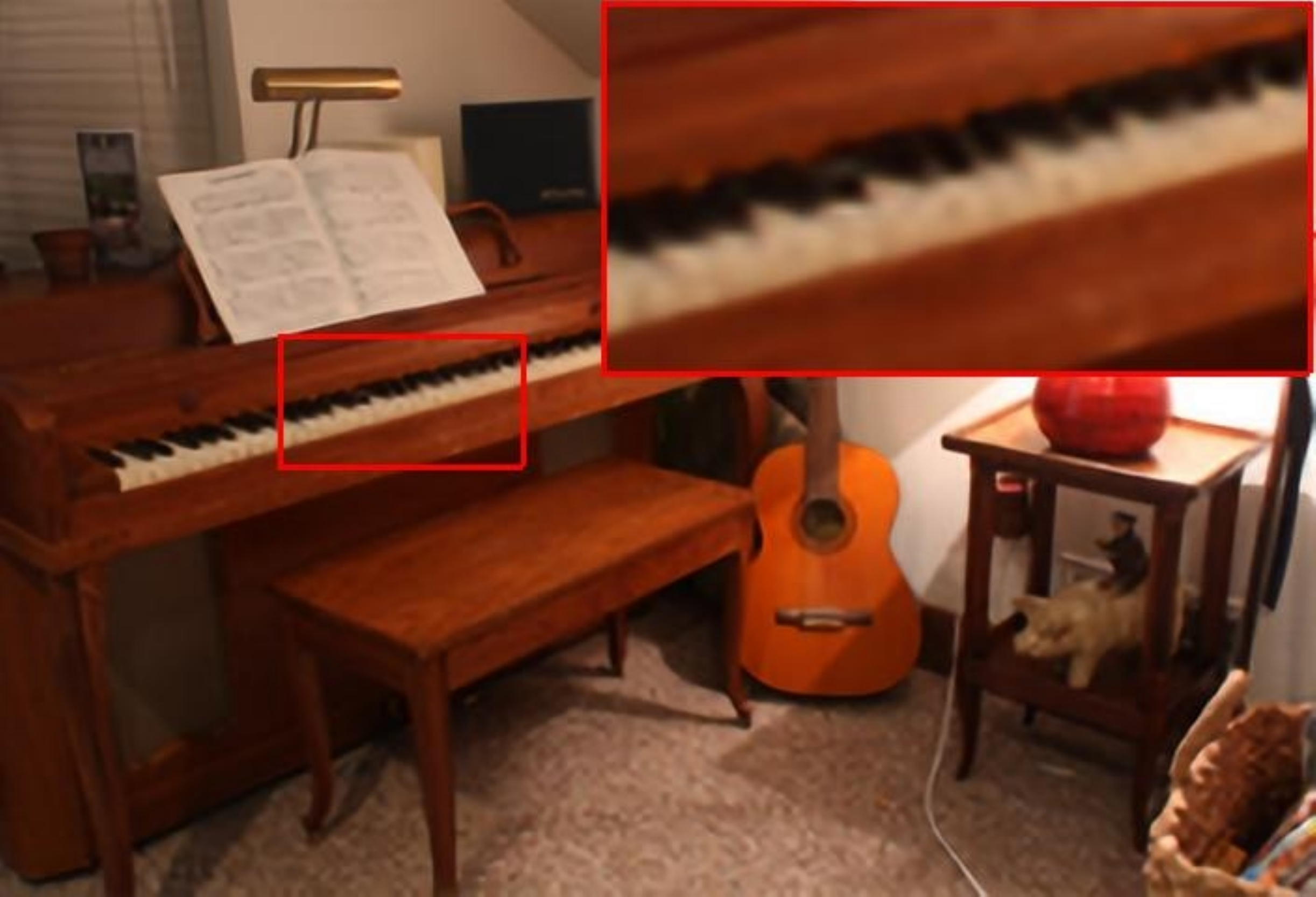} \\ \centering{\small{PASSR~\cite{passr}}} \end{minipage}
\begin{minipage}[t]{2.50cm}  \centering \includegraphics[width=2.50cm]{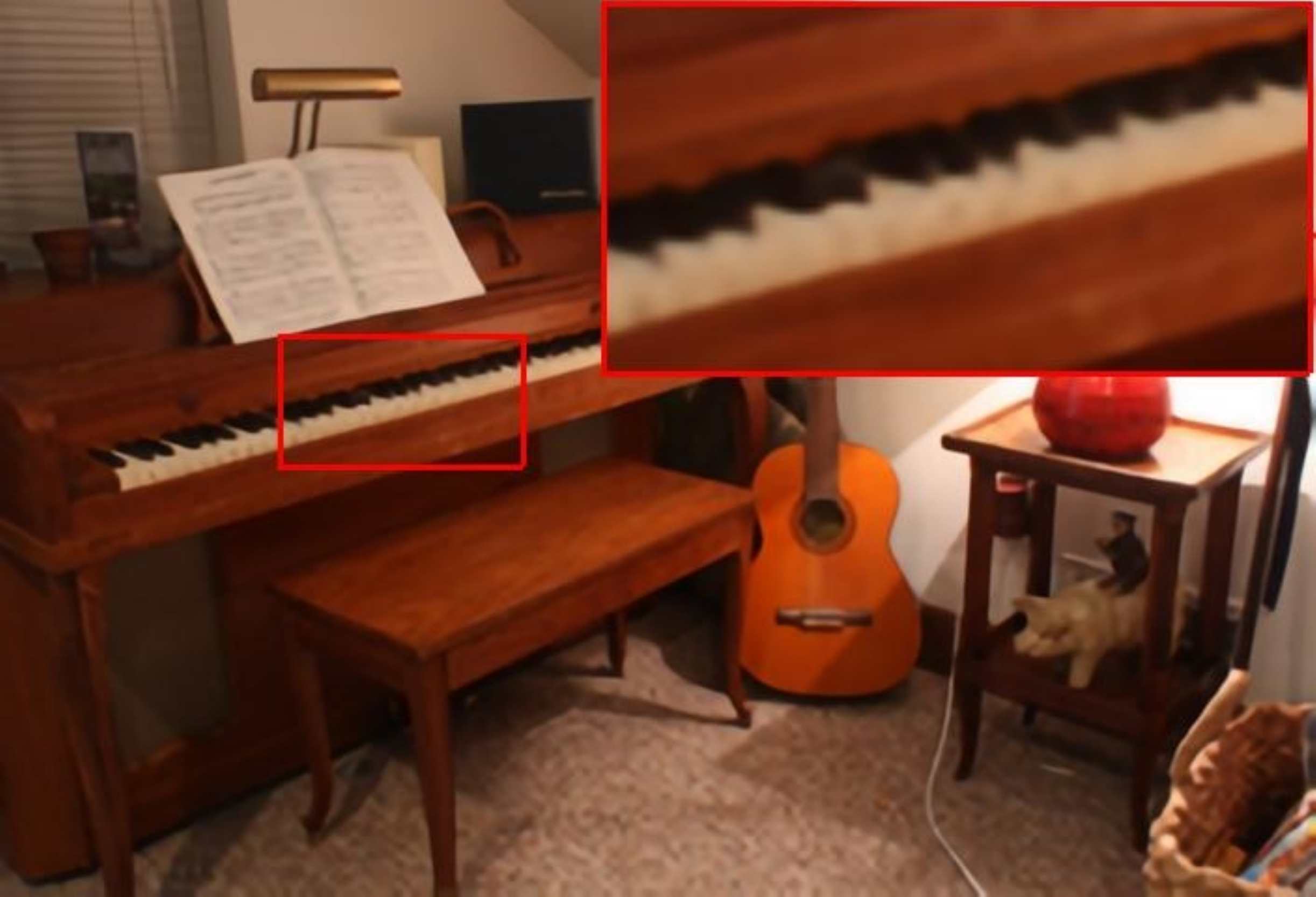} \\ \centering{\small{DASSR~\cite{our}}} \end{minipage}
   \begin{minipage}[t]{2.50cm}  \centering \includegraphics[width=2.50cm]{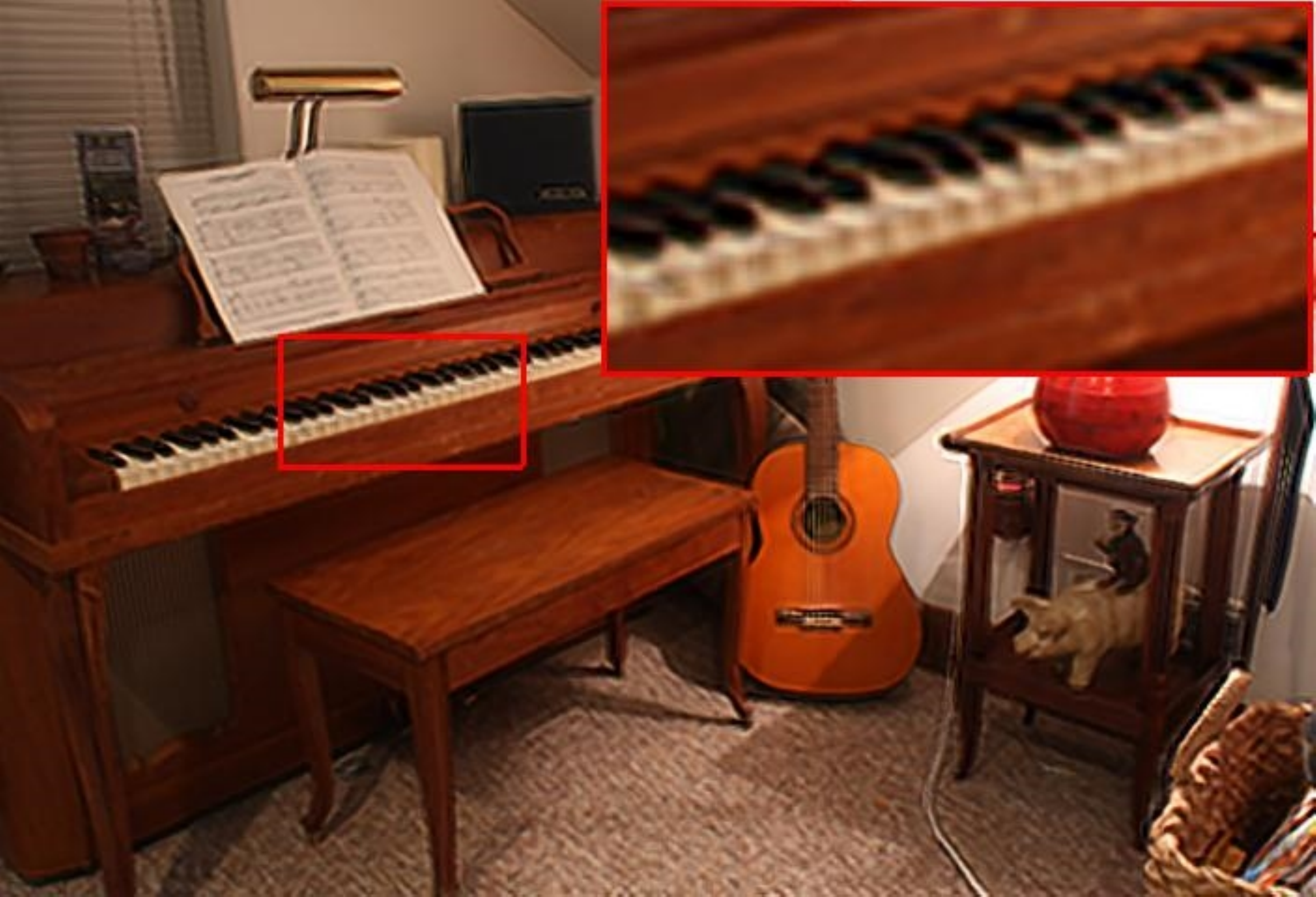} \\ \centering{\small{PSSR}} \end{minipage}
   \begin{minipage}[t]{2.50cm}  \centering \includegraphics[width=2.50cm]{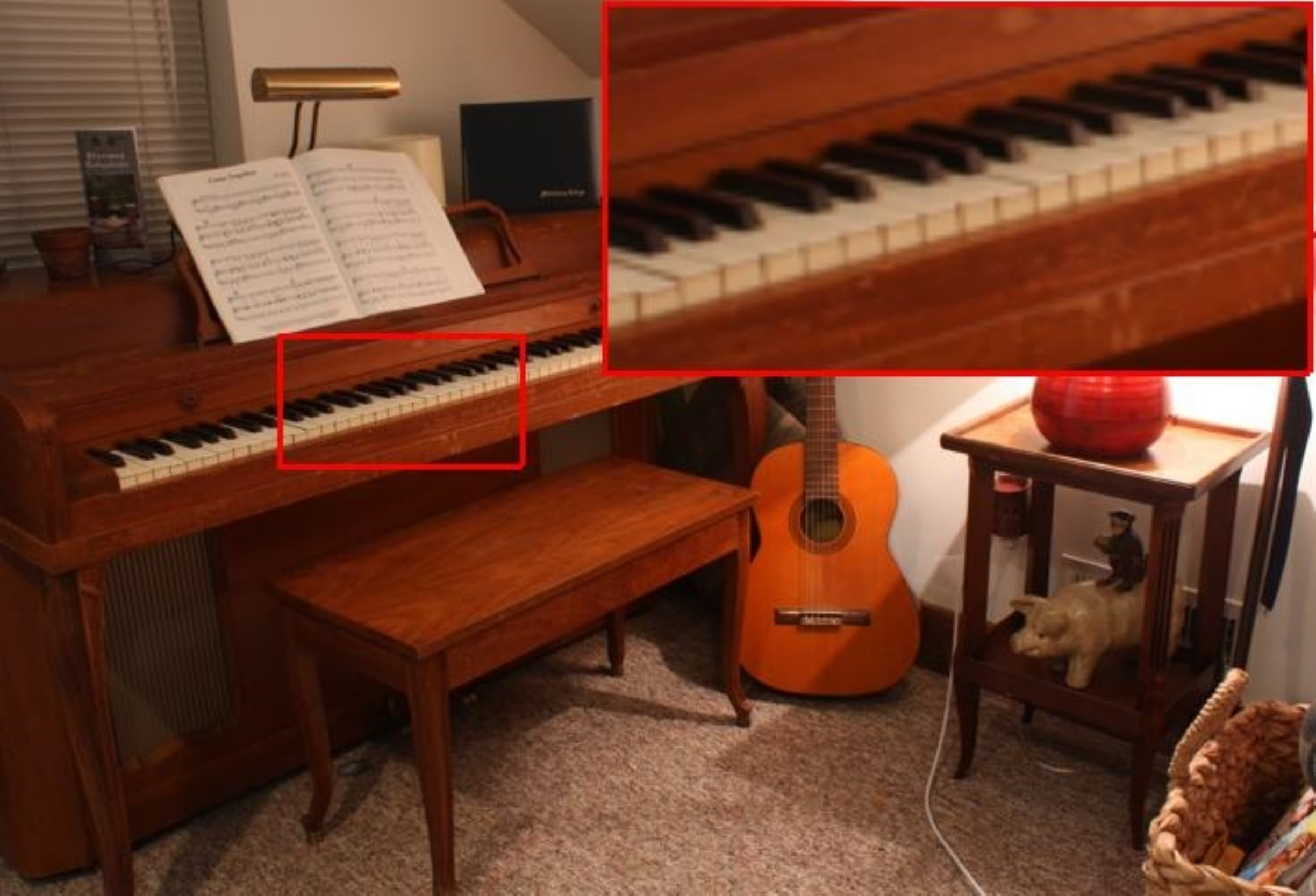} \\ \centering{\small{GT}} \end{minipage} }\  \vspace{-0.2em}
\caption{Comparison of state-of-the-art MSE-based SISR, GAN-based SISR, StereoSR, and our methods.}
\label{fstoasr} \end{figure*}

\subsection{Comparison with state-of-the-arts}
\textbf{Bicubic Degradation (BI).}
Existing StereoSR methods work for bicubic degradation (BI), and suppose LR image is generated by bicubic downscaling. Following their steps, we compare our PSSR with MSE-based SISR (EDSR~\cite{edsr}, SRFBN~\cite{srfbn}, DRN~\cite{drn}), GAN-based SISR (ESRGAN~\cite{esrgan}, SPSR~\cite{spsr}), and StereoSR (StereoSR~\cite{stereosr}, PASSRnet~\cite{passr}, DASSR~\cite{our}) methods on BI degradation in Table~\ref{tiqp1}. Some methods only public $\times$4 SR models.

Table~\ref{tiqp1} demonstrates several NR-IQA scores (including NIQE~\cite{niqe}, Perceptual Index (PI), the combination of Ma \emph{et al.}~\cite{maetal} and NIQE, and our StereoSRQA model) to evaluate the perceptual performance of different SR results. Lower NIQE/PI and higher StereoSRQA scores denote better subjective quality.

We can observe that some GAN-based SISR methods have lower NIQE/PI than groundtruth, which goes against the common sense. As stated in~\cite{pipal}, though NIQE and PI are relevant to human judgment, they cannot distinguish GAN generated noises and real details and prefer images with obvious unrealistic artifacts produced by GAN-based image restoration methods. Therefore, these perceptual evaluation metrics cannot fairly reflect the subjective SR performance. Even so, our PSSR achieves best NIQE/PI except the GAN-based SPSR, and makes significant progress on StereoSRQA. In Figure~\ref{f0} and~\ref{fstoasr}, the PSSR remarkably improves the subjective quality and restores clearer textures. 

We further adopt more CNN-based quality assessment models, such as two NR-IQA (SFA~\cite{sfa}, CNNIQA~\cite{cnniqa}), a NR-StereoQA~\cite{stereoqa}, and a FR-SISRQA (\cite{sisrqa}) models, to evaluate the StereoSR results on Middlebury in Table~\ref{tstoafr}.

To demonstrate the generalization ability of the PSSR, we deploy our perception-oriented training strategy on different SR networks (EDSR, StereoSR). Table~\ref{tiqp2} compares the results of the SR models trained with their original method and our method respectively.

\begin{table}[tb]
\caption{The average quality scores of CNN-based IQA methods for $\times$4 StereoSR on Middlebury.} \vspace{-0.2em}
\label{tstoafr}
\centering
\scalebox{0.75}[0.75]{
    \begin{tabular}{cc|ccccc}
\toprule \toprule &IQA &Bic &StereoSR~\cite{stereosr}& PASSRnet~\cite{passr}&DASSR~\cite{our}&PSSR\\ \hline
    \multirow{3}[3]{*}{NR} &
    SFA$\downarrow$&55.984&40.081&41.838&41.717&\textbf{32.213} \\
    &CNNIQA$\downarrow$&55.483 &37.052&34.673 &36.449&\textbf{27.889}\\
   &StereoQA$\downarrow$&29.439&19.707&22.291&21.019&\textbf{19.618}\\
    \hline
   FR&SISRQA$\uparrow$ &0.5185&0.5512&0.6090&0.5992&\textbf{0.6109}\\
\bottomrule \end{tabular} } \end{table}

\begin{table}[tb]
\caption{Perceptual scores (NIQE/PI/StereoSRQA) of applying the proposed training constraint on different SR models for $\times$ 4 StereoSR on Middlebury.}
\vspace{-0.5em}
 \label{tiqp2}  \centering
 \scalebox{0.72}[0.72]{ \begin{tabular}{c|cc|cc}
 \toprule \toprule
 &\multicolumn{2}{c}{EDSR~\cite{edsr}}&\multicolumn{2}{c}{StereoSR~\cite{stereosr}} \\
\hline
   &w/o $\mathcal{L}_{IQP}$ &w/ $\mathcal{L}_{IQP}$&w/o $\mathcal{L}_{IQP}$&w/ $\mathcal{L}_{IQP}$\\
\hline
    Middlebury&5.318/5.381/7.26&\textbf{4.322/4.270/8.61}&5.383/5.229/6.69&4.398/4.177/7.71\\
\hline
    KITTI 2012&5.835/5.723/7.17&\textbf{4.525/4.332/8.38}&4.669/4.088/7.14&3.745/3.100/7.47\\
\hline
    KITTI 2015&5.700/5.882/7.38&\textbf{4.826/4.783/8.05}&4.594/4.179/7.31&3.968/3.211/8.01\\
\hline
    Tsukuba&6.039/5.681/7.17&\textbf{4.973/4.325/7.46}&5.666/5.322/5.64&4.354/4.261/6.67\\
\hline
    SceneFlow&5.385/5.046/6.77&\textbf{4.767/4.561/7.49}&4.705/3.994/6.38&4.001/3.141/6.78\\
\bottomrule \end{tabular}} \end{table}

\begin{table}[htb]
\caption{$\times$4 StereoSR results (PSNR/SSIM/StereoSRQA) on BD degradation of models trained with $\mathcal{L}_{MSE}$ and $\mathcal{L}_{IQP}$.}\label{tblur}
\vspace{-0.5em}
  \centering  \scalebox{0.85}[0.85]{
    \begin{tabular}{cccc}
\toprule \toprule
     $\sigma$&Bicubic& $\mathcal{L}_{MSE}$& Our $\mathcal{L}_{IQP}$ \\
   \hline
1.0&26.225/0.8461/4.460&\textbf{28.598}/\textbf{0.8946}/6.957&27.277/0.8726/\textbf{7.286} \\
1.6&25.633/0.8298/3.934&\textbf{28.783}/\textbf{0.8938}/6.360&27.344/0.8724/\textbf{7.105}\\
2.6&24.442/0.7950/2.624&26.189/0.8428/4.694&\textbf{27.022}/\textbf{0.8675}/\textbf{6.676}\\
3.6&23.526/0.7661/1.852&24.221/0.7876/3.324&\textbf{26.856}/\textbf{0.8491}/\textbf{5.053}\\
\bottomrule    \end{tabular} } \end{table}

\begin{figure}[tb]
\centering
\includegraphics[width=0.49\textwidth]{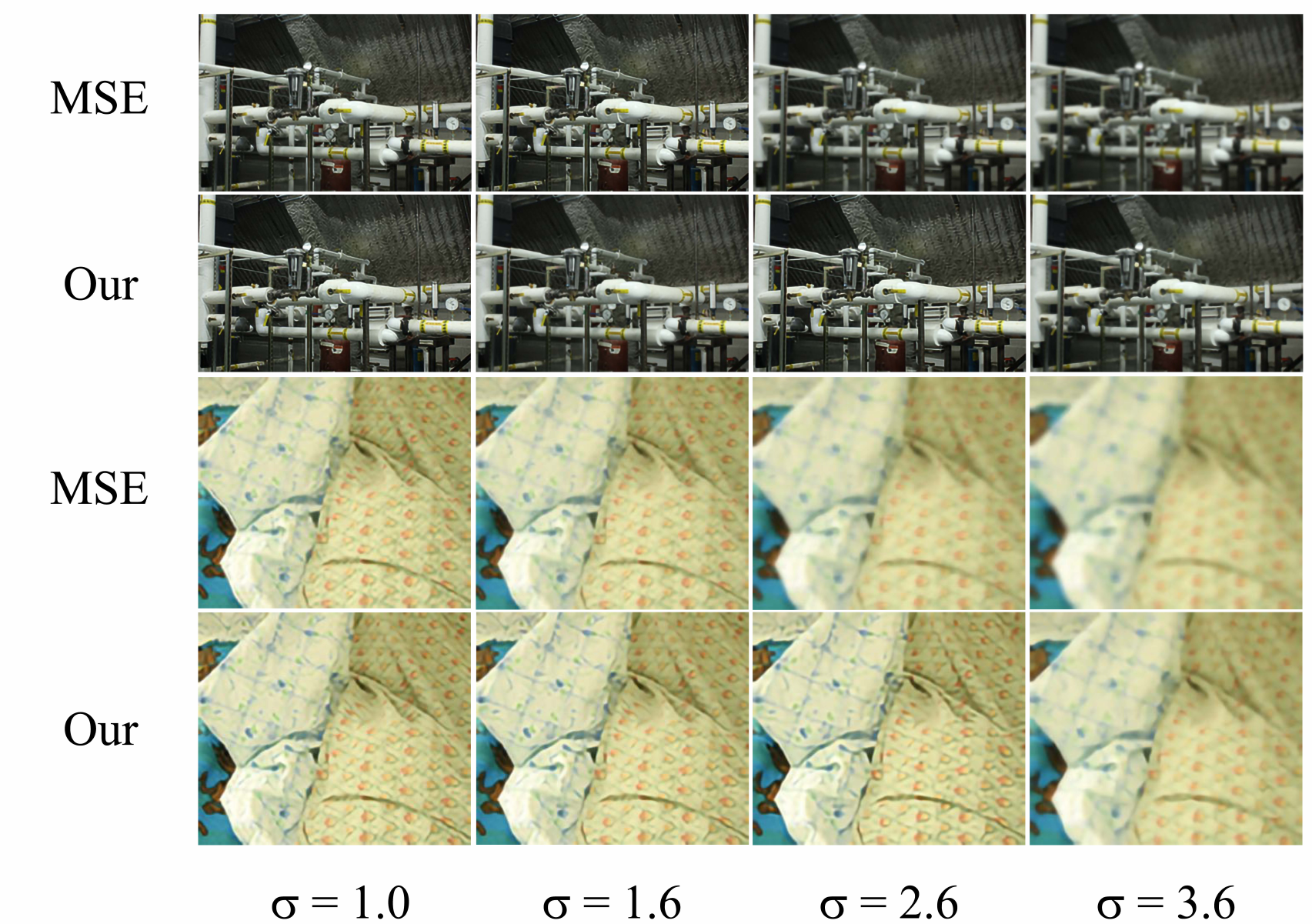}
\vspace{-0.5em}
\caption{Visual comparison of StereoSR models trained with MSE and our IQP losses on BD degradation with different kernel widths.}
\label{fblur} \end{figure}

\textbf{Blur-downscale Degradation (BD)}
To further illustrate the robustness of our PSSR, we retrain StereoSR models with MSE loss and our training method respectively on a more difficult blur downscale (BD) degradation. The LR image is blurred with 15 $\times$ 15 Gaussian kernel, kernel width $\sigma \in [0.8, 4.0]$. Table~\ref{tblur} and Figure~\ref{fblur} demonstrate the StereoSR results under BD degradation. Both StereoSR models work well for small blur kernel, and the MSE loss can maximize PSNR/SSIM, as we stated. However, the StereoSR model, trained with MSE loss, suffers from significant performance drop facing severe blur. In comparison, our PSSR keeps better subjective performance and even higher PSNR/SSIM for larger blur kernels, demonstrating the $\mathcal{L}_{IQP}$ is robust to more difficult degradation. 

\begin{table}[tb] 
\caption{The disparity estimation accuracy (end-point-error EPE) of $\times$4 SR results on A/B/C subsets in flyingthings3d.} \vspace{-0.5em}
 \label{tds} \centering  \scalebox{0.78}[0.78]{
\begin{tabular}{c|ccccccc}
     \toprule \toprule
     Database&\underline{HR}&Bicubic&EDSR*&StereoSR&PASSRnet&DASSR& PSSR \\
\hline
     A &\textcolor[rgb]{1,0,0}{2.238}&3.173&2.681&2.801&2.724&2.498&\textbf{2.428}\\
     B &\textcolor[rgb]{1,0,0}{1.098}&2.075&1.585&1.516&1.588&\textbf{1.432}&1.477\\
     C &\textcolor[rgb]{1,0,0}{2.624}&3.429&3.000&3.225&2.983&2.889&\textbf{2.885}\\
\bottomrule   \end{tabular} } \end{table}%

\begin{figure}[tb]
\centering \subfigure{
\begin{minipage}[t]{1.85cm} \centering \includegraphics[width=1.85cm]{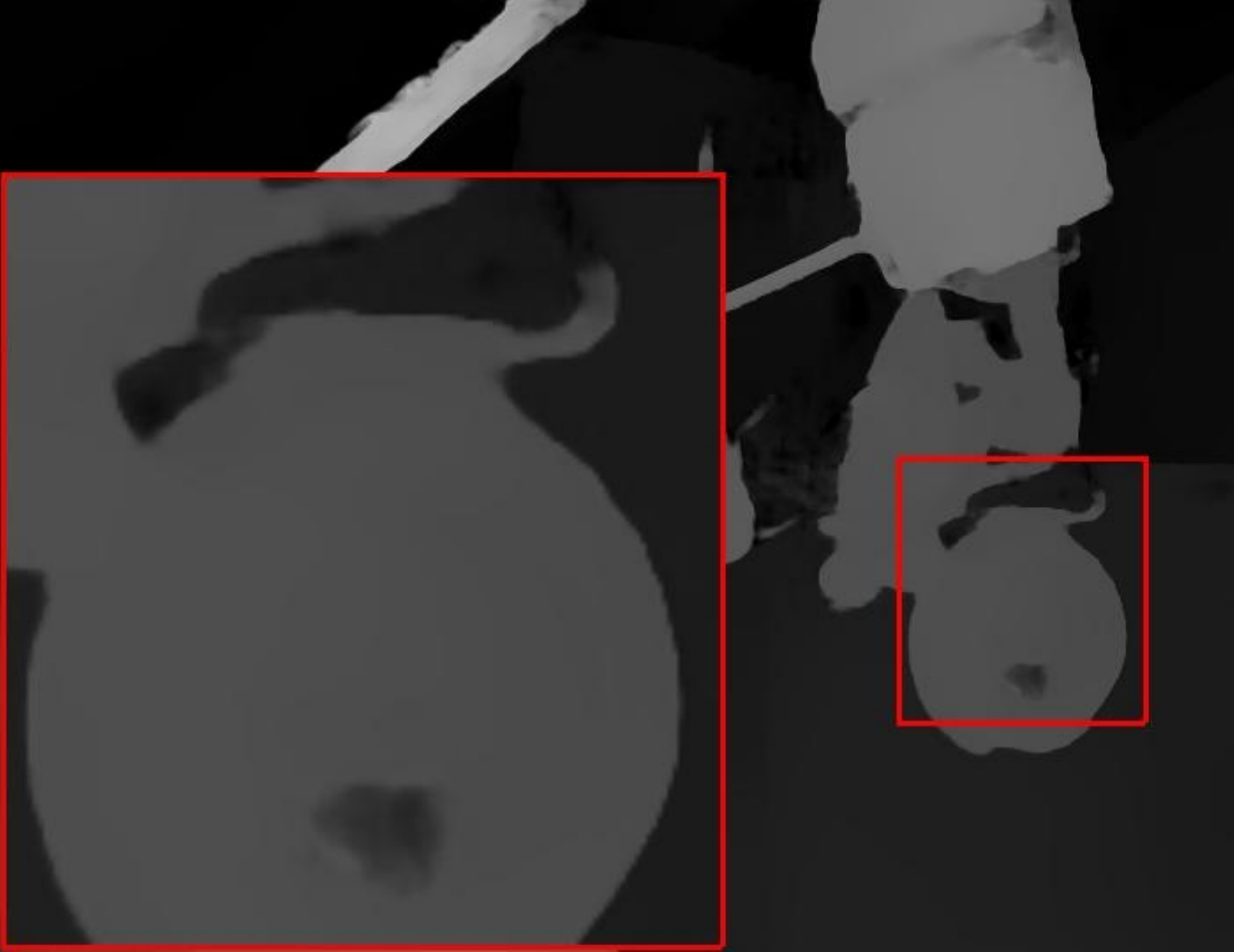} \\   \centering{\small{Bicubic}} \end{minipage}
\begin{minipage}[t]{1.73cm} \centering \includegraphics[width=1.85cm]{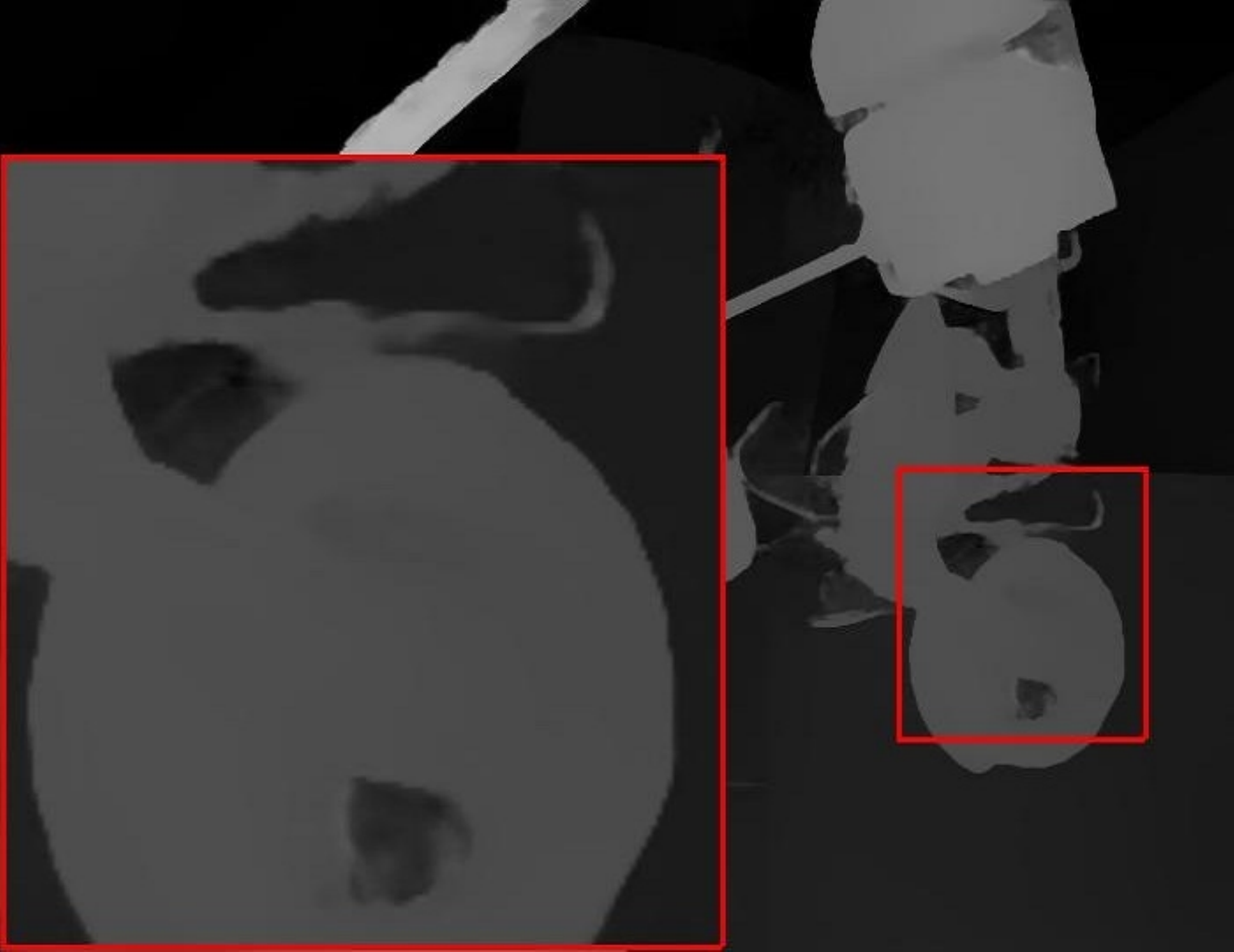} \\ \centering{\small{EDSR*~\cite{edsr}}}   \end{minipage}
\begin{minipage}[t]{2.10cm} \centering \includegraphics[width=1.85cm]{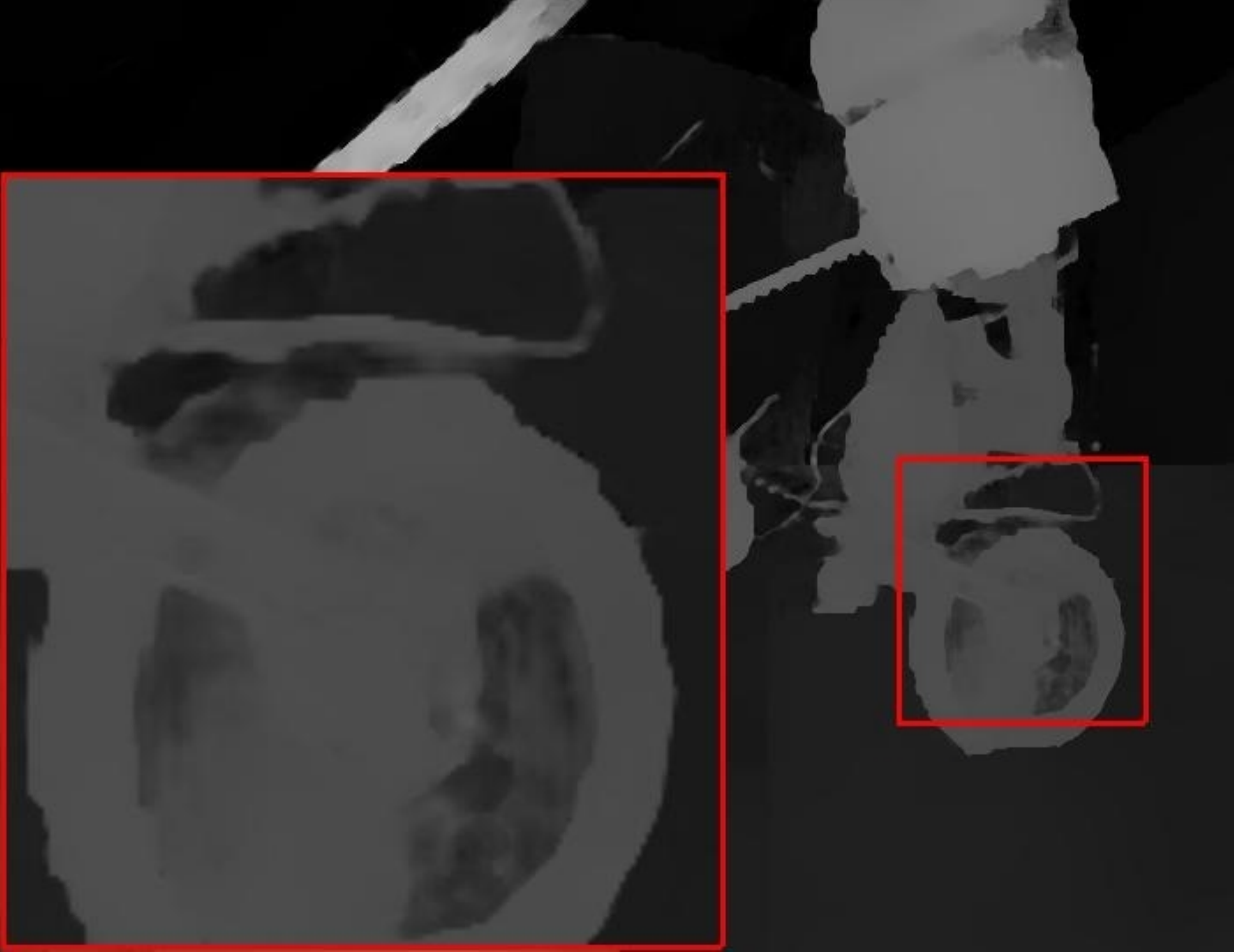} \\ \centering{\small{StereoSR~\cite{stereosr}}}  \end{minipage}
\begin{minipage}[t]{1.85cm} \centering \includegraphics[width=1.85cm]{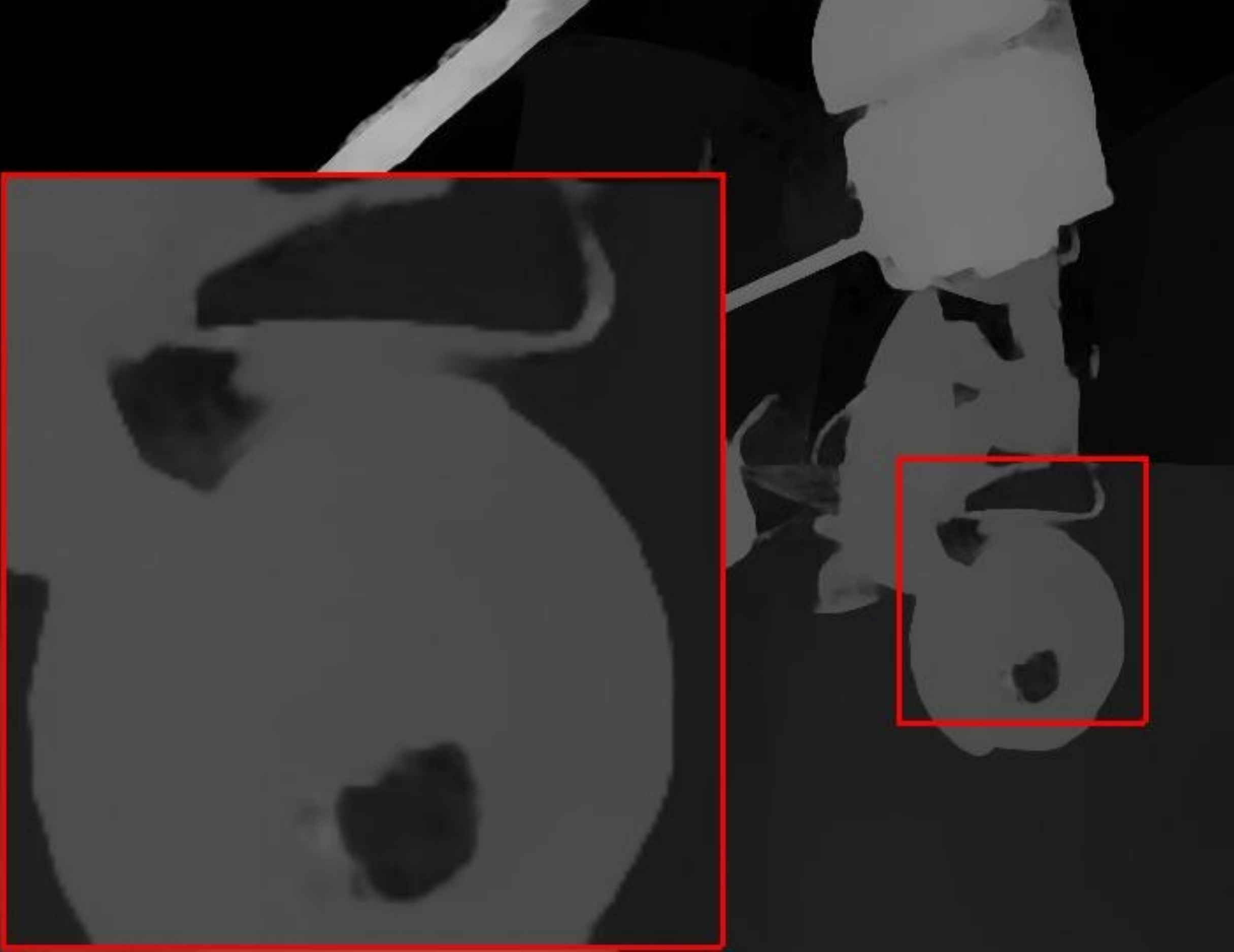} \\ \centering{\small{PASSR~\cite{passr}}}   \end{minipage}
}\ \centering  \subfigure{
\begin{minipage}[t]{1.85cm} \centering \includegraphics[width=1.85cm]{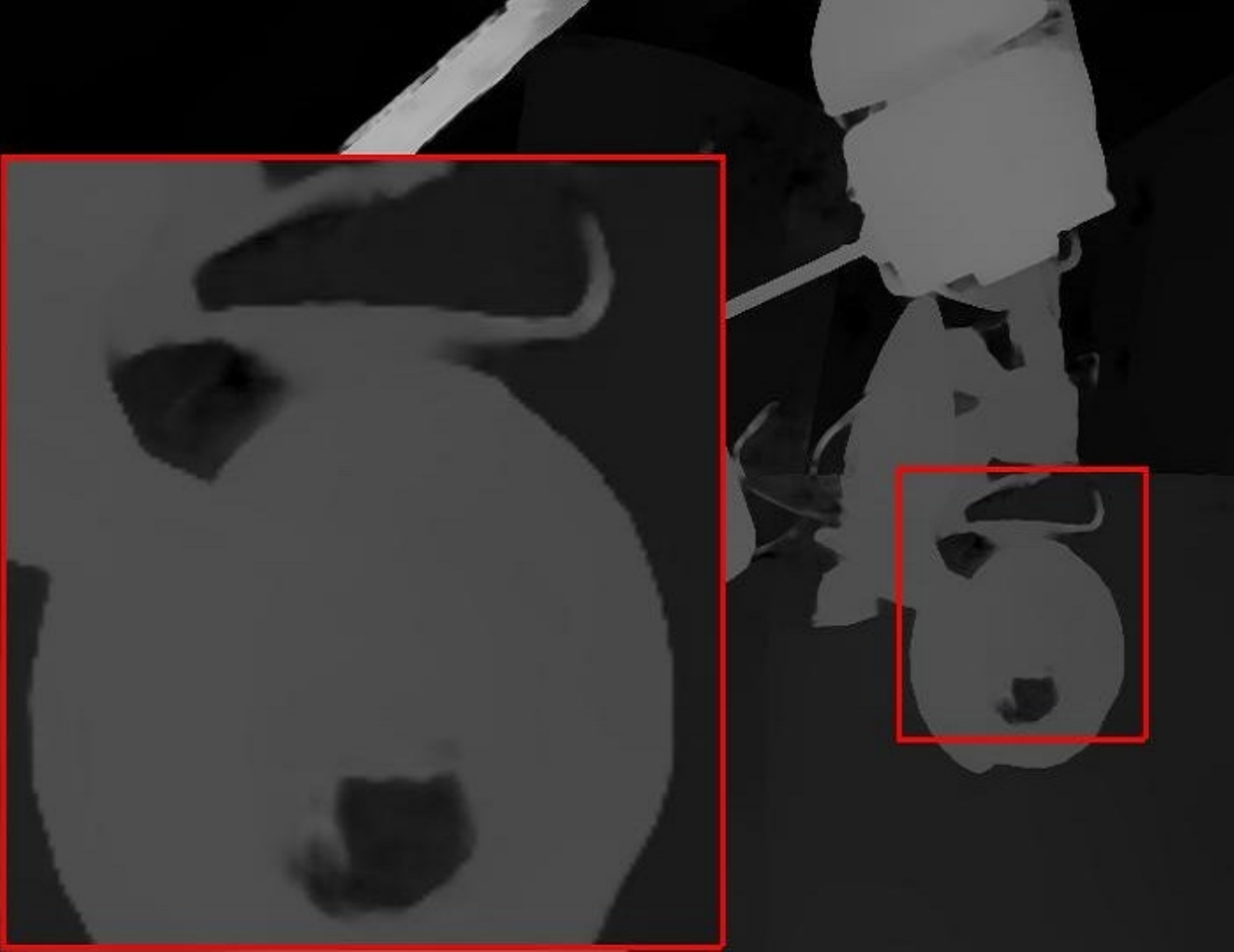} \\ \centering{\small{DASSR~\cite{our}}}   \end{minipage}
\begin{minipage}[t]{1.85cm} \centering \includegraphics[width=1.85cm]{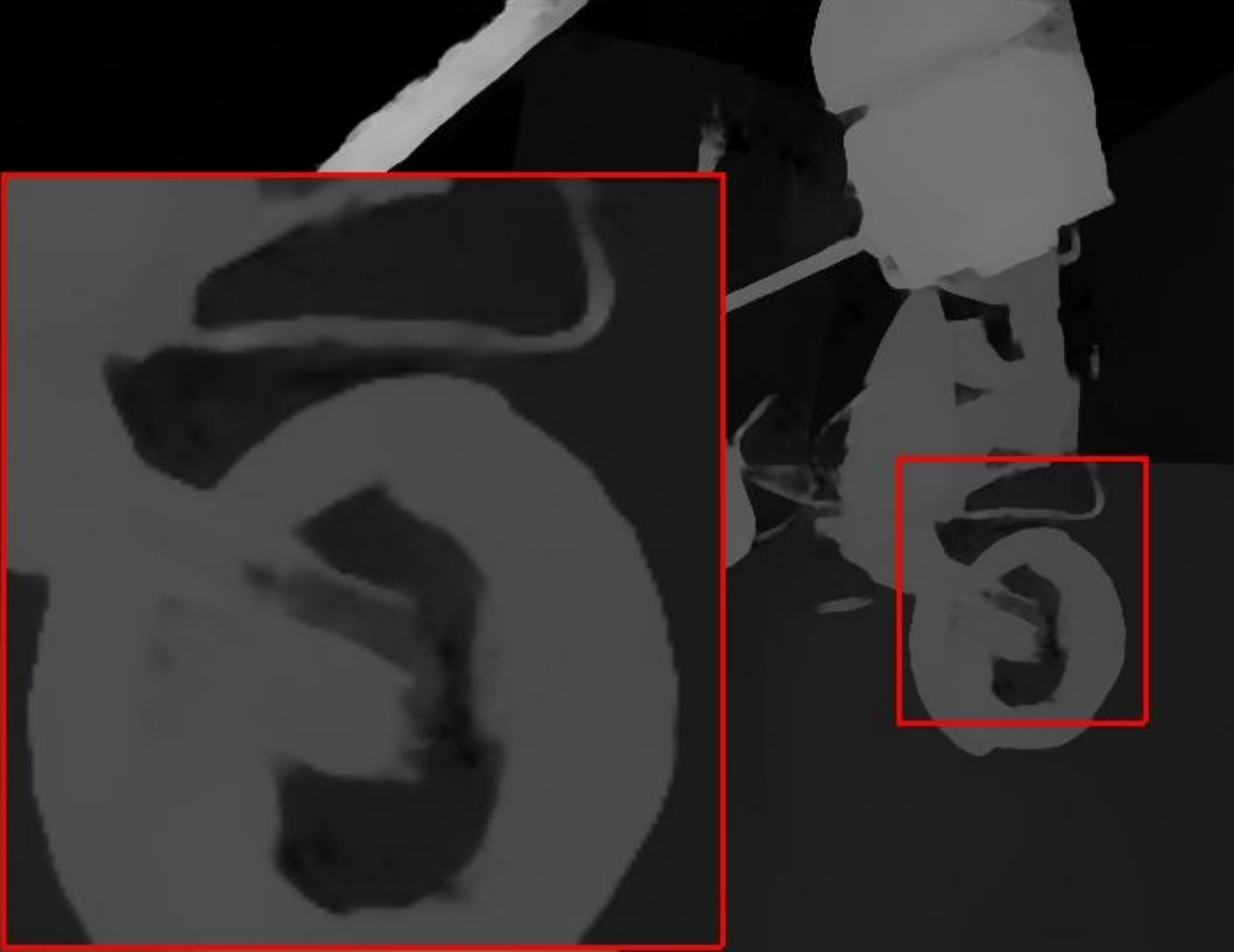} \\ \centering{\small{PSSR}}   \end{minipage}
\begin{minipage}[t]{1.85cm} \centering \includegraphics[width=1.85cm]{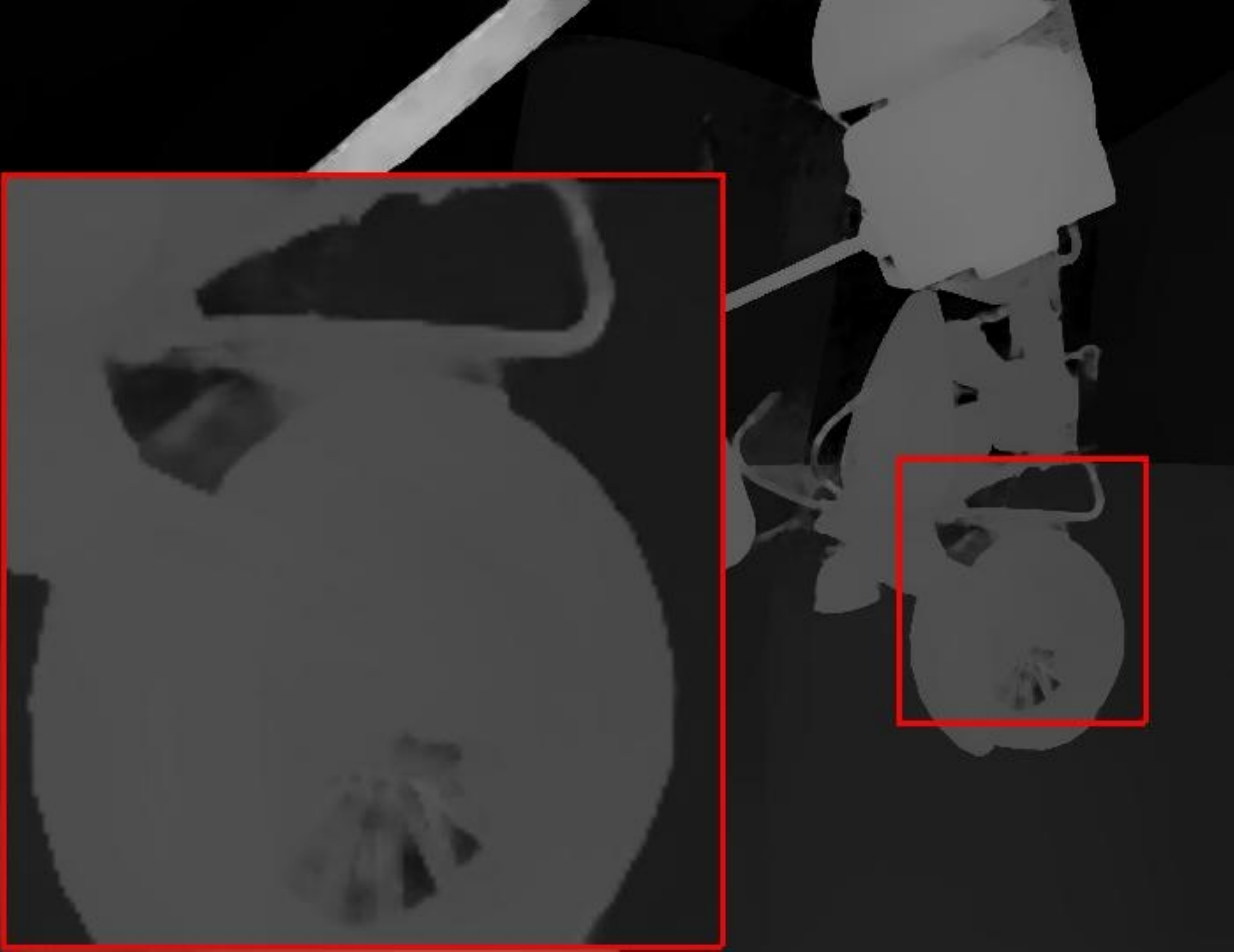} \\ \centering{\small{HR}}   \end{minipage}
\begin{minipage}[t]{1.85cm} \centering \includegraphics[width=1.85cm]{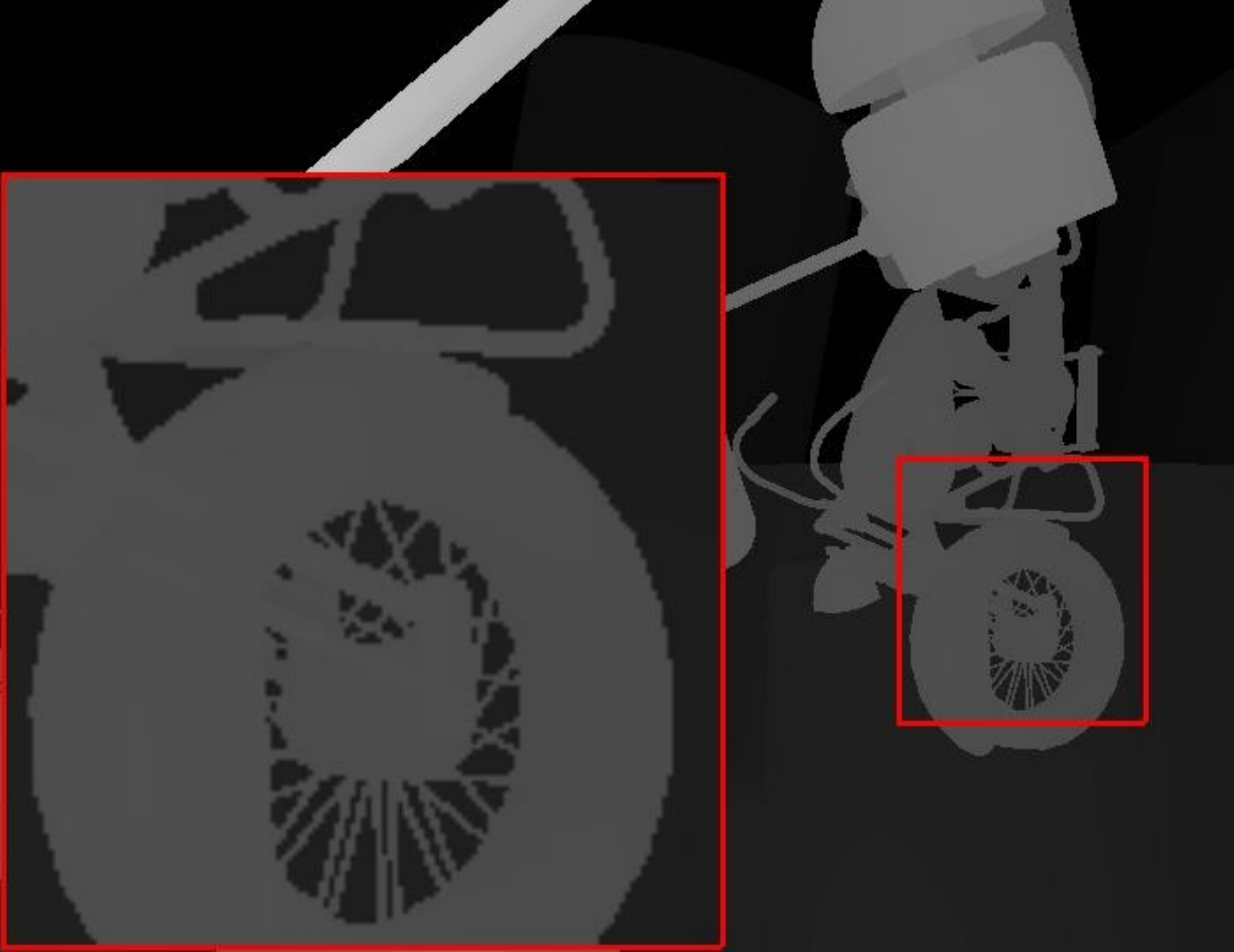} \\ \centering{\small{GroundTruth}}   \end{minipage} }\
\vspace{-0.2em}
\caption{Visual comparison of disparity accuracy.}
\vspace{-0.2em}
\label{dis} \end{figure}

\textbf{Disparity Estimation} 
Since disparity estimation is a non-negligible application of stereo image, we evaluate the quality of the super-resolved stereo images on disparity estimation task by following~\cite{our}. Table~\ref{tds} and Figure~\ref{dis} compare the accuracy of disparity, estimated by StereoNet~\cite{stereonet}. The PSSR can generate more textures and spatial details, which are valuable for predicting precise disparity. 
\subsection{Performance of StereoSRQA model}
By inputting the degraded stereo image pair or the difference map between the degraded and groundtruth stereo image pair to the StereoSRQA model, we can construct a no-reference and a full-reference StereoSRQA models (`NRStereoSRQA', `FRStereoSRQA'). Table~\ref{tiqa} compares FRStereoSRQA and NRStereoSRQA with both full-reference IQA (PSNR, SSIM~\cite{ssim}, IFC~\cite{ifc}, VIF~\cite{vif}, SISRQA~\cite{sisrqa}) and no-reference IQA (BRISQUE~\cite{brisque}, SSEQ~\cite{sseq}, NIQE~\cite{niqe}, CNNIQA~\cite{cnniqa}, SFA~\cite{sfa}, StereoQA~\cite{stereoqa}) methods, including CNN-based or non CNN-based, single image or stereo image based methods. Our FR-StereoSRQA achieves the best performance and is more consistent with human perception. 

\begin{table}[tb]
\caption{Comparison with state-of-the-art full-reference and no-reference IQA methods.}
\vspace{-0.2em}  \centering   \scalebox{0.88}[0.88]{
  \begin{tabular}{c|c|cccc}  \toprule \toprule
    &&RMSE$\downarrow$&SROCC$\uparrow$&PLCC$\uparrow$&KROCC$\uparrow$\\
  \hline
      \multirow{6}[0]{*}{FR}
      & PSNR &1.8865&0.7180&0.7275&0.5385\\
      & SSIM~\cite{ssim}&1.7363&0.7610&0.7754&0.5646\\
      & IFC~\cite{ifc}&0.4982&0.9820&0.9834&0.8807\\
      & VIF~\cite{vif}&0.6081&0.9738&0.9752&0.8567\\
      & SISRQA~\cite{sisrqa}&1.1208&0.9146&0.9131&0.7573\\
      & FRStereoSRQA&\textbf{0.3505}&\textbf{0.9910}&\textbf{0.9918}&\textbf{0.9170}\\
  \hline
      \multirow{7}[0]{*}{NR}
      &BRISQUE~\cite{brisque}&1.2705&0.8881&0.8868&0.7082\\
      &SSEQ~\cite{sseq}&1.4099&0.8495&0.8585&0.6872\\
      &NIQE~\cite{niqe}&1.3157&0.8632&0.8781&0.6725\\
      &CNNIQA~\cite{cnniqa}&1.6205&0.8066&0.8079&0.6078\\
      &SFA~\cite{sfa}&1.6296&0.6519&0.8054&0.5217\\
      &StereoQA~\cite{stereoqa}&1.8930&0.7007&0.7252&0.5403\\
      &NRStereoSRQA&\textbf{0.6249}&\textbf{0.9743}&\textbf{0.9738}&\textbf{0.8624} \\
  \bottomrule    \end{tabular} }  \label{tiqa}
\end{table}

To further illustrate the contribution of our Mid3D\_QA, Table~\ref{tqaondatabase} compares two SISRQA models and two StereoQA models, which are trained on their original database and our database respectively.
\begin{table}[tb]
\centering
\caption{Results of QA models trained with original databases and our Mid3D\_QA.} \vspace{-0.2em}
\label{tqaondatabase} \scalebox{0.75}[0.75]{ \begin{tabular}{c|c|cccc}
\toprule \toprule
 models&train set&RMSE$\downarrow$&SROCC$\uparrow$&PLCC$\uparrow$&KROCC$\uparrow$\\
 \hline
 SISRQA~\cite{sisrqa}&Ma \emph{et al.}~\cite{maetal}&1.1208&0.9146&0.9131&0.7573\\
 SISRQA\_mid&Mid3D\_QA&0.7656&0.9602&0.9598&0.8287\\
 \hline
 StereoQA~\cite{stereoqa}&LIVE Phase I~\cite{live1}&1.8930&0.7007&0.7252&0.5403\\
 StereoQA\_mid(our)&Mid3D\_QA&\textbf{0.6249}&\textbf{0.9743}&\textbf{0.9738}&\textbf{0.8624}\\
\bottomrule \end{tabular}} \end{table}

\section{Conclusion}
\label{seccon}
This paper proposes a perception-oriented stereo image super-resolution framework (PSSR). The proposed PSSR improves the visual quality for StereoSR by using the feedback guidance from the assessment of the StereoSR results. To provide accurate guidance, we construct the first quality assessment model (StereoSRQA) for StereoSR, and a special StereoSRQA database (Mid3D\_QA). In specific, we propose a image quality perceptual constraint to calculate the feature similarity, used to evaluate the perceptual quality of StereoSR results, between the groundtruth and the super-resolved stereo images. Extensive experiments demonstrate our PSSR achieves state-of-the-art performance in terms of the quality, practicality, and flexibly.

\bibliographystyle{ACM-Reference-Format}
\balance
\bibliography{sample-base}


\begin{thebibliography}{44}


\ifx \showCODEN    \undefined \def \showCODEN     #1{\unskip}     \fi
\ifx \showDOI      \undefined \def \showDOI       #1{#1}\fi
\ifx \showISBNx    \undefined \def \showISBNx     #1{\unskip}     \fi
\ifx \showISBNxiii \undefined \def \showISBNxiii  #1{\unskip}     \fi
\ifx \showISSN     \undefined \def \showISSN      #1{\unskip}     \fi
\ifx \showLCCN     \undefined \def \showLCCN      #1{\unskip}     \fi
\ifx \shownote     \undefined \def \shownote      #1{#1}          \fi
\ifx \showarticletitle \undefined \def \showarticletitle #1{#1}   \fi
\ifx \showURL      \undefined \def \showURL       {\relax}        \fi
\providecommand\bibfield[2]{#2}
\providecommand\bibinfo[2]{#2}
\providecommand\natexlab[1]{#1}
\providecommand\showeprint[2][]{arXiv:#2}

\bibitem[\protect\citeauthoryear{Chen, Su, Kwon, Cormack, and Bovik}{Chen
  et~al\mbox{.}}{2013}]%
        {live2}
\bibfield{author}{\bibinfo{person}{Ming~Jun Chen}, \bibinfo{person}{Che~Chun
  Su}, \bibinfo{person}{Do~Kyoung Kwon}, \bibinfo{person}{Lawrence~K. Cormack},
  {and} \bibinfo{person}{Alan~C. Bovik}.} \bibinfo{year}{2013}\natexlab{}.
\newblock \showarticletitle{Full-reference quality assessment of stereopairs
  accounting for rivalry}.
\newblock \bibinfo{journal}{\emph{Image Communication}} (\bibinfo{year}{2013}).
\newblock


\bibitem[\protect\citeauthoryear{Dong, Chen, He, and Tang}{Dong
  et~al\mbox{.}}{2014}]%
        {srcnn}
\bibfield{author}{\bibinfo{person}{Chao Dong}, \bibinfo{person}{Change~Loy
  Chen}, \bibinfo{person}{Kaiming He}, {and} \bibinfo{person}{Xiaoou Tang}.}
  \bibinfo{year}{2014}\natexlab{}.
\newblock \showarticletitle{Image Super-Resolution Using Deep Convolutional
  Networks}.
\newblock \bibinfo{journal}{\emph{IEEE Trans Pattern Anal Mach Intell}}
  \bibinfo{volume}{38}, \bibinfo{number}{2} (\bibinfo{year}{2014}),
  \bibinfo{pages}{295--307}.
\newblock


\bibitem[\protect\citeauthoryear{{Duan} and {Xiao}}{{Duan} and {Xiao}}{2019}]%
        {pscassr}
\bibfield{author}{\bibinfo{person}{C. {Duan}} {and} \bibinfo{person}{N.
  {Xiao}}.} \bibinfo{year}{2019}\natexlab{}.
\newblock \showarticletitle{Parallax-Based Spatial and Channel Attention for
  Stereo Image Super-Resolution}.
\newblock \bibinfo{journal}{\emph{IEEE Access}}  \bibinfo{volume}{7}
  (\bibinfo{year}{2019}), \bibinfo{pages}{183672--183679}.
\newblock


\bibitem[\protect\citeauthoryear{Geiger, {Lenz}, and Urtasun}{Geiger
  et~al\mbox{.}}{2012}]%
        {kt12}
\bibfield{author}{\bibinfo{person}{A. Geiger}, \bibinfo{person}{P. {Lenz}},
  {and} \bibinfo{person}{R. Urtasun}.} \bibinfo{year}{2012}\natexlab{}.
\newblock \showarticletitle{Are we ready for autonomous driving? The KITTI
  vision benchmark suite}. In \bibinfo{booktitle}{\emph{2012 IEEE Conference on
  Computer Vision and Pattern Recognition}}. \bibinfo{pages}{3354--3361}.
\newblock


\bibitem[\protect\citeauthoryear{Guo, Chen, Wang, Chen, Cao, Deng, Xu, and
  Tan}{Guo et~al\mbox{.}}{2020}]%
        {drn}
\bibfield{author}{\bibinfo{person}{Yong Guo}, \bibinfo{person}{Jian Chen},
  \bibinfo{person}{Jingdong Wang}, \bibinfo{person}{Qi Chen},
  \bibinfo{person}{Jiezhang Cao}, \bibinfo{person}{Zeshuai Deng},
  \bibinfo{person}{Yanwu Xu}, {and} \bibinfo{person}{Mingkui Tan}.}
  \bibinfo{year}{2020}\natexlab{}.
\newblock \showarticletitle{Closed-loop Matters: Dual Regression Networks for
  Single Image Super-Resolution}. In \bibinfo{booktitle}{\emph{IEEE Conference
  on Computer Vision and Pattern Recognition}}.
\newblock


\bibitem[\protect\citeauthoryear{Hui, Wang, and Gao}{Hui et~al\mbox{.}}{2018}]%
        {idn}
\bibfield{author}{\bibinfo{person}{Zheng Hui}, \bibinfo{person}{Xiumei Wang},
  {and} \bibinfo{person}{Xinbo Gao}.} \bibinfo{year}{2018}\natexlab{}.
\newblock \showarticletitle{Fast and Accurate Single Image Super-Resolution via
  Information Distillation Network}. In \bibinfo{booktitle}{\emph{Proceedings
  of the IEEE Conference on Computer Vision and Pattern Recognition}}.
  \bibinfo{pages}{723--731}.
\newblock


\bibitem[\protect\citeauthoryear{Jeon, Baek, Choi, and Kim}{Jeon
  et~al\mbox{.}}{2018a}]%
        {stereosr}
\bibfield{author}{\bibinfo{person}{D.~S. Jeon}, \bibinfo{person}{S. Baek},
  \bibinfo{person}{I. Choi}, {and} \bibinfo{person}{M.~H. Kim}.}
  \bibinfo{year}{2018}\natexlab{a}.
\newblock \showarticletitle{Enhancing the Spatial Resolution of Stereo Images
  Using a Parallax Prior}. In \bibinfo{booktitle}{\emph{2018 IEEE/CVF
  Conference on Computer Vision and Pattern Recognition}}.
  \bibinfo{pages}{1721--1730}.
\newblock


\bibitem[\protect\citeauthoryear{Jeon, Baek, Choi, and Kim}{Jeon
  et~al\mbox{.}}{2018b}]%
        {middlebury}
\bibfield{author}{\bibinfo{person}{D.~S. Jeon}, \bibinfo{person}{S. Baek},
  \bibinfo{person}{I. Choi}, {and} \bibinfo{person}{M.~H. Kim}.}
  \bibinfo{year}{2018}\natexlab{b}.
\newblock \showarticletitle{Enhancing the Spatial Resolution of Stereo Images
  Using a Parallax Prior}. In \bibinfo{booktitle}{\emph{2018 IEEE/CVF
  Conference on Computer Vision and Pattern Recognition}}.
  \bibinfo{pages}{1721--1730}.
\newblock


\bibitem[\protect\citeauthoryear{Jinjin, Haoming, Haoyu, Xiaoxing, Ren, and
  Chao}{Jinjin et~al\mbox{.}}{2020}]%
        {pipal}
\bibfield{author}{\bibinfo{person}{Gu Jinjin}, \bibinfo{person}{Cai Haoming},
  \bibinfo{person}{Chen Haoyu}, \bibinfo{person}{Ye Xiaoxing},
  \bibinfo{person}{Jimmy~S. Ren}, {and} \bibinfo{person}{Dong Chao}.}
  \bibinfo{year}{2020}\natexlab{}.
\newblock \showarticletitle{PIPAL: A Large-Scale Image Quality Assessment
  Dataset for Perceptual Image Restoration}. In
  \bibinfo{booktitle}{\emph{European Conference on Computer Vision}}.
\newblock


\bibitem[\protect\citeauthoryear{Johnson, Alahi, and Li}{Johnson
  et~al\mbox{.}}{2016}]%
        {perceptual}
\bibfield{author}{\bibinfo{person}{Justin Johnson}, \bibinfo{person}{Alexandre
  Alahi}, {and} \bibinfo{person}{Fei~Fei Li}.} \bibinfo{year}{2016}\natexlab{}.
\newblock \showarticletitle{Perceptual Losses for Real-Time Style Transfer and
  Super-Resolution}. In \bibinfo{booktitle}{\emph{European Conference on
  Computer Vision}}.
\newblock


\bibitem[\protect\citeauthoryear{Kang, Ye, Li, and Doermann}{Kang
  et~al\mbox{.}}{2014a}]%
        {cnniqa}
\bibfield{author}{\bibinfo{person}{Le Kang}, \bibinfo{person}{Peng Ye},
  \bibinfo{person}{Yi Li}, {and} \bibinfo{person}{David Doermann}.}
  \bibinfo{year}{2014}\natexlab{a}.
\newblock \showarticletitle{Convolutional Neural Networks for No-Reference
  Image Quality Assessment}. In \bibinfo{booktitle}{\emph{IEEE Conference on
  Computer Vision and Pattern Recognition}}.
\newblock


\bibitem[\protect\citeauthoryear{Kang, Ye, Li, and Doermann}{Kang
  et~al\mbox{.}}{2014b}]%
        {kang}
\bibfield{author}{\bibinfo{person}{Le Kang}, \bibinfo{person}{Peng Ye},
  \bibinfo{person}{Yi Li}, {and} \bibinfo{person}{David Doermann}.}
  \bibinfo{year}{2014}\natexlab{b}.
\newblock \showarticletitle{Convolutional Neural Networks for No-Reference
  Image Quality Assessment}. In \bibinfo{booktitle}{\emph{IEEE Conference on
  Computer Vision and Pattern Recognition}}.
\newblock


\bibitem[\protect\citeauthoryear{Khamis, Fanello, Rhemann, Kowdle, Valentin,
  and Izadi}{Khamis et~al\mbox{.}}{2018}]%
        {stereonet}
\bibfield{author}{\bibinfo{person}{Sameh Khamis}, \bibinfo{person}{Sean
  Fanello}, \bibinfo{person}{Christoph Rhemann}, \bibinfo{person}{Adarsh
  Kowdle}, \bibinfo{person}{Julien Valentin}, {and} \bibinfo{person}{Shahram
  Izadi}.} \bibinfo{year}{2018}\natexlab{}.
\newblock \showarticletitle{Stereonet: Guided hierarchical refinement for
  real-time edge-aware depth prediction}. In
  \bibinfo{booktitle}{\emph{Proceedings of the European Conference on Computer
  Vision}}. \bibinfo{pages}{573--590}.
\newblock


\bibitem[\protect\citeauthoryear{Kim, Lee, and Lee}{Kim et~al\mbox{.}}{2016}]%
        {vdsr}
\bibfield{author}{\bibinfo{person}{Jiwon Kim}, \bibinfo{person}{Jung~Kwon Lee},
  {and} \bibinfo{person}{Kyoung~Mu Lee}.} \bibinfo{year}{2016}\natexlab{}.
\newblock \showarticletitle{Accurate Image Super-Resolution Using Very Deep
  Convolutional Networks}. In \bibinfo{booktitle}{\emph{Computer Vision and
  Pattern Recognition}}. \bibinfo{pages}{1646--1654}.
\newblock


\bibitem[\protect\citeauthoryear{Kingma and Ba}{Kingma and Ba}{2014}]%
        {adam}
\bibfield{author}{\bibinfo{person}{Diederik~P Kingma} {and}
  \bibinfo{person}{Jimmy Ba}.} \bibinfo{year}{2014}\natexlab{}.
\newblock \showarticletitle{Adam: A Method for Stochastic Optimization}.
\newblock \bibinfo{journal}{\emph{arXiv: Learning}} (\bibinfo{year}{2014}).
\newblock


\bibitem[\protect\citeauthoryear{Ledig, Theis, Huszár, Caballero, Cunningham,
  Acosta, Aitken, Tejani, Totz, Wang, and Shi}{Ledig et~al\mbox{.}}{2017}]%
        {srgan}
\bibfield{author}{\bibinfo{person}{C. Ledig}, \bibinfo{person}{L. Theis},
  \bibinfo{person}{F. Huszár}, \bibinfo{person}{J. Caballero},
  \bibinfo{person}{A. Cunningham}, \bibinfo{person}{A. Acosta},
  \bibinfo{person}{A. Aitken}, \bibinfo{person}{A. Tejani}, \bibinfo{person}{J.
  Totz}, \bibinfo{person}{Z. Wang}, {and} \bibinfo{person}{W. Shi}.}
  \bibinfo{year}{2017}\natexlab{}.
\newblock \showarticletitle{Photo-Realistic Single Image Super-Resolution Using
  a Generative Adversarial Network}. In \bibinfo{booktitle}{\emph{2017 IEEE
  Conference on Computer Vision and Pattern Recognition}}.
  \bibinfo{pages}{105--114}.
\newblock


\bibitem[\protect\citeauthoryear{{Li}, {Jiang}, {Lin}, and {Jiang}}{{Li}
  et~al\mbox{.}}{2019b}]%
        {sfa}
\bibfield{author}{\bibinfo{person}{D. {Li}}, \bibinfo{person}{T. {Jiang}},
  \bibinfo{person}{W. {Lin}}, {and} \bibinfo{person}{M. {Jiang}}.}
  \bibinfo{year}{2019}\natexlab{b}.
\newblock \showarticletitle{Which Has Better Visual Quality: The Clear Blue Sky
  or a Blurry Animal?}
\newblock \bibinfo{journal}{\emph{IEEE Transactions on Multimedia}}
  \bibinfo{volume}{21}, \bibinfo{number}{5} (\bibinfo{year}{2019}),
  \bibinfo{pages}{1221--1234}.
\newblock


\bibitem[\protect\citeauthoryear{{Li}, {Han}, and {Chang}}{{Li}
  et~al\mbox{.}}{2019a}]%
        {cyclopean}
\bibfield{author}{\bibinfo{person}{S. {Li}}, \bibinfo{person}{X. {Han}}, {and}
  \bibinfo{person}{Y. {Chang}}.} \bibinfo{year}{2019}\natexlab{a}.
\newblock \showarticletitle{Adaptive Cyclopean Image-Based Stereoscopic
  Image-Quality Assessment Using Ensemble Learning}.
\newblock \bibinfo{journal}{\emph{IEEE Transactions on Multimedia}}
  \bibinfo{volume}{21}, \bibinfo{number}{10} (\bibinfo{year}{2019}),
  \bibinfo{pages}{2616--2624}.
\newblock


\bibitem[\protect\citeauthoryear{{Li}, {Yang}, {Liu}, {Yang}, {Jeon}, and
  {Wu}}{{Li} et~al\mbox{.}}{2019c}]%
        {srfbn}
\bibfield{author}{\bibinfo{person}{Z. {Li}}, \bibinfo{person}{J. {Yang}},
  \bibinfo{person}{Z. {Liu}}, \bibinfo{person}{X. {Yang}}, \bibinfo{person}{G.
  {Jeon}}, {and} \bibinfo{person}{W. {Wu}}.} \bibinfo{year}{2019}\natexlab{c}.
\newblock \showarticletitle{Feedback Network for Image Super-Resolution}. In
  \bibinfo{booktitle}{\emph{2019 IEEE/CVF Conference on Computer Vision and
  Pattern Recognition}}. \bibinfo{pages}{3862--3871}.
\newblock


\bibitem[\protect\citeauthoryear{Lim, Son, Kim, Nah, and Lee}{Lim
  et~al\mbox{.}}{2017}]%
        {edsr}
\bibfield{author}{\bibinfo{person}{Bee Lim}, \bibinfo{person}{Sanghyun Son},
  \bibinfo{person}{Heewon Kim}, \bibinfo{person}{Seungjun Nah}, {and}
  \bibinfo{person}{Kyoung~Mu Lee}.} \bibinfo{year}{2017}\natexlab{}.
\newblock \showarticletitle{Enhanced deep residual networks for single image
  super-resolution}. In \bibinfo{booktitle}{\emph{The IEEE conference on
  computer vision and pattern recognition workshops}},
  Vol.~\bibinfo{volume}{1}. \bibinfo{pages}{4}.
\newblock


\bibitem[\protect\citeauthoryear{Liu, Liu, Huang, and Bovik}{Liu
  et~al\mbox{.}}{2014}]%
        {sseq}
\bibfield{author}{\bibinfo{person}{Lixiong Liu}, \bibinfo{person}{Bao Liu},
  \bibinfo{person}{Hua Huang}, {and} \bibinfo{person}{Alan~Conrad Bovik}.}
  \bibinfo{year}{2014}\natexlab{}.
\newblock \showarticletitle{No-reference image quality assessment based on
  spatial and spectral entropies}.
\newblock \bibinfo{journal}{\emph{Signal Processing: Image Communication}}
  (\bibinfo{year}{2014}).
\newblock


\bibitem[\protect\citeauthoryear{{Liu}, {Van De Weijer}, and {Bagdanov}}{{Liu}
  et~al\mbox{.}}{2017}]%
        {rankiqa}
\bibfield{author}{\bibinfo{person}{X. {Liu}}, \bibinfo{person}{J. {Van De
  Weijer}}, {and} \bibinfo{person}{A.~D. {Bagdanov}}.}
  \bibinfo{year}{2017}\natexlab{}.
\newblock \showarticletitle{RankIQA: Learning from Rankings for No-Reference
  Image Quality Assessment}. In \bibinfo{booktitle}{\emph{IEEE International
  Conference on Computer Vision}}. \bibinfo{pages}{1040--1049}.
\newblock


\bibitem[\protect\citeauthoryear{Ma, Rao, Cheng, Chen, Lu, and Zhou}{Ma
  et~al\mbox{.}}{2020}]%
        {spsr}
\bibfield{author}{\bibinfo{person}{Cheng Ma}, \bibinfo{person}{Yongming Rao},
  \bibinfo{person}{Yean Cheng}, \bibinfo{person}{Ce Chen},
  \bibinfo{person}{Jiwen Lu}, {and} \bibinfo{person}{Jie Zhou}.}
  \bibinfo{year}{2020}\natexlab{}.
\newblock \showarticletitle{Structure-Preserving Super Resolution with Gradient
  Guidance}. In \bibinfo{booktitle}{\emph{Proceedings of the IEEE Conference on
  Computer Vision and Pattern Recognition}}.
\newblock


\bibitem[\protect\citeauthoryear{Ma, Yang, Yang, and Yang}{Ma
  et~al\mbox{.}}{2017}]%
        {maetal}
\bibfield{author}{\bibinfo{person}{Chao Ma}, \bibinfo{person}{Chih~Yuan Yang},
  \bibinfo{person}{Xiaokang Yang}, {and} \bibinfo{person}{Ming~Hsuan Yang}.}
  \bibinfo{year}{2017}\natexlab{}.
\newblock \showarticletitle{Learning a No-Reference Quality Metric for
  Single-Image Super-Resolution}.
\newblock \bibinfo{journal}{\emph{Computer Vision \& Image Understanding}}
  \bibinfo{volume}{158} (\bibinfo{year}{2017}), \bibinfo{pages}{1--16}.
\newblock


\bibitem[\protect\citeauthoryear{Mart\'{\i}n, Ashish, Paul, Eugene, Zhifeng,
  Craig, Greg, Andy, Jeffrey, Matthieu, Sanjay, Ian, Andrew, Geoffrey, Michael,
  Yangqing, Rafal, Lukasz, Manjunath, Josh, Dandelion, Rajat, Sherry, Derek,
  Chris, Mike, Jonathon, Benoit, Ilya, Kunal, Paul, Vincent, Vijay, Fernanda,
  Oriol, Pete, Martin, Martin, Yuan, and Xiaoqiang}{Mart\'{\i}n
  et~al\mbox{.}}{2015}]%
        {tensorflow}
\bibfield{author}{\bibinfo{person}{Abadi Mart\'{\i}n}, \bibinfo{person}{Agarwal
  Ashish}, \bibinfo{person}{Barham Paul}, \bibinfo{person}{Brevdo Eugene},
  \bibinfo{person}{Chen Zhifeng}, \bibinfo{person}{Citro Craig},
  \bibinfo{person}{S.~Corrado Greg}, \bibinfo{person}{Davis Andy},
  \bibinfo{person}{Dean Jeffrey}, \bibinfo{person}{Devin Matthieu},
  \bibinfo{person}{Ghemawat Sanjay}, \bibinfo{person}{Goodfellow Ian},
  \bibinfo{person}{Harp Andrew}, \bibinfo{person}{Irving Geoffrey},
  \bibinfo{person}{Isard Michael}, \bibinfo{person}{Jia Yangqing},
  \bibinfo{person}{Jozefowicz Rafal}, \bibinfo{person}{Kaiser Lukasz},
  \bibinfo{person}{Kudlur Manjunath}, \bibinfo{person}{Levenberg Josh},
  \bibinfo{person}{Man\'{e} Dandelion}, \bibinfo{person}{Monga Rajat},
  \bibinfo{person}{Moore Sherry}, \bibinfo{person}{Murray Derek},
  \bibinfo{person}{Olah Chris}, \bibinfo{person}{Schuster Mike},
  \bibinfo{person}{Shlens Jonathon}, \bibinfo{person}{Steiner Benoit},
  \bibinfo{person}{Sutskever Ilya}, \bibinfo{person}{Talwar Kunal},
  \bibinfo{person}{Tucker Paul}, \bibinfo{person}{Vanhoucke Vincent},
  \bibinfo{person}{Vasudevan Vijay}, \bibinfo{person}{Vi\'{e}gas Fernanda},
  \bibinfo{person}{Vinyals Oriol}, \bibinfo{person}{Warden Pete},
  \bibinfo{person}{Wattenberg Martin}, \bibinfo{person}{Wicke Martin},
  \bibinfo{person}{Yu Yuan}, {and} \bibinfo{person}{Zheng Xiaoqiang}.}
  \bibinfo{year}{2015}\natexlab{}.
\newblock \bibinfo{title}{TensorFlow: Large-Scale Machine Learning on
  Heterogeneous Systems}.
\newblock
\newblock


\bibitem[\protect\citeauthoryear{Mayer, Ilg, H{\"a}usser, Fischer, Cremers,
  Dosovitskiy, and Brox}{Mayer et~al\mbox{.}}{2016}]%
        {sceneflow}
\bibfield{author}{\bibinfo{person}{N. Mayer}, \bibinfo{person}{E. Ilg},
  \bibinfo{person}{P. H{\"a}usser}, \bibinfo{person}{P. Fischer},
  \bibinfo{person}{D. Cremers}, \bibinfo{person}{A. Dosovitskiy}, {and}
  \bibinfo{person}{T. Brox}.} \bibinfo{year}{2016}\natexlab{}.
\newblock \showarticletitle{A Large Dataset to Train Convolutional Networks for
  Disparity, Optical Flow, and Scene Flow Estimation}. In
  \bibinfo{booktitle}{\emph{IEEE International Conference on Computer Vision
  and Pattern Recognition}}.
\newblock
\urldef\tempurl%
\url{http://lmb.informatik.uni-freiburg.de/Publications/2016/MIFDB16}
\showURL{%
\tempurl}
\newblock
\shownote{arXiv:1512.02134.}


\bibitem[\protect\citeauthoryear{{Mayer}, Ilg, Husser, Fischer, Cremers,
  Dosovitskiy, and Brox}{{Mayer} et~al\mbox{.}}{2016}]%
        {kt15}
\bibfield{author}{\bibinfo{person}{N. {Mayer}}, \bibinfo{person}{E. Ilg},
  \bibinfo{person}{P. Husser}, \bibinfo{person}{P. Fischer},
  \bibinfo{person}{D. Cremers}, \bibinfo{person}{A. Dosovitskiy}, {and}
  \bibinfo{person}{T. Brox}.} \bibinfo{year}{2016}\natexlab{}.
\newblock \showarticletitle{A Large Dataset to Train Convolutional Networks for
  Disparity, Optical Flow, and Scene Flow Estimation}. In
  \bibinfo{booktitle}{\emph{2016 IEEE Conference on Computer Vision and Pattern
  Recognition}}. \bibinfo{pages}{4040--4048}.
\newblock


\bibitem[\protect\citeauthoryear{{Mittal}, {Soundararajan}, and
  {Bovik}}{{Mittal} et~al\mbox{.}}{2013}]%
        {niqe}
\bibfield{author}{\bibinfo{person}{A. {Mittal}}, \bibinfo{person}{R.
  {Soundararajan}}, {and} \bibinfo{person}{A.~C. {Bovik}}.}
  \bibinfo{year}{2013}\natexlab{}.
\newblock \showarticletitle{Making a “Completely Blind” Image Quality
  Analyzer}.
\newblock \bibinfo{journal}{\emph{IEEE Signal Processing Letters}}
  \bibinfo{volume}{20}, \bibinfo{number}{3} (\bibinfo{year}{2013}),
  \bibinfo{pages}{209--212}.
\newblock


\bibitem[\protect\citeauthoryear{Moorthy, Su, Mittal, and Bovik}{Moorthy
  et~al\mbox{.}}{2013}]%
        {live1}
\bibfield{author}{\bibinfo{person}{Anush~Krishna Moorthy},
  \bibinfo{person}{Che~Chun Su}, \bibinfo{person}{Anish Mittal}, {and}
  \bibinfo{person}{Alan~Conrad Bovik}.} \bibinfo{year}{2013}\natexlab{}.
\newblock \showarticletitle{Subjective evaluation of stereoscopic image
  quality}.
\newblock \bibinfo{journal}{\emph{Signal Processing Image Communication}}
  \bibinfo{volume}{28}, \bibinfo{number}{8} (\bibinfo{year}{2013}),
  \bibinfo{pages}{870--883}.
\newblock


\bibitem[\protect\citeauthoryear{Peris, Martull, Maki, Ohkawa, and Fukui}{Peris
  et~al\mbox{.}}{2012}]%
        {tsukuba}
\bibfield{author}{\bibinfo{person}{M. Peris}, \bibinfo{person}{S. Martull},
  \bibinfo{person}{A. Maki}, \bibinfo{person}{Y. Ohkawa}, {and}
  \bibinfo{person}{K. Fukui}.} \bibinfo{year}{2012}\natexlab{}.
\newblock \showarticletitle{Towards a simulation driven stereo vision system}.
  In \bibinfo{booktitle}{\emph{Proceedings of the 21st International Conference
  on Pattern Recognition}}. \bibinfo{pages}{1038--1042}.
\newblock


\bibitem[\protect\citeauthoryear{Shao, Lin, Gu, Jiang, and Srikanthan}{Shao
  et~al\mbox{.}}{2013}]%
        {nbu}
\bibfield{author}{\bibinfo{person}{Feng Shao}, \bibinfo{person}{Weisi Lin},
  \bibinfo{person}{Shanbo Gu}, \bibinfo{person}{Gangyi Jiang}, {and}
  \bibinfo{person}{Thambipillai Srikanthan}.} \bibinfo{year}{2013}\natexlab{}.
\newblock \showarticletitle{Perceptual Full-Reference Quality Assessment of
  Stereoscopic Images by Considering Binocular Visual Characteristics}.
\newblock \bibinfo{journal}{\emph{IEEE Transactions on Image Processing A
  Publication of the IEEE Signal Processing Society}} \bibinfo{volume}{22},
  \bibinfo{number}{5} (\bibinfo{year}{2013}), \bibinfo{pages}{1940--1953}.
\newblock


\bibitem[\protect\citeauthoryear{Sheikh and Bovik}{Sheikh and Bovik}{2006}]%
        {vif}
\bibfield{author}{\bibinfo{person}{H.~R. Sheikh} {and} \bibinfo{person}{A.~C.
  Bovik}.} \bibinfo{year}{2006}\natexlab{}.
\newblock \showarticletitle{Image information and visual quality}.
\newblock \bibinfo{journal}{\emph{IEEE Transactions on Image Processing}}
  \bibinfo{volume}{15}, \bibinfo{number}{2} (\bibinfo{year}{2006}),
  \bibinfo{pages}{430--444}.
\newblock


\bibitem[\protect\citeauthoryear{Sheikh, Member, IEEE, Bovik, and
  Fellow}{Sheikh et~al\mbox{.}}{2006}]%
        {ifc}
\bibfield{author}{\bibinfo{person}{Hamid~Rahim Sheikh},
  \bibinfo{person}{Member}, \bibinfo{person}{IEEE},
  \bibinfo{person}{Alan~Conrad Bovik}, {and} \bibinfo{person}{Fellow}.}
  \bibinfo{year}{2006}\natexlab{}.
\newblock \showarticletitle{An Information Fidelity Criterion for Image Quality
  Assessment Using Natural Scene Statistics}.
\newblock \bibinfo{journal}{\emph{IEEE Transactions on Image Processing}}
  \bibinfo{volume}{14}, \bibinfo{number}{12} (\bibinfo{year}{2006}),
  \bibinfo{pages}{2117--2128}.
\newblock


\bibitem[\protect\citeauthoryear{Simonyan and Zisserman}{Simonyan and
  Zisserman}{2014}]%
        {vgg19}
\bibfield{author}{\bibinfo{person}{Karen Simonyan} {and}
  \bibinfo{person}{Andrew Zisserman}.} \bibinfo{year}{2014}\natexlab{}.
\newblock \showarticletitle{Very Deep Convolutional Networks for Large-Scale
  Image Recognition}.
\newblock \bibinfo{journal}{\emph{Computer Science}} (\bibinfo{year}{2014}).
\newblock


\bibitem[\protect\citeauthoryear{Song, Ko, and Kuo}{Song et~al\mbox{.}}{2014}]%
        {mcl}
\bibfield{author}{\bibinfo{person}{Rui Song}, \bibinfo{person}{Hyunsuk Ko},
  {and} \bibinfo{person}{C.~C.~Jay Kuo}.} \bibinfo{year}{2014}\natexlab{}.
\newblock \showarticletitle{MCL-3D: a database for stereoscopic image quality
  assessment using 2D-image-plus-depth source}.
\newblock \bibinfo{journal}{\emph{Journal of Information ence \& Engineering}}
  \bibinfo{volume}{31}, \bibinfo{number}{5} (\bibinfo{year}{2014}),
  \bibinfo{pages}{1593--1611}.
\newblock


\bibitem[\protect\citeauthoryear{Tao, Zhu, Pan, Wen, and Meng}{Tao
  et~al\mbox{.}}{2015}]%
        {brisque}
\bibfield{author}{\bibinfo{person}{Sun Tao}, \bibinfo{person}{Xingjie Zhu},
  \bibinfo{person}{J.~S. Pan}, \bibinfo{person}{Jiajun Wen}, {and}
  \bibinfo{person}{Fanqiang Meng}.} \bibinfo{year}{2015}\natexlab{}.
\newblock \showarticletitle{No-Reference Image Quality Assessment in Spatial
  Domain}.
\newblock \bibinfo{journal}{\emph{Advances in Intelligent Systems \&
  Computing}}  \bibinfo{volume}{329} (\bibinfo{year}{2015}),
  \bibinfo{pages}{381--388}.
\newblock


\bibitem[\protect\citeauthoryear{Wang, Wang, Liang, Lin, Yang, An, and
  Guo}{Wang et~al\mbox{.}}{2019}]%
        {passr}
\bibfield{author}{\bibinfo{person}{Longguang Wang}, \bibinfo{person}{Yingqian
  Wang}, \bibinfo{person}{Zhengfa Liang}, \bibinfo{person}{Zaiping Lin},
  \bibinfo{person}{Jungang Yang}, \bibinfo{person}{Wei An}, {and}
  \bibinfo{person}{Yulan Guo}.} \bibinfo{year}{2019}\natexlab{}.
\newblock \showarticletitle{Learning parallax attention for stereo image
  super-resolution}. In \bibinfo{booktitle}{\emph{Proceedings of the IEEE
  Conference on Computer Vision and Pattern Recognition}}.
  \bibinfo{pages}{12250--12259}.
\newblock


\bibitem[\protect\citeauthoryear{Wang, Yu, Wu, Gu, Liu, Dong, Loy, Qiao, and
  Tang}{Wang et~al\mbox{.}}{2018}]%
        {esrgan}
\bibfield{author}{\bibinfo{person}{Xintao Wang}, \bibinfo{person}{Ke Yu},
  \bibinfo{person}{Shixiang Wu}, \bibinfo{person}{Jinjin Gu},
  \bibinfo{person}{Yihao Liu}, \bibinfo{person}{Chao Dong},
  \bibinfo{person}{Chen~Change Loy}, \bibinfo{person}{Yu Qiao}, {and}
  \bibinfo{person}{Xiaoou Tang}.} \bibinfo{year}{2018}\natexlab{}.
\newblock \showarticletitle{{ESRGAN:} Enhanced Super-Resolution Generative
  Adversarial Networks}.
\newblock \bibinfo{journal}{\emph{CoRR}}  \bibinfo{volume}{abs/1809.00219}
  (\bibinfo{year}{2018}).
\newblock


\bibitem[\protect\citeauthoryear{Wei, Zhou, Zhibo, Chen, and Weiping}{Wei
  et~al\mbox{.}}{2019}]%
        {stereoqa}
\bibfield{author}{\bibinfo{person}{Wei}, \bibinfo{person}{Zhou},
  \bibinfo{person}{Zhibo}, \bibinfo{person}{Chen}, {and}
  \bibinfo{person}{Weiping}.} \bibinfo{year}{2019}\natexlab{}.
\newblock \showarticletitle{Dual-Stream Interactive Networks for No-Reference
  Stereoscopic Image Quality Assessment.}
\newblock \bibinfo{journal}{\emph{IEEE transactions on image processing : a
  publication of the IEEE Signal Processing Society}} (\bibinfo{year}{2019}).
\newblock


\bibitem[\protect\citeauthoryear{Wu, Ma, Liang, Dong, Shi, and Lin}{Wu
  et~al\mbox{.}}{2020}]%
        {tip20}
\bibfield{author}{\bibinfo{person}{J. Wu}, \bibinfo{person}{J. Ma},
  \bibinfo{person}{F. Liang}, \bibinfo{person}{W. Dong}, \bibinfo{person}{G.
  Shi}, {and} \bibinfo{person}{W. Lin}.} \bibinfo{year}{2020}\natexlab{}.
\newblock \showarticletitle{End-to-End Blind Image Quality Prediction With
  Cascaded Deep Neural Network}.
\newblock \bibinfo{journal}{\emph{IEEE Transactions on Image Processing}}
  \bibinfo{volume}{29} (\bibinfo{year}{2020}), \bibinfo{pages}{7414--7426}.
\newblock


\bibitem[\protect\citeauthoryear{{Yan}, {Bare}, {Ma}, {Li}, and {Tan}}{{Yan}
  et~al\mbox{.}}{2019}]%
        {sisrqa}
\bibfield{author}{\bibinfo{person}{B. {Yan}}, \bibinfo{person}{B. {Bare}},
  \bibinfo{person}{C. {Ma}}, \bibinfo{person}{K. {Li}}, {and}
  \bibinfo{person}{W. {Tan}}.} \bibinfo{year}{2019}\natexlab{}.
\newblock \showarticletitle{Deep Objective Quality Assessment Driven Single
  Image Super-Resolution}.
\newblock \bibinfo{journal}{\emph{IEEE Transactions on Multimedia}}
  \bibinfo{volume}{21}, \bibinfo{number}{11} (\bibinfo{year}{2019}),
  \bibinfo{pages}{2957--2971}.
\newblock


\bibitem[\protect\citeauthoryear{{Yan}, {Ma}, {Bare}, {Tan}, and {Hoi}}{{Yan}
  et~al\mbox{.}}{2020}]%
        {our}
\bibfield{author}{\bibinfo{person}{B. {Yan}}, \bibinfo{person}{C. {Ma}},
  \bibinfo{person}{B. {Bare}}, \bibinfo{person}{W. {Tan}}, {and}
  \bibinfo{person}{S.~C.~H. {Hoi}}.} \bibinfo{year}{2020}\natexlab{}.
\newblock \showarticletitle{Disparity-Aware Domain Adaptation in Stereo Image
  Restoration}. In \bibinfo{booktitle}{\emph{IEEE/CVF Conference on Computer
  Vision and Pattern Recognition}}. \bibinfo{pages}{13176--13184}.
\newblock


\bibitem[\protect\citeauthoryear{Zhang, Zuo, and Zhang}{Zhang
  et~al\mbox{.}}{2018}]%
        {srmd}
\bibfield{author}{\bibinfo{person}{Kai Zhang}, \bibinfo{person}{Wangmeng Zuo},
  {and} \bibinfo{person}{Lei Zhang}.} \bibinfo{year}{2018}\natexlab{}.
\newblock \showarticletitle{Learning a single convolutional super-resolution
  network for multiple degradations}.
\newblock \bibinfo{journal}{\emph{IEEE Conference on Computer Vision and
  Pattern Recognition}}, \bibinfo{pages}{3262--3271}.
\newblock


\bibitem[\protect\citeauthoryear{{Zhou Wang}, {Bovik}, {Sheikh}, and
  {Simoncelli}}{{Zhou Wang} et~al\mbox{.}}{2004}]%
        {ssim}
\bibfield{author}{\bibinfo{person}{{Zhou Wang}}, \bibinfo{person}{A.~C.
  {Bovik}}, \bibinfo{person}{H.~R. {Sheikh}}, {and} \bibinfo{person}{E.~P.
  {Simoncelli}}.} \bibinfo{year}{2004}\natexlab{}.
\newblock \showarticletitle{Image quality assessment: from error visibility to
  structural similarity}.
\newblock \bibinfo{journal}{\emph{IEEE Transactions on Image Processing}}
  \bibinfo{volume}{13}, \bibinfo{number}{4} (\bibinfo{year}{2004}),
  \bibinfo{pages}{600--612}.
\newblock


\end{thebibliography}

\end{document}